\documentclass[sigconf,letterpaper]{acmart}

\setcopyright{none}
\renewcommand\footnotetextcopyrightpermission[1]{} %

\usepackage{gensymb}
\usepackage{booktabs}  %
\usepackage{pbox}
\usepackage{subfigure}
\usepackage{epsfig,endnotes}
\usepackage{epstopdf}
\usepackage{color}
\usepackage{graphicx}
\usepackage{xspace}
\usepackage{flushend}  %
\usepackage{enumitem}
\usepackage{multirow}
\usepackage{tikz}
\usepackage[ruled,vlined,linesnumbered]{algorithm2e}
\usepackage{amsmath}

\usepackage{url}

\usepackage{breakurl}
\usepackage{hyperref}

\usepackage{xurl}
\usepackage{xcolor,colortbl}

\usepackage{filecontents}
\usepackage{soul}
\usepackage{xcolor}

\usepackage{cleveref}
\crefname{section}{§}{§§}
\Crefname{section}{§}{§§}

\usepackage{amsmath}

\usepackage{diagbox}

\newif\ifcomment
\commenttrue

\newif\ifwatermark
\watermarkfalse

\ifwatermark
      \usepackage{draftwatermark}
      \SetWatermarkScale{.7}
      \SetWatermarkText{\shortstack{In Submission:\\Do Not Distribute}}
      \SetWatermarkColor[rgb]{1,0.88,0.88}
\fi

\ifcomment
    \newcounter{MVNumberOfComments}
    \stepcounter{MVNumberOfComments}
    \newcommand{\mvnote}[1]{\textcolor{violet}{\small \bf [MV\#\arabic{MVNumberOfComments}\stepcounter{MVNumberOfComments}: #1]}}
    
    \newcounter{YXNumberOfComments}
    \stepcounter{YXNumberOfComments}
    \newcommand{\yxnote}[1]{\textcolor{orange}{\small \bf [YX\#\arabic{YXNumberOfComments}\stepcounter{YXNumberOfComments}: #1]}}

    \newcounter{AKNumberOfComments}
    \stepcounter{AKNumberOfComments}
    \newcommand{\aknote}[1]{\textcolor{red}{\small \bf [AK\#\arabic{AKNumberOfComments}\stepcounter{AKNumberOfComments}: #1]}}  
    
    \newcounter{CWNumberOfComments}
    \stepcounter{CWNumberOfComments}
    \newcommand{\cwnote}[1]{\textcolor{brown}{\small \bf [CW\#\arabic{CWNumberOfComments}\stepcounter{CWNumberOfComments}: #1]}} 
    
    \newcommand{\del}[1]{{{\color{red}\st{#1}}}}
\else
    \newcommand\mvnote[1]{} 
    \newcommand\yxnote[1]{}    
    \newcommand\aknote[1]{} 
    \newcommand\cwnote[1]{} 

    \newcommand{\del}[1]{}
\fi

\newcommand{\eg}{{e.g.,}\xspace}
\newcommand{\ie}{{\it i.e.,}\xspace}

\smallskip
\smallskip

\newcommand{\pirdns}{{PDNS}\xspace}
\newcommand{\RR}{{ReR}\xspace}
\newcommand{\RRs}{{ReRs}\xspace}
\newcommand{\pRR}{{\pirdns \xspace \RR}\xspace}
\newcommand{\pRRs}{{\pirdns \xspace \RRs}\xspace}
\newcommand{\row}{{slot}\xspace}
\newcommand{\rows}{{slots}\xspace}

\usepackage{calc}
\newlength\myheight
\newlength\mydepth
\settototalheight\myheight{Xygp}
\definecolor{ao(english)}{rgb}{0.0, 0.5, 0.0}

\begin{document}
\fancyhead{}

\date{}

\title{Collusion Resistant DNS With Private Information Retrieval}
\author{
Yunming Xiao$^{1}$, Peizhi Liu$^{2}$, Ruijie Yu$^{2}$, Chenkai Weng$^{3}$, Matteo Varvello$^{4}$, Aleksandar Kuzmanovic$^{2}$\\
$^{1}$ The Chinese University of Hong Kong, Shenzhen \quad
$^{2}$ Northwestern University\\
$^{3}$ Arizona State University \quad
$^{4}$ Nokia Bell Labs
}

\begin{abstract}
There has been a growing interest in Internet user privacy, demonstrated by the popularity of privacy-preserving products such as Telegram and Brave, and the widespread adoption of HTTPS. The Domain Name System (DNS) is a key component of Internet-based communication and its privacy has been neglected for years. Recently, DNS over HTTPS (DoH) has improved the situation by fixing the issue of in-path middleboxes. Further progress has been made with proxy-based solutions such as Oblivious DoH (ODoH), which separate a user's identity from their DNS queries. However, these solutions rely on non-collusion assumptions between DNS resolvers and proxies -- an assumption difficult to guarantee in practice. To address this, we explore integrating single-server Private Information Retrieval (PIR) into DNS to enable encrypted query processing without relying on trust assumptions. However, applying PIR to DNS is challenging due to its hierarchical nature -- particularly, interactions with recursive resolvers can still leak information. Navigating performance and privacy trade-offs, we propose PDNS, a DNS extension leveraging single-server PIR to strengthen privacy guarantees. We have implemented a prototype of PDNS and compared its performance against state-of-the-art solutions via trace-driven experiments. The results show that PDNS achieves acceptable performance (2x faster than DoH over Tor with similar privacy guarantees) and strong privacy guarantees today, mainly at the cost of its scalability, which specialized hardware for PIR can address in the near future. 
\end{abstract}

\maketitle

\section{Introduction}
\label{sec:intro}

The Domain Name System (DNS) is the phonebook of the Internet~\cite{cloudflare_blog} which maps IP addresses like ``151.101.195.5'' to human-friendly names like ``cnn.com''. At the birth of the Web, security and privacy were not contemplated, leaving DNS traffic as plaintext. This means that \textit{any} middlebox placed between a DNS client and  \textit{recursive resolver} (\RR) could monitor user activity, potentially building accurate user profiles~\cite{DBLP:conf/pet/GuhaF07}. Twenty-eight years later, DNS-over-TLS (DoT)~\cite{rfc_dot} and DNS-over-HTTPS (DoH)~\cite{rfc_doh} solve this limitation by means of end-to-end encryption. DoT and DoH have been gradually supported both by clients (\eg browsers like  Chromium~\cite{chrome_doh} and Firefox~\cite{firefox_doh}) and \RRs~\cite{google_public_doh, cloudflare_dns}. 

End-to-end encryption protects a user's privacy from eavesdroppers but not from a \RR. ODNS~\cite{schmitt2019oblivious} is a recent solution -- a variant of which, ODoH~\cite{odoh}, has already deployed by Cloudflare -- to address such problem by detaching a user identity from a DNS request. This is achieved by adding a proxy between DNS client and \RR such that: $(i)$ the proxy is blind with respect to an encrypted DNS query, $(ii)$ the \RR is blind with respect to the client's identity (IP address). Assuming a non-colluding proxy and \RR (\ie the proxy and resolver do not cooperate to share information), user privacy is enforced. However, non-collusion is hard to enforce and verify in reality. For example,  both proxy and \RR can be subjects of a subpoena, at which point privacy is again sacrificed. Or, in the case of proxies and \RRs located in different countries, there may be insufficient legal bounds to prevent collusion. Adopting a complex proxy setup, such as using the Tor network, can address the non-collusion issue, but it comes with a significant performance trade-off. 

An alternative approach to protect users from the above privacy infringement would be either removing the \RRs from DNS~\cite{no-recursive-resolver-14},  or having \RRs operating \textit{in the blind}, \ie by resolving domains without knowing what they are. The former option exhibits high performance penalties to users, amplifies workloads on the authoritative name servers (ANSes), and raises additional security concerns. The latter option seems counter-intuitive, but in reality, several techniques exist which allow similar operations. These techniques fall in the branch of Private Information Retrieval (PIR), which is achieved by various cryptographic tools such as homomorphic encryption~\cite{gentryfhe,bgvfhe,Fan_somewhatpractical,regevfhe}. Indeed, private DNS is often cited as a motivating example in PIR research, but no practical implementation currently exists. 

The goal of this work is to explore the feasibility of integrating PIR into DNS and fill the gap between PIR and DNS research. We do so by introducing \pirdns, a Privacy-Preserving DNS designed to \textit{augment} rather than replace DNS, in a spirit similar to DoH and ODoH. To achieve our vision, we had to solve the following challenges.

\vspace{0.05in}
\noindent \textbf{Cache population}: PIR protocols assume that a database (or cache in DNS context) is either given or can be privately populated. This is not the case for DNS where the \RR is responsible for populating its cache based on the user request. Clearly, a \textit{blind} \RR cannot perform such operation which should be tackled by the client instead. Meanwhile, the client cannot update the \RR cache either, or it would invalidate the system privacy. 
To address this issue, we propose EDNS-PR, a custom EDNS(0)~\cite{rfc_edns} extension which allows a client to communicate the IP address of its \RR in presence of cache misses, so that an ANS can privately populate the \RR's cache. 

Concerns may arise regarding the privacy exposure for clients performing iterative DNS lookups. 
Indeed, \pirdns is not yet the final sanctuary. Essentially, \pirdns trades the privacy exposure at the \RR, or the \RR and the proxy combined in the case of ODoH, for \textit{partial} privacy exposure at the ANSes. We argue that this trade-off raises the level of privacy preservation because, with the current DNS system, the \RR alone or in collusion with proxies can collect all the information of every individual user. Distributing information exposure to millions of ANSes significantly increases the difficulty of gathering information about an individual user or group. The impact of privacy exposure to ANSes is further reduced by the fact that ANSes are likely already gaining information about users from users' subsequent traffic~\cite{d-dns-1-20}, such as HTTP(S) requests for webpages directed to domains within their region. Thus queries to ANSes add no \textit{additional} information than they already know. Moreover, the caching by \pRRs and our shortcut mechanism (see \cref{sec:cache-miss}) ensure that only a small portion of user activities reach the ANSes. Therefore, we consider \pirdns an overall privacy enhancement compared to existing solutions, providing resistance to \textit{realistic} collusion between two or several parties, \eg proxies and \RRs, but not among millions of ANSes.

\vspace{0.05in}
\noindent\textbf{Security challenges}: The previous construction imposes new security challenges for DNS. Attackers can either congest ANSes, or launch \textit{reflection} attacks to congest or poison the cache of \RRs. We leverage the security properties of Spiral with digital signatures to allow ANSes to \textit{validate} cache misses when needed, \ie when suspecting a potential attack. 

\vspace{0.05in}
\noindent \textbf{PIR selection and optimization}: Out of all the available PIR categories, we suggest utilizing the single-server stateless PIR schemes for DNS, as they do not require a non-collusion agreement, bear low costs for cache updates, and offer satisfactory running times for query processing. We benchmark multiple schemes and find that Spiral~\cite{spiral-pir} offers %
the fastest running time, the shortest query size, and high-quality open-source implementation. 
To integrate Spiral into DNS, we researched the optimal DNS cache configuration for PIR, and implemented performance enhancements leveraging multi-threading and low-level instruction support.
 
\vspace{0.05in}
We implement a proof-of-concept of PDNS, including a client and \RR, and extend the popular BIND9~\cite{bind-popular} to support EDNS-PR as our own extension of EDNS(0) %
at the ANS. In our experiments, \pirdns answer queries 2x faster than DoH over Tor -- a privacy-preserving anonymous network -- even on a \textit{large} cache (512MB, up to 13M DNS records). \pirdns is also faster than ODoH (208ms versus 272ms) with a \textit{small} cache (64MB, up to 1.6M DNS records), and adds 180ms with a large cache. %
We envision that the advent of specialized hardware for PIR would reduce \pirdns' query duration to 70ms (even on much larger caches), thus making its performance comparable with DoH. 

Such competitive performance and strong privacy guarantees do not come for free. \pirdns requires extra computational resources at a \RR to handle queries fast. Our benchmarking on an 8-core 3.0GHz AMD EPYC shows that a \pRR can only handle few queries per second, while DoH can handle hundreds of queries per second. This implies a higher deployment cost for an operator, which can be absorbed via a subscription model for privacy-oriented customers, as currently done by VPN providers. Indeed, Our analysis shows that a monthly subscription fee of \$5 per user is sufficient to make \pirdns financially viable today without accelerators (see Appendix~\ref{sec:appendix:deployment-cost}), a reasonable cost given that people may pay more for privacy-preserving network services such as VPN. %
This cost is likely to be significantly reduced along with substantial performance improvements in \pirdns with near-future hardware accelerators~\cite{f1,IntelHEXLFPGA}, making \pirdns competitive with DoH. Further, participating ANSes need to support DoH, which causes a significant bandwidth increase. Nevertheless, such DoH adoption is not only meant for \pirdns but also beneficial to current DNS, as it amends an existing user privacy violation~\cite{hal-adot-operational-considerations-02, encrypting-rr-auth}. DoH for ANSes has been proposed independently from \pirdns~\cite{ietf-dprive-opportunistic-adotq-02}.

One final question remains: \textit{what are the incentives for the adoption of \pirdns?} For users, the extra privacy provided justifies the minor performance penalty. For the \RR, the extra cost is justified by unprecedented privacy guarantees, which could be offered at a premium. Participating ANSes also have an incentive to support PDNS, as the additional traffic is offset by the increased privacy they can provide to their supporting web domains (which are either their customers or directly operated by them), a valuable asset for competing domains especially when offering sensitive content. 
Notably, \pirdns can be adopted incrementally, as detailed in \cref{sec:deep:deployment}. %

\section{Background and Motivation}
\label{sec:background}

\subsection{DNS and Privacy}
\label{sec:background:dns}
DNS clients send queries to a \RR,  either run by an ISP or by public providers such as Google~\cite{google_dns} and Cloudflare~\cite{cloudflare_dns}. The \RR uses a cache to speed up DNS queries; cache misses trigger \textit{iterative} DNS lookups to the ANSes for the root, top-level domain, and final zones (``root/TLD/final ANS'' for short), 
before returning an answer to the user while updating the \RR's cache. %
The DNS RFC~\cite{rfc_dns} specifies to send DNS queries either via UDP (DoUDP) or TCP (DoTCP). UDP was adopted in most cases because of better performance given due to its absence of connection handshake~\cite{kosek2022measuring}. 

The original DNS protocol does not use encryption, potentially exposing user privacy to in-network eavesdroppers. %
Recent IETF standards DoT~\cite{rfc_dot} and DoH~\cite{rfc_doh} require the client to establish an encrypted session with the \RR. While DoT/DoH protects user privacy from in-network eavesdroppers, the \RRs still have full visibility of the DNS queries from their users. This represents a considerable privacy breach, especially in presence of public \RR with massive user bases like Google and Cloudflare. Distributing the DNS queries to multiple resolvers helps to protect the user privacy to a certain extent~\cite{d-dns-1-20, d-dns-2-20, d-dns-3-23, hounsel2021encryption}; yet the best solution can only achieve k-anonymity, which is still vulnerable to various attacks~\cite{k-anonymity-vulnerable-07}.

Oblivious-DoH (ODoH)~\cite{odoh, rfc9230} introduces an \textit{oblivious} proxy between user and \RR. Assuming the \RR does not collude with the oblivious proxy, user identity and DNS queries are disjoint. However, non-collusion is hard to enforce and verify. 
To mitigate the non-collusion problem (we restrict our discussion of collusion in realistic scenarios, see \cref{sec:background:challenge}), it requires carefully distributing queries across multiple proxies or using a long chain of proxies~\cite{muODNS-23, rivera2020leveraging, dohot-muffett, dohot-cloudflare}. For instance, DoH over Tor (DoHoT) relays the encrypted DNS queries over Tor~\cite{dohot-muffett, dohot-cloudflare}, taking advantage of onion routing which builds up a chain of proxies and hence significantly increases the difficulty of collusion~\cite{castillo2017contributions}.\footnote{Note that DoUDP/DoTCP over Tor is less secure than DoHoT, as they are vulnerable to DNS pollution by a single entity -- the Tor network's exit node -- which could return malicious DNS responses, such as phishing sites. } However, DoHoT incurs significant delays, \eg 4x slower than DoH~\cite{odoh} and 2x slower than \pirdns (\cref{sec:eval:performance}), %
and is error-prone given its reliance on volunteers~\cite{patel2016tor, tor-code-audit-17}. It further faces additional usability issues. For instance, it is banned in certain regions due to censorship and/or association with dark net activities. As a result, users in those regions are unable to use such DNS privacy solutions. %
$\mu$ODNS~\cite{muODNS-23} adopts a similar multi-hop setting and %
assumes a dedicated and trusted proxy exists in the proxy chain. Nevertheless, this setup is costly and too complicated for ordinary Internet users. %

In~\cite{no-recursive-resolver-14}, the authors have recently proposed the provocative idea of eliminating all \RRs, thereby addressing DNS privacy issues -- including the non-collusion problem since it does not apply anymore -- associated with them. This approach has several concerning shortcomings. First and foremost, it introduces DNS ``flattening'';  DNS is hierarchical to provide \textit{speed}, as a closeby resolver would respond if it can, \textit{scalability}, as distributed caching avoids redundant queries, and \textit{reliability}, as distributed caching allows to cope with failures at ANSes. As discussed in~\cite{no-recursive-resolver-14}, such flattening would increase the overall DNS load by a few times, potentially becoming unsustainable for ANSes of popular or root zones~\cite{cache-effect-19}. Second, this approach is currently impractical as many ANSes, \eg Akamai~\cite{akamai-dns}, adopt complex rate limiting solutions to prevent traffic from non \RRs. Finally, ANSes currently do not support HTTPS (see \cref{sec:deep:discussion}) thus requiring~\cite{no-recursive-resolver-14} to rely on DoUDP. This implies that pervasive traffic monitoring is possible under such approach, which indeed deteriorates the overall DNS privacy. A more robust solution should involve implementing DoH across all ANSes (``\RR-Less DoH''), which would further increase the overall DNS load and still raise new security issues at ANSes.

\subsection{Private Information Retrieval}
\label{sec:background:pir}

PIR protocols~\cite{pir:CKGM, incremental-pir, corrigan2022single, checklist-pir, corrigan2020private, dpf-pir, seal-pir} are advanced cryptographic techniques which allow a client to fetch an item from a remote database, \eg the \RR cache in the DNS scenario, without letting the server know which item it is. %
At a high level, PIR protocols are divided into \textit{single-server} and \textit{multi-server}, referring to how many server-side components they rely on. Overall, multi-server PIR provides efficient data transmission between the user and the servers, and lightweight computation, but it requires at least two non-colluding servers~\cite{dpf-pir,corrigan2020private, incremental-pir,checklist-pir}. It also requires intensive synchronization between servers for the maintenance of identical databases. \textit{Single-server} PIR protocols only require one untrusted server but rely on heavier cryptographic operations~\cite{spiral-pir,simplepir,seal-pir}, which make them slower than multi-server PIR. Despite the latter, we argue that a robust privacy-preserving DNS should not rely on a non-collusion agreement, and thus discard multi-server PIR solutions.

PIR protocols can also be \textit{stateful} or \textit{stateless}. Stateful protocols~\cite{checklist-pir,simplepir,corrigan2020private} require the user to maintain a state, which contains information to generate PIR queries. The state is fetched from the PIR server and expires whenever the database is updated.  With stateless protocols~\cite{seal-pir,spiral-pir,dpf-pir}, the user does not store and update any database-related state except for the query keys. %
Given that a DNS cache changes frequently, stateless PIR should be preferred. Appendix~\ref{sec:appendix:pir} offers more details on single-server stateless PIR.

\newcommand{\No}{\cellcolor{red!25}No\xspace}
\newcommand{\Yes}{\cellcolor{green!25}Yes\xspace}
\newcommand{\NoYes}{\cellcolor{orange!25}No*\xspace}
\newcommand{\YesNo}{\cellcolor{orange!25}Yes*\xspace}
\newcommand{\NA}{\cellcolor[gray]{0.75}N/A\xspace}
\newcommand{\NoEmtpy}{\cellcolor{red!25}\xspace}
\newcommand{\YesEmtpy}{\cellcolor{green!25}\xspace}
\newcommand{\YesNoEmtpy}{\cellcolor{orange!25}\xspace}
\newcommand{\NAEmtpy}{\cellcolor[gray]{0.75}\xspace}
\begin{table}[t]
\footnotesize
\setlength\tabcolsep{3pt}
\caption{Privacy properties of various DNS solutions. }%
\label{tab:solutions}
\begin{tabular}{c | c c c}
\toprule
    \multirow{3}{*}{\diagbox[width=2.6cm]{Solution}{Property}}%
    & \multirow{3}{*}{\parbox{1.4cm}{\footnotesize \centering Defend Pervasive Monitoring}} 
    & \multirow{3}{*}{\parbox{1.3cm}{\footnotesize \centering Protect Individual Profile}} 
    & \multirow{3}{*}{\parbox{1.6cm}{\footnotesize \centering Survive Non-Collusion Violation}}  \\\\\\
\toprule
    
    DoUDP~\cite{rfc_dns} / DoTCP~\cite{rfc_dotcp} & \No & \No & \NA \\\hline
    DoT~\cite{rfc_dot} / DoH~\cite{rfc_doh} & \Yes & \No & \NA \\\hline
    
    \multirow{2}{*}{\parbox{2.6cm}{\centering DoT/DoH + Resolver Rotation~\cite{rivera2020leveraging, hounsel2021encryption}}} 
    & \YesEmtpy & \YesEmtpy & \NAEmtpy \\
    & \multirow{-2}{*}{\parbox{1.2cm}{\centering \Yes}} 
    & \multirow{-2}{*}{\parbox{1.0cm}{\centering \Yes}} 
    & \multirow{-2}{*}{\parbox{1.2cm}{\centering \NA}} \\\hline
    
    ODNS~\cite{schmitt2019oblivious} / ODoH~\cite{odoh} & \Yes & \Yes & \No \\\hline

    DoHoT~\cite{dohot-muffett, dohot-cloudflare} / $\mu$ODNS~\cite{muODNS-23}
    & \Yes 
    & \Yes 
    & \Yes \\\hline\midrule%

    \RR-Less + DoUDP~\cite{no-recursive-resolver-14} & \No & \No & \NA \\\hline
    \RR-Less + DoH/DoT & \Yes & \Yes & \NA \\\hline\midrule
    
    DNS + Multi-Server PIR & \Yes & \Yes & \No \\\hline
    \textbf{DNS + Single-Server PIR} & \textbf{\Yes} & \textbf{\Yes} & \textbf{\Yes} \\
    
\bottomrule
\end{tabular}
\vspace{-0.15in}
\end{table}

\subsection{Goals and Challenges}
\label{sec:background:challenge}
\noindent
\textbf{Goals} -- 
Table~\ref{tab:solutions} summarizes the privacy-preserving properties of state-of-the-art DNS solutions. Among the existing options, only DoHoT and the hypothetical \RR-Less DoH offer comprehensive privacy protections, %
including collusion resistance, which is the primary focus of this paper. However, it is noteworthy that DoHoT suffers from performance inefficiencies and may be vulnerable to %
usability constraints (\eg in certain countries/regions), while \RR-Less DoH compromises DNS security and reliability. Single-server PIR has the potential to deliver a new solution with the same comprehensive privacy protections to DNS users while outperforming DoHoT in performance and usability, and \RR-Less DoH in compatibility and practicality, but it comes with several extra challenges we discuss below.

\vspace{0.05in}
\noindent
\textbf{Challenge 1: cache population} -- PIR guarantees that the \RR cannot identify the queried domains. This also implies that PIR prevents a \RR from populating its cache, which invalidates its function. A strawman solution consists of bypassing PIR in presence of cache misses, \eg resorting to regular or ODoH as discussed in Appendix~\ref{sec:appendix:cachs-miss}. %
However, these solutions put user privacy at risk. We propose a slight DNS modification wherein clients directly resolve DNS cache misses, and final ANSes populate a \RR's cache (\cref{sec:cache-miss}).

\vspace{0.05in}
\noindent
\textbf{Challenge 2: compatibility with existing DNS} -- Our goal is to enhance DNS with comprehensive privacy, rather than a complete overhaul. Specifically, we envision a design change similar to DoH, with minimal modifications required for clients and servers (\RR and ANS). While this constrains the design space, it also reduces the barriers to adoption.

\vspace{0.05in}
\noindent
\textbf{Challenge 3: performance} --  The recent DNS evolution in the interest of user privacy %
has caused a slowdown in DNS queries. For instance, DoH requires at least three times the query time of DoUDP because of the handshakes to establish an encrypted channel. The handshake can be avoided if the HTTPS connection is re-used. However, this does not apply to DoH with proxy rotation which provides better privacy guarantees (see Table~\ref{tab:solutions}). In our measurements (see \cref{sec:eval:performance}), the median query duration for DoH is 69~ms, versus 25~ms for DoUDP, and it grows to about 272~ms for ODoH. Some previous studies~\cite{bottger2019empirical,odoh} report similar results while others~\cite{DBLP:conf/imc/ChhabraM0BW21, doh-reuse-verify} report much higher values depending on the user location and distance to the \RR.

The introduction of PIR in DNS brings further slowdowns due to its additional complexity. On the one hand, this is expected and understood by users as a trade-off for additional privacy, as commonly experienced in privacy tools like Tor or VPNs. On the other hand, a key challenge is to conceive a design that minimizes such overhead and achieves query times competitive with the state of the art. To address this challenge, we carefully select the PIR scheme, and modify its implementation where performance bottlenecks are detected. Further, we explore optimizations in DNS record storage and transmission.

\vspace{0.05in}
\noindent
\textbf{Non-goals and limitations} -- Our aim is to enhance DNS privacy in a practical manner, rather than eliminating all privacy leaks, as doing so may require a complete overhaul of DNS. It is also unattainable to claim full privacy given the ever-evolving nature of privacy demands and definitions. 
Importantly, this paper focuses exclusively on DNS privacy and does not address vulnerabilities associated with other Internet activities \textit{before} or \textit{after} the DNS queries, such as search engines, Web browsing, etc~\cite{patel2016tor, private-browsing-10, private-browsing-18, snatch-24, lightweb-pir-web-23, tiptoe-private-search-23, TOCS25-Snatch}. 

Moreover, our proposal focuses on exploring the potential of integrating DNS and PIR to enhance DNS privacy, particularly in terms of resisting collusion. Note that, \pirdns does not introduce a new or complete solution to the longstanding issue of DNS end-to-end authentication, \ie enabling users to verify the validity of a retrieved DNS record on their own. Instead, \pirdns relies on existing protocols: it is compatible with DNSSEC~\cite{rfc-dnssec} and more recent proposals like RHINE~\cite{rhine-e2e-23}. Nonetheless, \pirdns aims to minimize violations of end-to-end authentications to the greatest extent within our capability. These violations at the \RR can occur in two ways: modifying or dropping a cached DNS record. In the first scenario, the \RR may substitute a domain’s IP address with a self-controlled one to conduct phishing attacks. However, such a fake IP can be identified by the users when they make subsequent actions, \eg sending web requests to the IP address. Assuming a covert adversary model~\cite{covert-adversery-10} for the \RR, this type of detectable misbehavior would damage the \RR's reputation or result in legal consequences. In the second scenario, the \RR might deliberately drop cached DNS records to force users to query final ANSes, thereby repopulating the \RR's cache, which could expose user activity. While this is more difficult to detect on the user's side, we propose solutions to mitigate this issue (see \cref{sec:deep:discussion}, ``Delayed Response Forwarding'' in \cref{sec:cache-miss}, and \cref{sec:security}).

In addition, regarding the collusion resistance, we restrict our discussion to realistic scenarios involving a limited number of parties, such as a few \RRs and proxies, but not impractical scenarios, \eg involving millions of colluding ANSes.  
Finally, \pirdns’s guarantee relies on having a \textit{sufficient} number of active users using the system, \eg at least a few at a time. If only one user is active and experiences a cache miss, it becomes possible to determine the content of that user's query through a timing attack (see \cref{sec:cache-miss}). 
It is important to note that DoHoT also has similar requirements; that is, a certain number of honest nodes is necessary to ensure privacy.

\section{\pirdns Overview}
\label{sec:overview}
Figure~\ref{fig:dns-overview} visualizes the workflow of \pirdns both at a high level (on the left) and reporting its key PIR ``primitives'' (on the right). In the remainder of this section, we define our privacy and threat models. Next, we define some fundamental PIR primitives which we use to formulate \pirdns's workflow.

\subsection{Privacy and Threat Models}
\label{sec:overview:models}
\noindent
\textbf{Privacy model} -- DNS involves three main actors that can violate user privacy: \RRs, ANSes, and any in-network device capable of intercepting DNS traffic. 
We assume \RRs act as \textit{covert adversaries}~\cite{covert-adversery-10} who may deviate from the protocol arbitrarily as long as their malicious actions remain undetected. For instance, they might track and inspect DNS queries or deliberately/selectively drop DNS records from their cache in an attempt to infer which user might query a domain when the record is re-cached. However, actions like modifying the IP addresses of cached domains can be detected, \eg when users send subsequent requests to those addresses. The exposure of such misbehaviors can damage the \RRs' reputation, discouraging users from utilizing their services, and may also lead to financial or legal penalties for the \RRs.
We also assume that DNS traffic can be intercepted by third parties, \ie middleboxes interposing between DNS clients and both \RRs and ANSes. However, we assume that the attackers cannot break cryptographic primitives and are not ubiquitous in the WAN.

With respect to ANSes, their role requires some further discussion. ANSes operate at different levels in DNS, \eg from \textit{non-final} ANSes which are responsible for large domains like \texttt{.\@com}, to \textit{final} ANSes which are responsible for one or just a few domains. Queries to non-final ANSes are less sensitive as they reveal only partial information -- under the assumption that DNS query minimization~\cite{rfc-query-minimization-21, query-minimization-1-19, query-minimization-2-23} is used, \eg avoid forwarding the full domain at each step. %

The privacy leak of a query increases as we approach the final ANS, since the full domain name is required in each query. However, it has to be noted that such ANSes are either operated by the same organization as the target domain, or by an organization contracted by the domain provider, \eg when leveraging Amazon Route53~\cite{awsroute53}. It follows that such DNS queries do not leak any extra private information about a user than what the subsequent traffic directed to the domain~\cite{d-dns-1-20}, \eg HTTP(S) in case of a webpage. Finally, ANSes are not in the position to gain access to a full individual user profile.
In conclusion, we assume that ANSes cannot be \textit{fully} trusted but they are not a critical DNS actor with respect to user privacy, differently from ReRs.

\begin{figure}[t]
    \centering
    \includegraphics[width=\linewidth, trim=12mm 2mm 41mm 3mm, clip=true]{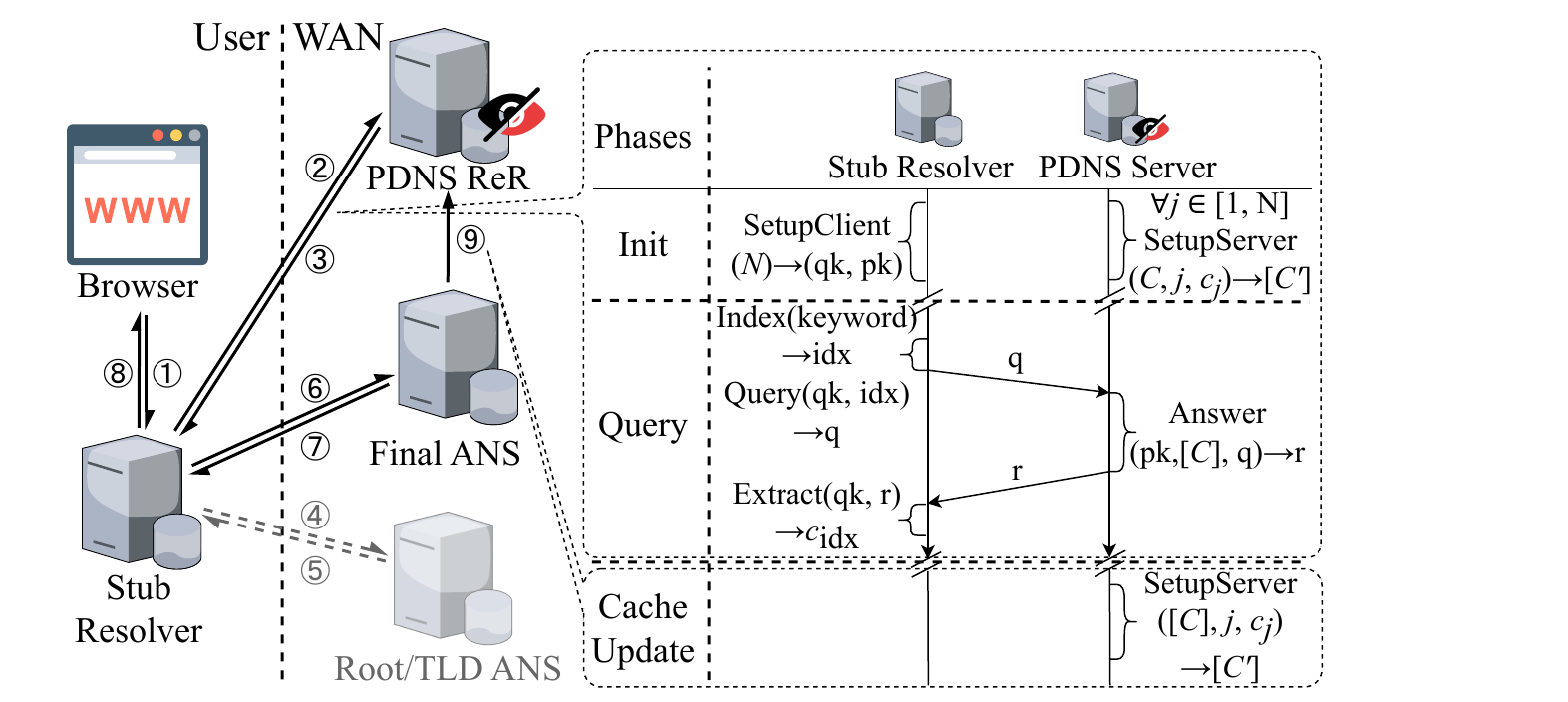}
    \vspace{-0.25in}
    \caption{Visualization of \pirdns and its workflow.}
    \label{fig:dns-overview}
    \vspace{-0.2in}
\end{figure}

\vspace{0.05in}
\noindent
\textbf{Threat model} -- Many threats exist for DNS today, \eg amplification, snooping, and DoS attacks~\cite{dns-dos-ccs08, dns-amplification-revisit, dns-snoop-imc20, akamai-dns}. Overall, existing solutions to counter such attacks are still viable in \pirdns. However, \pirdns departs from the regular DNS workflow requiring its users to directly perform iterative DNS lookups (steps {\large \textcircled{\small 4}} -- {\large \textcircled{\small 7}} in Figure~\ref{fig:dns-overview}) in presence of cache misses. While this is not a threat per se, it invalidates a common practice adopted by large ANS providers %
which limit requests from non-well-known \RRs to protect against potential DoS attacks~\cite{akamai-dns}. Such rate limitation works in the current DNS where users are supposed to perform their queries recursively (see \cref{sec:background:dns}), but would fail in \pirdns when handling cache misses. Note that this applies even more to solutions like \RR-Less~\cite{no-recursive-resolver-14}, which fully bypass \RRs. 

Finally, ANSes in \pirdns are tasked to populate a \RR's cache, and could be misused to DoS a \RR via ``reflection''. This attack can be performed both by a malicious ANS or by an attacker disguising as an ANS. \cref{sec:security} presents a defense mechanism to handle both reflection and DoS attacks. This mechanism relies on validating \pirdns cache misses, and can thus not be applied to solutions like \RR-Less~\cite{no-recursive-resolver-14}.

\subsection{PIR Primitives} 
\label{sec:overview:pir}
PIR schemes assume a key-value database $\mathcal{C}=(c_1,\dots,c_N)$ of size $N$ where the $i$-th key-value pair is defined as $(i,c_i)$. All entries $c_i$ are of the same length. Later we will explain how our construction allows for variable-length DNS records. We here define several PIR \textit{primitives} which are the founding blocks of most single-server stateless PIR schemes. Assuming that the database size $N$ is known to both user and server. $[\mathcal{C}]$ is denoted as an encoding of $\mathcal{C}$ and is initially filled with zeros. The way a server encodes the database is specific to the PIR scheme. Define the following primitives for the user setup and database construction.

\begin{itemize}[leftmargin=0.3cm, topsep=0.2pt, itemsep=0in]   
    \item ${\sf SetupUser}(N)\rightarrow ({\sf qk},{\sf pk})$: Given as input the database size, the user executes the \textit{SetupUser} primitive which outputs a query key ${\sf qk}$ and a public key ${\sf pk}$. The user stores ${\sf qk}$ as a private key and sends ${\sf pk}$ to the server. This step is only needed the first time a user connects to a server, or when the user generates a new pair of keys. \cref{sec:discussion} discusses how to share ${\sf pk}$ in a real-world deployment consisting of multiple clusters of \RRs. 

    \item ${\sf SetupServer}([\mathcal{C}], j,c)\rightarrow([\mathcal{C}'])$: Given as input the encoded database, an index $1\leq j\leq N$ and an entry $c$, the server executes the \textit{SetupServer} primitive which replaces the $j$-th entry in $[\mathcal{C}]$ with $c$ and outputs it as $[\mathcal{C}']$. The server executes this alone when updating the database. %
\end{itemize}
\vspace{0.05in}

\noindent After setting up the users and constructing the database, the following primitives are used for PIR queries.
\begin{itemize}[leftmargin=0.3cm, topsep=0.2pt, itemsep=0in]
    \item ${\sf Index}({\sf keyword})\rightarrow {\sf idx}$: Given as input a keyword of the target record, \ie the domain name in the DNS scenario, the user or server executes the \textit{Index} primitive, which hashes the keyword to an index ${\sf idx} \in [1,N]$. 
    
    \item ${\sf Query}({\sf qk},{\sf idx})\rightarrow [{\sf q}]$: Given as input a query key and an index, the user executes the \textit{Query} primitive which outputs a query $[{\sf q}]$. It is a ciphertext that encrypts ${\sf idx}$. 
    
    \item ${\sf Answer}({\sf pk},[\mathcal{C}],[{\sf q}])\rightarrow[{\sf r}]$: Given as input the public key of the user, the encoded database, and a query, the server executes the \textit{Answer} primitive, which outputs a response $[{\sf r}]$. It is a ciphertext that encrypts $c_{\sf idx}$.
    
    \item ${\sf Extract}({\sf qk},[{\sf r}])\rightarrow c_{\sf idx}$:  Given as input the query key and response, the user executes the \textit{Extract} primitive which decrypts the ciphertext ${\sf r}$ and outputs $c_{\sf idx}$.
\end{itemize}

\subsection{Workflow}
\label{sec:overview:arch}
\pirdns workflow consists of three main parts: initialization, query, and cache update (see Figure~\ref{fig:dns-overview}). Define $N$ as the maximum number of PIR cache entries and $X$ as the number of DNS records. The PIR cache entries are of the same length and each of them may be filled with multiple DNS records.

\vspace{0.05in}
\noindent\textbf{Initialization} -- Given as input $X_0$ initial DNS records, the \pRR constructs a PIR DNS cache $\mathcal{C}:=(c_1,\dots,c_N)$ and executes the ${\sf SetupServer}$ primitive to obtain an encoded cache $[\mathcal{C}]$. Upon registering to a \pRR, the user executes ${\sf SetupUser}(N)\rightarrow({\sf qk},{\sf pk})$ to derive the query key ${\sf qk}$ and public key ${\sf pk}$. The user sends ${\sf pk}$ to the \RR, which needs it to answer private DNS queries. This is a per-user key that can be shared across multiple \RRs, \eg in the case of a cloud DNS with multiple machines for load balancing (see \cref{sec:discussion}).

\vspace{0.05in}
\noindent\textbf{Query} -- A DNS query in \pirdns implies the following steps:

\begin{enumerate}[leftmargin=0.5cm, topsep=0.2pt, itemsep=0in]
    \item A user who wants to visit a domain $d$ executes ${\sf Index}({\sf d})\rightarrow {\sf idx}$. ${\sf idx}$ is a hash result of the domain ${\sf d}$ and it points to a specific slot in \RR's cache, where the DNS record for ${\sf d}$ might be located. User executes ${\sf Query}({\sf qk},{\sf idx})\rightarrow [{\sf q}]$ and sends the encrypted query $[{\sf q}]$ to the \pRR \ ({\large \textcircled{\small 2}}).
    
    \item The \pRR executes ${\sf Answer} ({\sf pk},[\mathcal{C}],[{\sf q}]) \rightarrow [{\sf r}]$. The output $[{\sf r}]$ is a ciphertext that encrypts the corresponding cache \row, and is kept secret from \RR. The \RR sends $[{\sf r}]$ to user ({\large \textcircled{\small 3}}).
    
    \item User executes ${\sf Extract}({\sf qk},[{\sf r}])\rightarrow c_{\sf idx}$. If $c_{\sf idx}$ contains a valid DNS record for the domain $d$, the DNS query is terminated. Otherwise, the user performs an iterative DNS lookup ({\large \textcircled{\small 4}} -- {\large \textcircled{\small 7}}).  Note that \pirdns attempts to speed up such iterative DNS lookup by providing in $c_{\sf idx}$ the NS-record of $d$, or the IP address of the ANS for ${\sf d}$ (thereby skipping {\large \textcircled{\small 4}} and {\large \textcircled{\small 5}}, see \cref{sec:cache-miss}). The ANS could optionally invoke an authenticity request which asks the user to prove the existence of cache miss (see \cref{sec:security}).
    
\end{enumerate}

\vspace{0.05in}
\noindent\textbf{Cache update} -- The cache update happens after a cache miss is triggered and the user finishes an iterative DNS lookup for a domain $d$. The final ANS for ${\sf d}$ populates the \pRR's cache by sending its most recent DNS record for ${\sf d}$ ({\large \textcircled{\small 9}}). \pRR constructs a new entry $c_j$ that contains the new record, and executes ${\sf SetupServer}([\mathcal{C}], j,c_j)\rightarrow([\mathcal{C}'])$ to obtain a new encoded cache. As long as the final ANS does not ``collude'' with the \RR, the user's privacy is maintained. This %
is not a violation of our privacy model as final ANSes have access to the user queries anyway (\cref{sec:overview:models}); instead, rather than being a case of collusion, it is merely a unidirectional information transfer from the final ANS to the \RR.

\section{\pirdns Deep Dive}
\label{sec:design}

\subsection{DNS Cache Construction}
\label{sec:design:hash:table}
PIR schemes assume a key-value cache $\mathcal{C}=(c_1,\dots,c_N)$, \ie a hash table where a standard hash function $H$ -- known by both server and users -- realizes ${\sf Index}(\cdot)$ (see \cref{sec:overview:pir}). A hash table is similar to the data structure used by current \RRs for their cache, \eg the popular BIND9~\cite{bind}. Yet, one critical difference is that, with PIR, the capacity of the hash table is determined beforehand and all slots -- no matter if occupied by DNS records or placeholders -- have to be encoded into $[\mathcal{C}]$ via the ${\sf SetupServer}$ primitive (see \cref{sec:overview:pir}). At run time, DNS records are inserted by re-encoding the content at a \row indicated by the hash function.

Generally speaking, the query time of PIR increases as the capacity of the hash table increases (see %
Appendix~\ref{sec:appendix:pir}). 
There is thus an incentive to reduce its size which in turn leads to more hash \textit{collisions}, \ie multiple entries hashing to the same \row. PIR schemes often use Cuckoo hashing -- where colliding entries are hashed with a second function -- to minimize the hash collision rate~\cite{ali2021communication,mughees2022vectorized}. The drawback of Cuckoo hashing is that the users need to send multiple queries, one per hash function used, which increases the query time by at least $2x$. %

Recent PIR schemes such as Spiral~\cite{spiral-pir} offer faster query time in presence of larger cache \rows, \eg \cref{sec:impl:benchmark} shows the optimal performance when considering 16KB \rows where roughly 264 (IPv6) to 431 (IPv4) \pirdns records per slot (see Figure~\ref{fig:cache-design}) could be stored. Accordingly, instead of reducing hash collisions via Cuckoo hashing, we leverage hash collisions to purposely build large cache \rows, which reduces the query time at the expense of more data to be returned to the user since the entire \row is returned (even if only containing placeholder data and no actual DNS records). To do so, we adopt a slight modification of \textit{chaining}~\cite{michael2002high} (see Figure~\ref{fig:cache-design}), where colliding entries in a \row are stored in a priority queue instead of the classic linked list. We order each priority queue using DNS record expiration times such that, once the queue overflows, the record that is most likely to expire is evicted.

\subsection{Handling Cache Misses}
\label{sec:cache-miss}

In presence of cache misses, the user performs the iterative DNS lookup -- although minimizing traffic to non-final ANSes as discussed next. After answering the DNS query from the user, the final ANS forwards the response to the \pRR (whose IP was provided by the user) to populate its cache without leaking the IP address of the requesting user. This traffic is randomly delayed to avoid the \RR correlating a previous query with a record update. %
Note that the final ANS might return a different response to the \RR occasionally %
when client and \RR are geographically distant.

\begin{figure}[t]
    \centering
    \includegraphics[width=0.9\linewidth, clip=true]{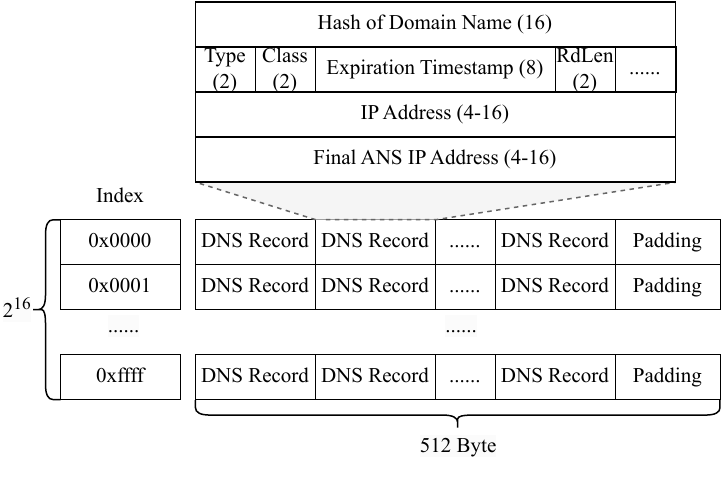}
    \vspace{-0.3in}
    \caption{\pirdns cache as a hash table with $2^{16}$ \rows.} %
    \label{fig:cache-design}
    \vspace{-0.2in}
\end{figure}

\vspace{0.05in}
\noindent\textbf{Minimizing iterative traffic} -- To shortcut the iterative lookup, and mitigate the privacy leak to root and TLD ANSes, we merge NS and A/AAAA records by appending the IP address of the final ANS at the end of the A/AAAA record (see Figure~\ref{fig:cache-design}). The main drawback of this approach is that it consumes precious cache space (\eg 4 extra bytes per IPv4 address of the ANS added)  which negatively affects the query time (see \cref{sec:impl:benchmark}). To regain some cache space, we replace the domain name field (variable length of maximum 256 bytes) with a hash value of the domain name -- a fixed length of 16 bytes digest. 
In addition, embedding the ANS's IP address in the DNS record may introduce reliability concerns if ANSes occasionally change their IP addresses. However, such events are extremely rare, and the impact is acceptable, only requiring clients to repeat the iterative lookup.

For simplicity, we only consider A/AAAA/NS DNS records in this paper.  Note that \pirdns is compatible with other types of DNS records since they are essentially strings with different lengths of up to 512B, which can be easily fit in large slots (tens of KBs, see \cref{sec:impl:benchmark}) of \pirdns cache.

It is noteworthy that the above shortcut only applies to expired cached domains and will not work with uncached domains, which instead require full iterative DNS lookup at the client. We adopt this approach in \pirdns because DNS record expiration is the main cause of the cache miss~\cite{DBLP:journals/ccr/CallahanAR13}.%

To identify cache misses at the client, another modification of the DNS record is required. Currently, an A/AAAA record contains: domain name, type, class, time-to-live (TTL), rdlength, and rdata~\cite{rfc_dns}. The TTL field indicates for how long the received DNS record is valid. DNS records stored at the \RR also contain a timestamp of when the record was resolved. Each time a DNS query is matched in the cache, the \RR checks whether this record is expired. To allow the client to perform this operation, we replace the TTL field with a timestamp indicating when the record expires, \ie timestamp at insertion plus TTL, as illustrated in Figure~\ref{fig:cache-design}. 

The client communicates the IP address of its \pRR to the ANS using a solution \`a la EDNS-PR, our own extension of EDNS(0)~\cite{rfc_edns} similar to EDNS Client Subnet (ECS)~\cite{rfc-ecs} but with a different EDNS(0) \texttt{OPTION-CODE}. ECS is a DNS extension that allows a \RR to share the subnet of a client's IP address to an ANS for improved geolocation, enabling the \RR to resolve a domain to the closest available IP to the client. In \pirdns, EDNS-PR includes the IP address of the \RR so that the ANS knows where to send the copied response to populate the \RR's cache.%
\footnote{Note that we use the full IP address for the \pRR in EDNS-PR, instead of the subnet as suggested in ECS RFC~\cite{rfc-ecs}. The ECS RFC suggested using a subnet to protect the privacy of the user's IP address being communicated (and unfortunately mostly in plaintext with DoUDP today~\cite{rfc_dns}), whereas in our proposal we send the IP address of the \pRR (using encryption, see next section), which is not violating user's privacy.}

\vspace{0.05in}
\noindent\textbf{Delayed response forwarding} -- 
\pRR may perform a timing attack to correlate the user who finished the DNS query right before it receives the populated cache entry from the authoritative server. To mitigate this attack, the cache population traffic from the authoritative server to \pRR should be randomly delayed. %
For example, assume two users $A$ and $B$ send a query at time $t$ and $t+\Delta$ ($\Delta>0)$, respectively. 
Define a random variable $X$ as the delay of the response forwarding to \RR determined by an authoritative server. We let the authoritative server sample the delay from a distribution $\mathcal{D}$, \ie $X \leftarrow \mathcal{D}$. 
If ${\sf Prob}(X > \Delta) > 0$, meaning that the response for $A$'s query might arrive after B finishes its query, then it might pose difficulty for the \pRR to decide which query should this response correlate to. %
A simple solution is to use a uniform distribution from 0 to $2\Delta$, \ie $\mathcal{D} = U(0, 2\Delta)$. It follows that the probabilities of response occurrences for both user queries at $\Delta +s$ where $0 < s < \Delta$ are identical, \ie ${\sf Prob}(X = \Delta + s) = {\sf Prob}(X = s)$. This translates to a 50\% chance for \pRR to fail to execute a timing attack.  Appendix~\ref{sec:appendix:timing-attack} shows -- via a formal investigation -- that by sampling the random wait time from appropriate distributions, \pirdns can effectively defend timing attack.

\subsection{Communication Encryption}
\label{sec:deep:discussion}

DoT or DoH are rarely supported by ANSes, \eg we tested 100 popular ANSes for different zones and found no support. Given that \pirdns relies on iterative DNS lookups performed by its clients to handle cache misses (see \cref{sec:cache-miss}), the lack of encryption on this communication channel fails to protect against in-network eavesdroppers, thus violating our privacy model (see \cref{sec:overview:models}). Therefore, ANSes participating in \pirdns need to adopt DoT/DoH to %
satisfy our privacy requirements. %

On the path between user and \pRR, PIR offers chosen-plaintext attack (CPA) security~\cite{katz2020introduction}, meaning that, although the query content is kept secret from an attacker, \textit{integrity} and \textit{authentication} are not provided. It follows that an attacker can disrupt the service or even impersonate the \RR and attempt other attacks like phishing. 
Therefore, \pirdns adopts HTTPS between client and \RR to guarantee integrity and authentication. We favor HTTPS over TLS since it makes it hard for third parties to distinguish between DNS and HTTP traffic given they both use port 443~\cite{csikor2021privacy}. 
It is also important to note that in proxy-based mechanisms (like ODoH), HTTPS connection reuse requires no correlation between client-proxy and proxy-\RR connections in order not to affect user privacy~\cite{odoh}. This constraint does not apply to \pirdns as \pRR is blind to DNS query contents.

Finally, when the final ANS populates the cache of \RR, it should ideally use DoH to prevent malicious attackers from impersonating its identity through techniques like IP spoofing. This also helps avoid the false accusations of misbehavior of \RR, \eg selectively dropping DNS records for target users (\cref{sec:security}), which could otherwise occur due to the unreliability of UDP transmission.

\subsection{Gradual deployment}
\label{sec:deep:deployment}

Similar to DoH, universal adoption of PDNS is not a requirement. We envision three scenarios: 
\begin{itemize}[leftmargin=0.3cm, topsep=0.2pt, itemsep=0in]   
\item Ad hoc networks: operators in specialized environments such as military networks have full control over the entire network. This allows for end-to-end PDNS implementation, covering clients, ReRs, and ANSes.
\item ReR-based approach: clients install a local proxy which intercepts DNS queries and converts them into PDNS queries for a predefined subset of domains. ReR proactively maintains a curated list of domains to be resolved privately, ensuring enhanced privacy for specific use cases. While PDNS can be bootstrapped without ANS support, it faces two challenges: $(i)$ it contradicts DNS caching principles, and $(ii)$ limited domain sets reduce privacy guarantees.
\item Organic adoption: some ANSes begin supporting \pirdns to attract privacy-conscious users and gain a competitive edge over other domains offering similar services. Over time, this could drive widespread adoption of \pirdns, following a trajectory similar to the adoption of HTTPS and DoH. Meanwhile, cloud ANS providers like Route53 offer PDNS as a premium service for targeted use cases, balancing costs with enhanced privacy benefits.
\end{itemize}
Note that the latter two scenarios can coexist, facilitating gradual \pirdns deployment at both \RRs and ANSes.

\section{System Security}
\label{sec:security}

\noindent\textbf{Defend DoS and reflection attacks} -- \pirdns's mechanism to handle cache misses is susceptible to 1) DoS on ANSes, and 2) a ``reflection'' attack towards the \RR (see \cref{sec:overview:models}). We here present a solution to protect against both attacks. 

ANSes should only accept direct queries for a domain ${\sf d}$ from users who can \textit{prove} an actual cache miss occurred at the \RR. In order to prove the existence of a cache miss, the user forwards to the ANS the encrypted query and its response $({\sf q}, {\sf r})$ along with its secret key ${\sf qk}$. To prevent a malicious user from counterfeiting $({\sf q}, {\sf r})$, we require the \pRR to sign an additional message containing the user IP, query timestamp, and the pair $({\sf q},{\sf r})$. With this information, the ANS can check whether ${\sf q}$ encrypts a query for ${\sf d}$, and ${\sf r}$ indeed refers to a recent cache miss from the contacting IP. 

The above solution has two major limitations though. First, sharing the secret key ${\sf qk}$ can be dangerous, as a misbehaving ANS could collude with \pRR causing a privacy violation. Second, according to our performance evaluation (see \cref{sec:impl:benchmark}), this approach would bloat direct DNS queries with about 39KB extra traffic: 36.5KB from $({\sf q},{\sf r})$, about 2KB from the certificate, and few bytes for timestamp and IP address. It also requires additional computation at ANSes for verification, but the overhead is moderate: the challenge increases CPU usage by only 17\%, even if all queries require verification (see \cref{sec:impl:benchmark} and Appendix~\ref{sec:appendix:benchmark}).

We solve both issues by asking for cache misses proofs only for \textit{frequent} requests, \ie within a domain's TTL, which might indicate a potential attack. To do so, the ANS keeps track of the time at which it populated a domain record at a given \pRR. Further, in presence of such (rare) proof request the user would send a new query for the same domain to the \RR using \textit{backup} key pairs; this query is needed to generate a proof without leaking the user's main private key. After a successful proof, the user runs the \textit{SetupUser} primitive ($<$0.15~sec, see Appendix~\ref{sec:appendix:benchmark}) to generate new backup key pairs. The ANS answers the direct DNS query right away regardless of whether the proof was requested, unless a pending proof for this IP already exists. However, it holds on the reception of the proof to update the \pRR.  

It is noteworthy that the above process only applies to the additional traffic to ANSes because of the adoption of \pirdns. It can be distinguished from existing traffic to the ANSes by the presence of the custom EDNS-PR extension flag. Existing traffic, \eg measurement queries for performance monitoring, remains unaffected and continues to operate under the current DoS defense mechanisms in place for ANSes today.

\vspace{0.05in}
\noindent\textbf{Identify misbehavior of \RRs} --  \RRs acting as covert adversaries may selectively drop DNS records from their caches for specific users to gain access to their queries. This can be mitigated through the cache miss proof process introduced above. Specifically, suppose a \RR avoids caching a domain $d$ for user $u$. If $d$ was previously queried by $u$ or other users within the valid cache period, $u$'s query would trigger a cache miss and initiate an iterative query to the ANS. The ANS would then challenge $u$, as the record should not have expired in the \RR's cache. But $u$ can submit a cache miss proof. This would in tern reveal \RR’s misbehavior. 

Note that such accusations may occasionally result in false positives due to reliability issues, such as \RRs rebooting because of power outages or software bugs. A robust credit system for \pRRs, combined with additional verification processes for handling various incidents, should help mitigate these false positives. We leave further exploration of these measures for future work.

\vspace{0.05in}
\noindent\textbf{Validate ANSes} --  An attacker may impersonate an ANS and either launch a DoS attack at a \pRR or pollute its cache. However, in \pirdns, only legitimate ANSes can populate the cache of a \RR, allowing for simple access control. %
\pRRs should only accept DNS records from %
verified ANSes. The verification can be done via regular DNS lookups, or more strictly DNSSEC~\cite{rfc-dnssec} avoiding DNS hijacking~\cite{dns-hijacking-19}, confirming with top-level ANSes that the IP address of a sender matches that of the final ANS of the domain name in a DNS record.

\section{Implementation}
\label{sec:impl}

\subsection{Implementation Details}
\label{sec:impl:choice}
\noindent\textbf{PIR scheme selection} -- %
Appendix~\ref{sec:appendix:pir-selection} lists three state-of-the-art single-server PIR solutions we have benchmarked for different cache and slot sizes using a single core of 3.0~GHz AMD EPYC CPU. SimplePIR~\cite{simplepir} achieves the fastest query processing time overall. However, it is stateful, meaning that a user has to download a state (with size comparable to the square root of the database size) from the \RR whenever its DNS cache is updated. Given that a \RR's cache updates frequently, we discard this solution.

We instead favor single-server stateless PIRs:  SealPIR~\cite{seal-pir} and Spiral~\cite{spiral-pir}. SealPIR is slightly faster than Spiral when considering smaller cache and slot sizes, while Spiral outperforms SealPIR when assuming \rows with large sizes.  This is because Spiral does not provide optimal parameters for caches with small \row size; each \row with a size smaller than 256B simply gets padded to 256B. Spiral also requires less traffic than SealPIR (0.1x to 0.12x) for both the {\em Query} and {\em Answer} primitives. Though Spiral has a slightly longer update time than SealPIR, this is a server-only operation requiring less than $<1$~ms, and is thus not a decisive factor. %
Therefore, we select Spiral as the underlying PIR protocol for building the proof-of-concept of \pirdns. But we do note that \pirdns is open to any future single-server stateless PIR protocols.

\vspace{0.05in}
\noindent\textbf{Spiral optimizations} -- %
We have applied some changes to optimize the performance of Spiral.
Specifically, Spiral's {\em Answer} primitive operates on 4 chunks of ciphertext for \rows smaller than $2^{15}$ bytes; as the \row size increases, the number of chunks grows as a factor of 4. It follows that (at least) 4 concurrent threads can be used to parallelize and thus speed up the operations on each chunk. We leverage this observation to extend the current Spiral implementation to support increased multi-threading. 
We also enable the Intel Advanced Vector Extensions 2~\cite{avx2} when running Spiral. This allows a 2x speedup of the {\em Answer} primitive in our benchmark.

\begin{figure*}[t]
  \centering
  \subfigure[{\centering Query duration as a function of number of slots with 512B slot size.}]{\psfig{figure=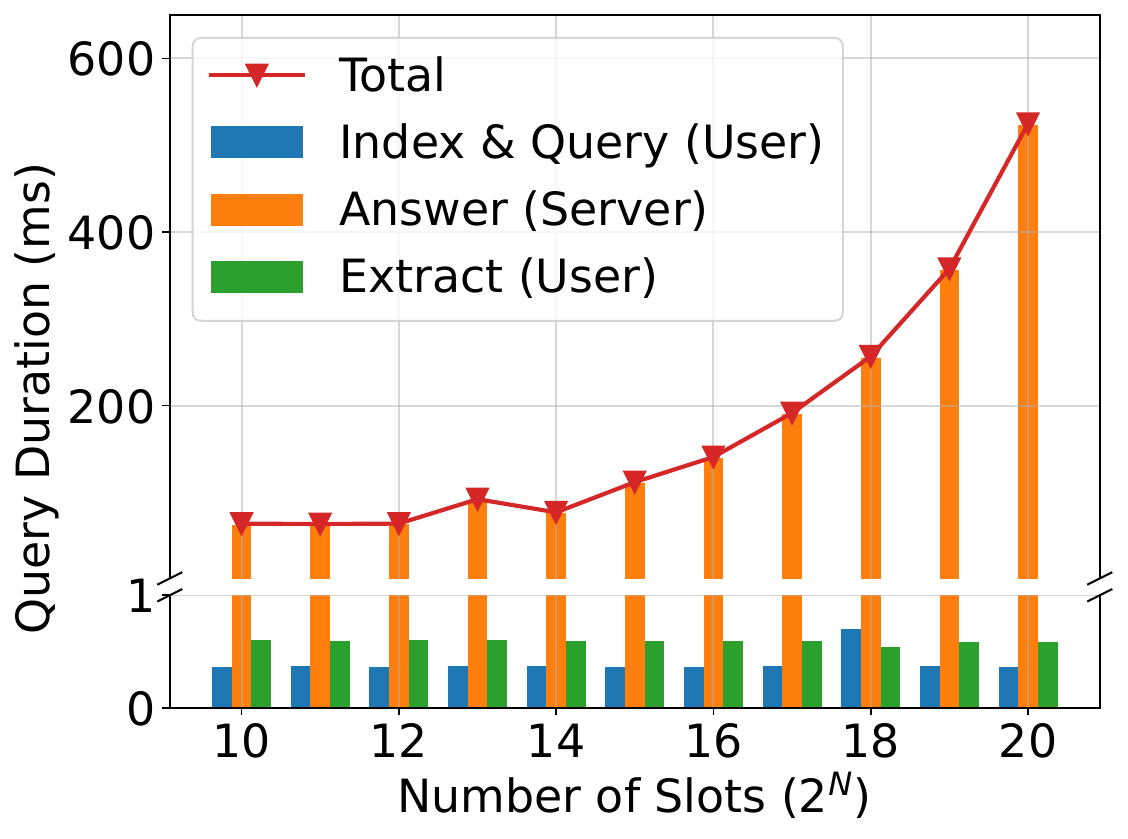, width=0.232\linewidth, trim=0mm 0mm 0mm 0mm, clip=true}\label{fig:benchmark:query-time-cache-size}}
  \hspace{0.1em}
  \subfigure[{\centering Query duration as a function of slot size given a 512~MB cache.}]{\psfig{figure=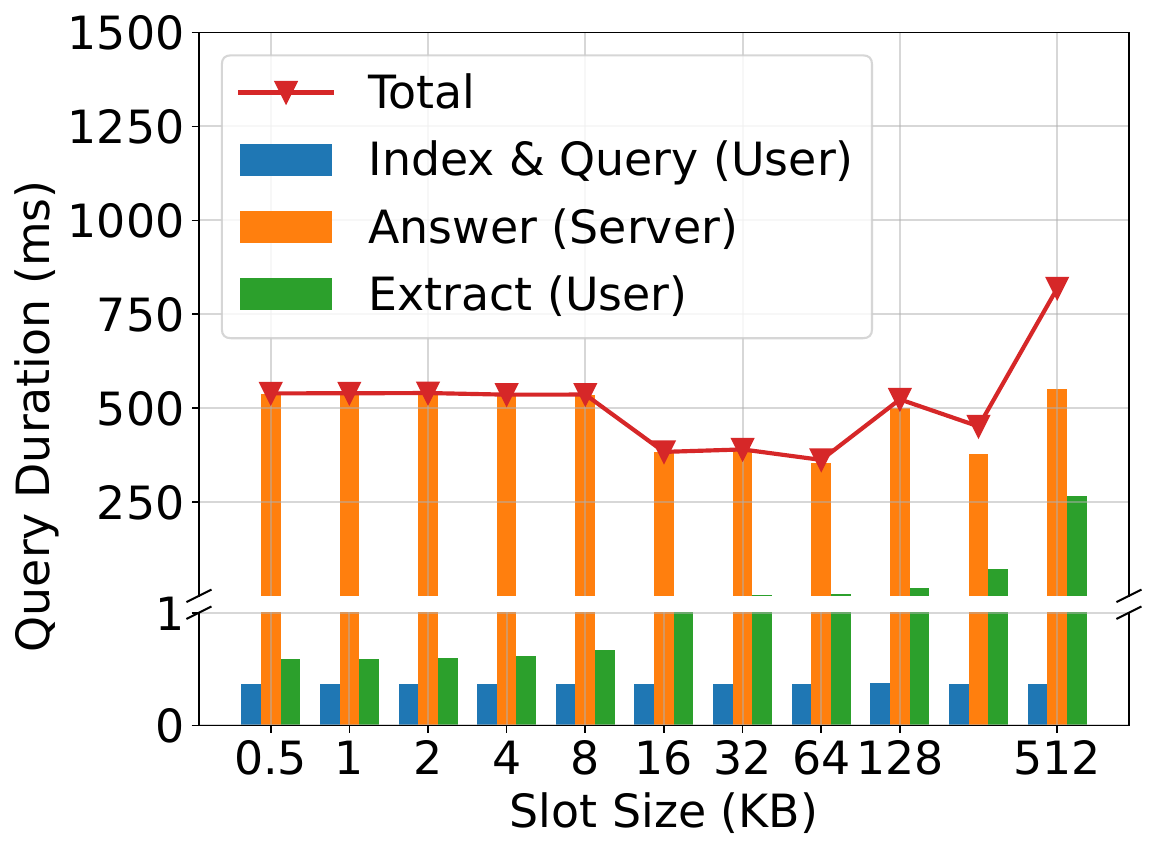, width=0.24\linewidth, trim=0mm 0mm 0mm 0mm, clip=true}\label{fig:benchmark:query-time-number-entry}}
  \hspace{0.1em}
  \subfigure[{\centering Query and answer traffic as a function of slot size given a 512~MB cache.}]{\psfig{figure=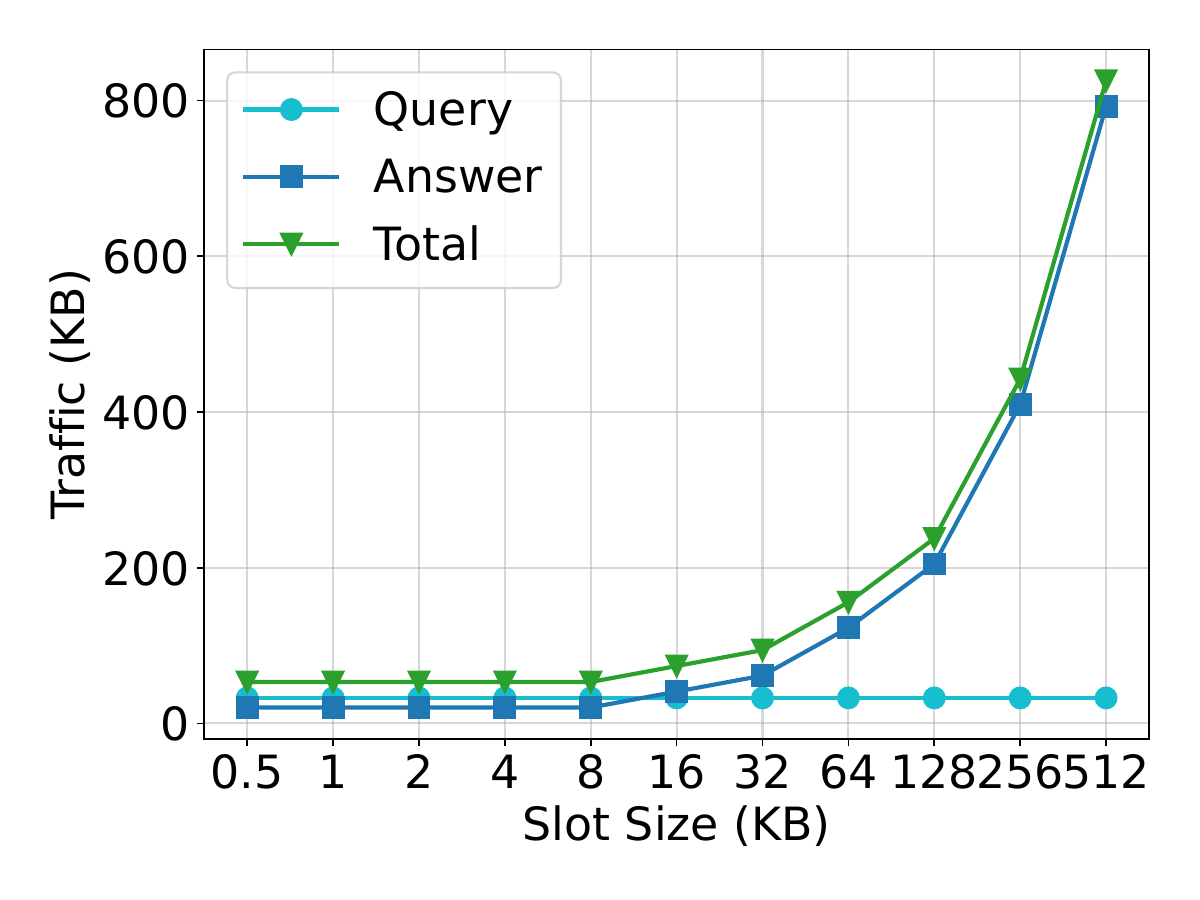, width=0.24\linewidth, trim=3mm 6mm 6mm 6mm, clip=true}\label{fig:benchmark:query-comm-number-entry}}
  \hspace{0.1em}
  \subfigure[{\centering CPU usage of ReRs as a function of query rate. }]{\psfig{figure=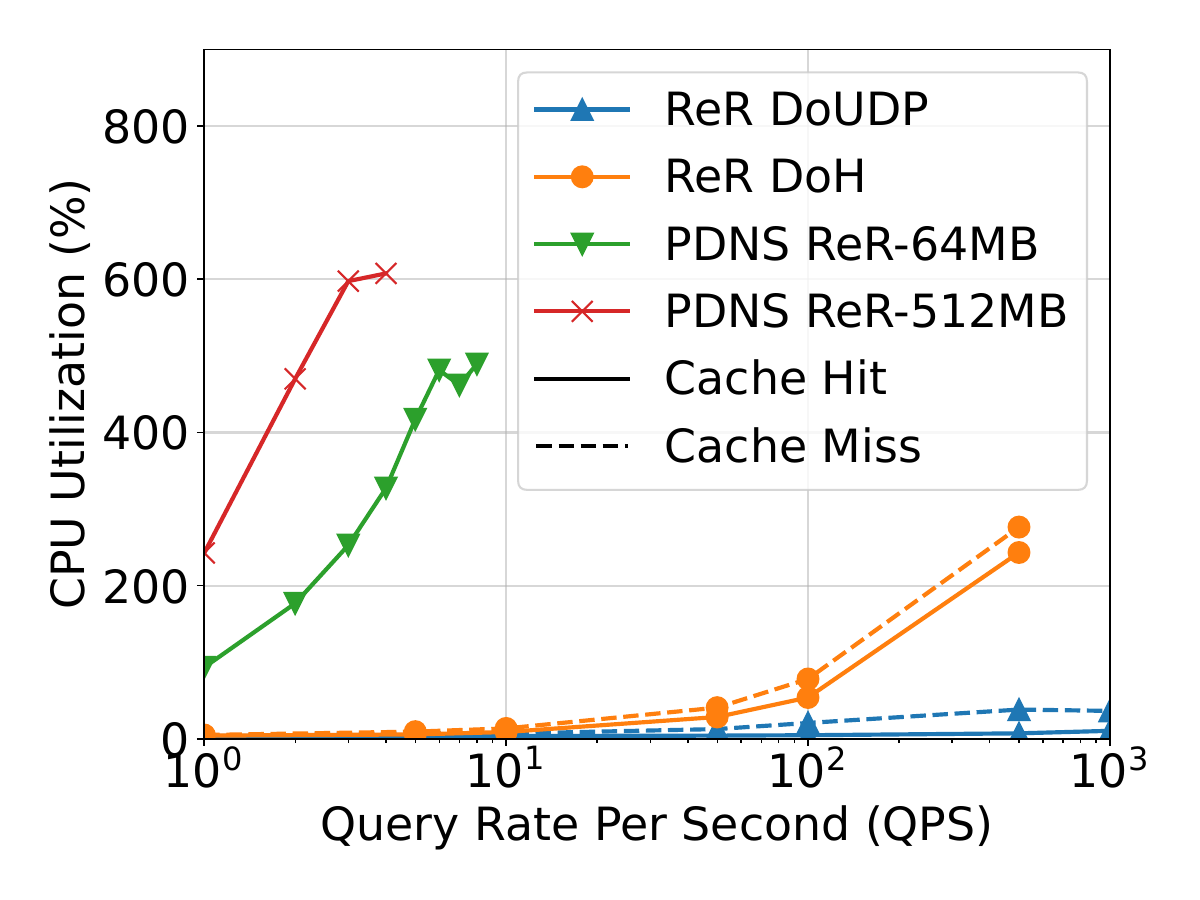, 
  width=0.24\linewidth, trim=3mm 6mm 6mm 6mm, clip=true}\label{fig:benchmark:cpu-rr}}
  \vspace{-0.15in}
  \caption{Benchmarking results for \pirdns. Network latency is negligible. }%
  \vspace{-0.15in}
  \label{fig:benchmark}
\end{figure*}

\vspace{0.05in}
\noindent\textbf{\pRR and client} -- We develop a custom \pRR using the Rust-based Spiral PIR repository~\cite{blyss}. We opt for this approach, instead of extending an existing resolver, since \pRR's workflow departs from a regular \RR, mainly due to the lack of iterative lookup to handle cache misses. We also leverage the Spiral repository to develop the \pirdns client. While integrating the \pirdns client into the OS would provide better performance, we opt for a DNS proxy for better flexibility and ease of adoption. The proxy intercepts outgoing DNS queries, transforms them into PIR queries, receives responses from \pRR, and decodes the results. In presence of a cache miss, the proxy also performs (shortcut) iterative DNS lookups. In the end, the proxy constructs DNS responses itself using answers from either \RR or ANS, in presence of a cache miss, and returns them to the OS. 

\vspace{0.05in}
\noindent\textbf{Authoritative name server} --  \pirdns requires three main changes at a participating ANS. First, support for EDNS-PR, our small extension of EDNS(0) to share the IP address of the \RR when a cache miss has occurred. Second, a routine to forward a DNS record to the \RR indicated in EDNS-PR. Third, a mechanism to challenge clients for proof of cache misses when needed (see \cref{sec:security}). We implement the above features as a patch (about 200 lines of code) of the popular Berkeley Internet Name Domain (BIND9)~\cite{bind},  a fully-fledged open-source resolver used by many ANSes worldwide~\cite{bind-popular}.

\subsection{Benchmarking}
\label{sec:impl:benchmark}
We benchmark all three components of \pirdns: client, \RR, and ANS on machines equipped with an 8-core 3.0GHz AMD EPYC CPU and 8~GB RAM. This is an upgrade setup from~\cite{odoh} because \pirdns is more computationally intensive compared to ODoH. %
Note that while PIR operations happen on the \textit{encoded} cache, we use the size of the \textit{plaintext} cache as a reference since it directly relates to the number of DNS records it can store. We ignore network latency in this benchmarking; we will instead introduce realistic latencies in the evaluation (\cref{sec:eval}). We assume \pirdns was previously setup and ready to use, \ie we ignore the one-time cost of {\em SetupServer} and {\em SetupClient} (as defined in \cref{sec:overview:pir}). The interested reader can find benchmarking for both setup PIR primitives as well as resource consumption in Appendix~\ref{sec:appendix:benchmark}. 

\vspace{0.05in}
\noindent\textbf{Query duration} --  Each \pirdns query involves four PIR primitives: {\em Index}, {\em Query} and {\em Extract} at the user, and {\em Answer} at the \pRR. Figure~\ref{fig:benchmark:query-time-cache-size} shows the query duration for each primitive assuming (plaintext) caches composed of between $2^{10}$ (1K) and $2^{20}$ (1M) 512-Bytes \rows, \ie total cache sizes ranging between 0.5~MB and 512~MB. These values are chosen because Spiral's authors selected optimal low-level cryptographic parameters for this range based on a heuristic algorithm.  %
It follows that each 512B-\row can hold about 13 DNS records with average size of 38~Bytes (see Figure~\ref{fig:cache-design}). The PIR primitives at the user ({\em Index}, {\em Query}, and {\em Extract}) require sub-millisecond durations and, as expected, are independent from the cache size. The total query duration is dominated by the {\em Answer} PIR (at the server side) which requires between 80 ms -- on a small cache with about 1,000 \rows and up to about 13K DNS records -- and up to 500~ms -- on a large cache which might contain several million DNS records. %

Next, we fix the cache size to 512~MB and investigate query duration as the number of \rows increases, \ie the size of each \row decreases. For example, a 512~MB cache can be composed of either $2^{10}$ \rows with a size $S=512$KB, or $2^{20}$ \rows with $S=512$B. Figure~\ref{fig:benchmark:query-time-number-entry} shows that the duration of the {\em Index} and {\em Query} primitives (at the user) are unaffected by the \row size. Conversely, the duration of the {\em Extract} primitive, also at the user, decreases as the \row size decreases, although non-linearly, \eg from 266~ms ($S=512$KB) down to 21~ms ($S=128$KB).  %
This nonlinearity comes from the better optimizations of Spiral for large slot sizes. %

For {\em Answer} (server side), the impact of the slot size is non-trivial due to the conflicts and compromises between multiple underlying cryptographic primitives. %
Its performance is best when $S=64$KB and worsens when the slot size either increases or decreases. Given the {\em Answer} primitive has an overall much higher duration than all other primitives,  the total duration of the PIR query is the fastest (362~ms) for $S=64KB$ to which it corresponds $2^{13}$ slots in a large 512~MB cache.  This is a $\sim$40\% drop from 539~ms when assuming a small slot ($S=512$B) and a $\sim$55\% drop from 819~ms with a large slot ($S=512$KB).

\vspace{0.05in}
\noindent\textbf{Query and answer traffic} -- First, we focus on the traffic between \pirdns client and \RR. Figure~\ref{fig:benchmark:query-comm-number-entry} shows the traffic for both queries (\pirdns client) and answers (\RR), assuming a cache size of 512~MB and increasing \row sizes (decreasing numbers of \rows). Refer to Appendix~\ref{sec:appendix:benchmark} for an analysis of the impact of different cache sizes. At the client, each query consumes a constant 32~KB independently of the \row size, since the query is the encryption of an index. At the \RR, the traffic increases as the \row size increases since a full \row is returned.  For example, the answer to a query contains 20KB when there are $2^{16}$ \rows ($S=8KB$), versus nearly 800KB when there are $2^{10}$ \rows ($S=512KB$). Considering both query duration (Figure~\ref{fig:benchmark:query-time-number-entry}) and traffic (Figure~\ref{fig:benchmark:query-comm-number-entry}), we select a cache \textit{shape} for \pirdns of  $2^{15}$ \rows with a size of $16KB$, for a total cache size of 512MB.

Next, we focus on the traffic between \pirdns client and an ANS in presence of cache misses. Since \pirdns require DoH, this increases the traffic from around 200B, \ie a DoUDP query from \RR to an ANS, to about $\sim$7KB~\cite{doh-size, bottger2019empirical} because of the TLS handshake -- $\sim$1KB is needed instead if HTTPS connection is reused. Next, the ANS also needs to duplicate the query response to the \pRR with DoH, which doubles the costs.  %
Overall, this is a significant bandwidth increase for an ANS mostly due to DoH. However, the current adoption of DoUDP between \RRs and ANSes is yet-another violation of user privacy when EDNS(0) is adopted~\cite{hal-adot-operational-considerations-02, encrypting-rr-auth}. Proposals have been made to encrypt the channel between \RR and the ANS~\cite{ietf-dprive-opportunistic-adotq-02}. If this proposal is adopted, then \pirdns doubles the ANS traffic. %

\vspace{0.05in}
\noindent\textbf{Scalability} -- \pirdns's intensive CPU usage is expected to limit its \textit{query rate}, \ie the concurrent number of queries per second (QPS) it can handle. We compare the query rate of \pirdns with both DoUDP and DoH provided by BIND9~\cite{bind}. We do not report numbers for ODoH since equivalent to DoH for the \RR. For each protocol, we increase the query rate until the query duration increases by 50\%, and then report the previous CPU usage and rate. Figure~\ref{fig:benchmark:cpu-rr} shows that DoUDP and DoH scale much better as they can serve roughly 1,000~QPS (CPU usage of 32\%) and 500~QPS (CPU usage of 200\%) before significantly delaying their respective query response times. In comparison, \pRR reaches a query rate of 4 and 8~QPS for respectively a large (512~MB) and small cache (64~MB).\footnote{While there is no broad consensus on the memory size typically deployed at ReRs, we believe our setup is feasible for most providers. One DNS trace in \cite{schomp2016towards} shows that over 200M queries were reduced to approximately 2M unique hostnames queried by over 8,000 users at a university over one week. This is only slightly more than what our small cache (64MB, 1.6M hostnames) can accommodate. Considering record replacement (a common practice for ReRs), the small cache is sufficient. A large cache (512MB, 13M hostnames) can support significantly more users.} While the \RR's CPU is not fully utilized (500-600\%), adding even one extra query would increase the query duration by more than 50\%. 

It has to be noted that the uprising specialized hardware would substantially improve \pirdns scalability. In fact, assuming that the query resolution reduces from few hundreds ms down to 1~ms~\cite{f1}, then \pirdns would be able to reach about 1,000~QPS without HTTPS or hundreds of QPS with HTTPS, similar to DoH in Figure~\ref{fig:benchmark:cpu-rr}. %
The figure also differentiates between queries triggering a cache hit or a miss. At their maximum query rate, the iterative process associated with a cache miss costs an extra 33\% CPU usage for both DoUDP and DoH. No impact is observed for \pirdns as cache misses are resolved at the client and not the \RR. 

Finally, we also benchmark the CPU usage at the ANS, using the same methodology adopted for the \RR %
(see Appendix \ref{sec:appendix:benchmark}). 
Our results show that the CPU usage at ANS for answering DoUDP queries is negligible: only 7\% with a rate of 1,000 QPS. DoH is instead much more challenging, but still reaches a rate of 500 QPS with a CPU usage of 240\%. Note that the combination of extra bandwidth and CPU might explain the current lack of adoption of DoH among ANSes (see \cref{sec:deep:discussion}). %
In addition to DoH, \pirdns further adds 50\% CPU usage to support EDNS-PR queries, \ie privately populate DNS records at a \RR, and another 50\% to validate proof of cache miss for suspicious queries.

\section{\pirdns Evaluation}
\label{sec:eval}
This section evaluates \pirdns. Given that an actual \pirdns deployment is challenging, \eg it requires participating ANSes and users, we resort to the next most realistic setup. We set up a test-bed consisting of a \pirdns client, \RR, and a participating final ANS. Next, we simulate real network latencies and DNS queries we have collected in the wild. We then evaluate \pirdns from four perspectives: query duration, privacy guarantees, resilience to attacks, and impact on Web performance. When possible, we compare \pirdns against state-of-the-art solutions: DoUDP, DoH, ODOH, and DoHoT. We further include a hypothetical \RR-Less DNS assuming both DoUDP and DoH. Due to space constraints, please refer to Appendix~\ref{sec:eval:meth} for the detailed evaluation setup. %

\begin{figure*}[t]
  \centering
  \subfigure[{\centering Boxplots of query duration of different DNS protocols.}]{\psfig{figure=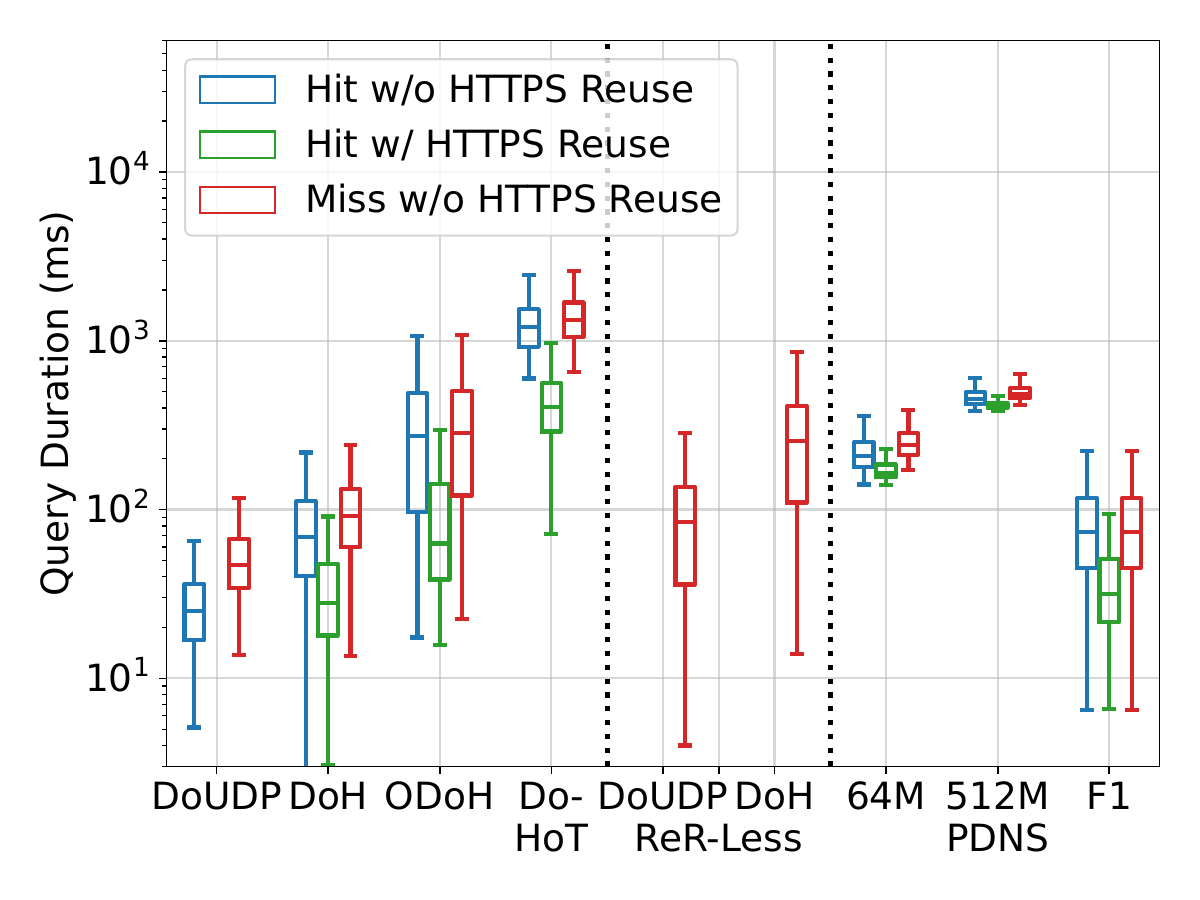, 
  width=0.235\linewidth, trim=7mm 8mm 6mm 6mm, clip=true}\label{fig:eval:query-time}}
  \hspace{0.1em}
  \subfigure[{\centering Reflection traffic as a function of attackers.}]{\psfig{figure=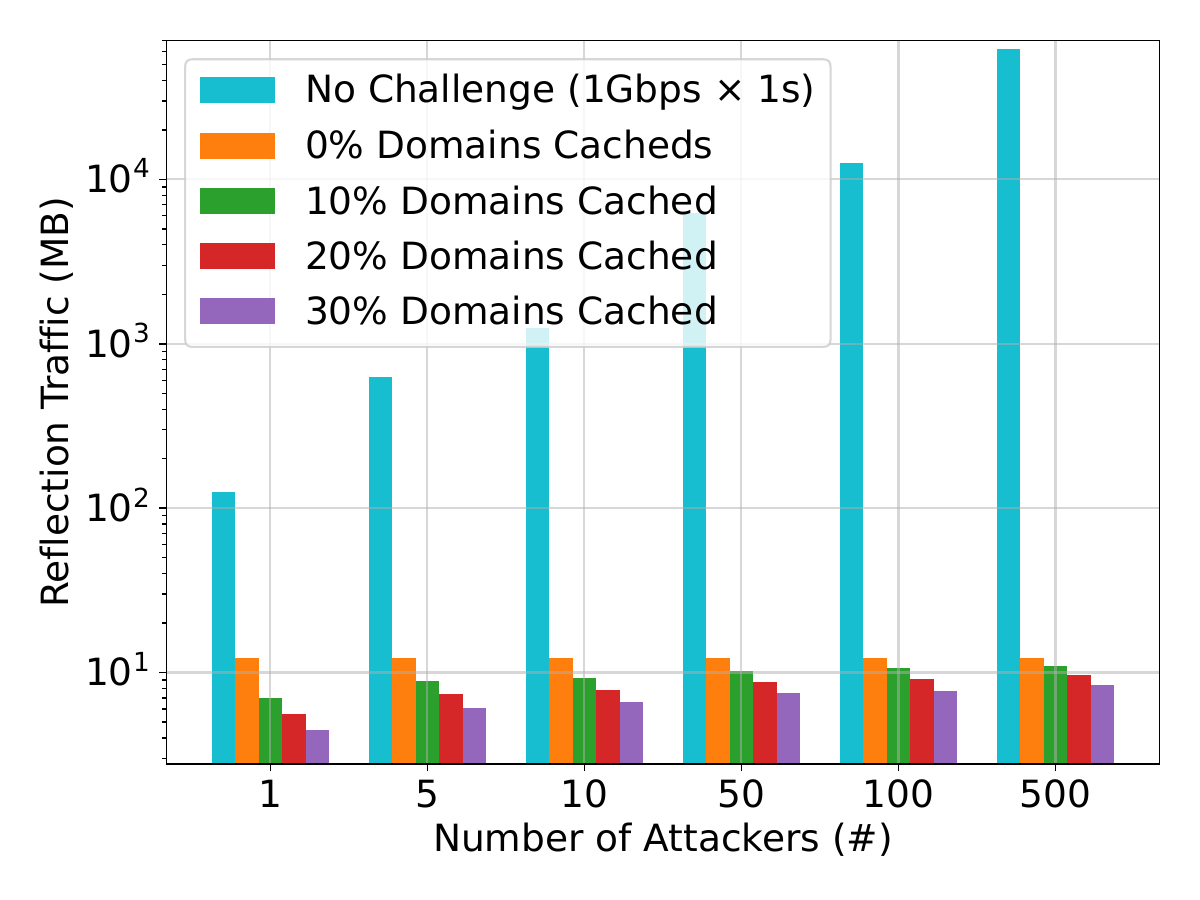, 
  width=0.235\linewidth, trim=7mm 8mm 6.5mm 6mm, clip=true}\label{fig:eval:attack}}
  \hspace{0.1em}
  \subfigure[{\centering CDF of domain frequency rank differences.}]{\psfig{figure=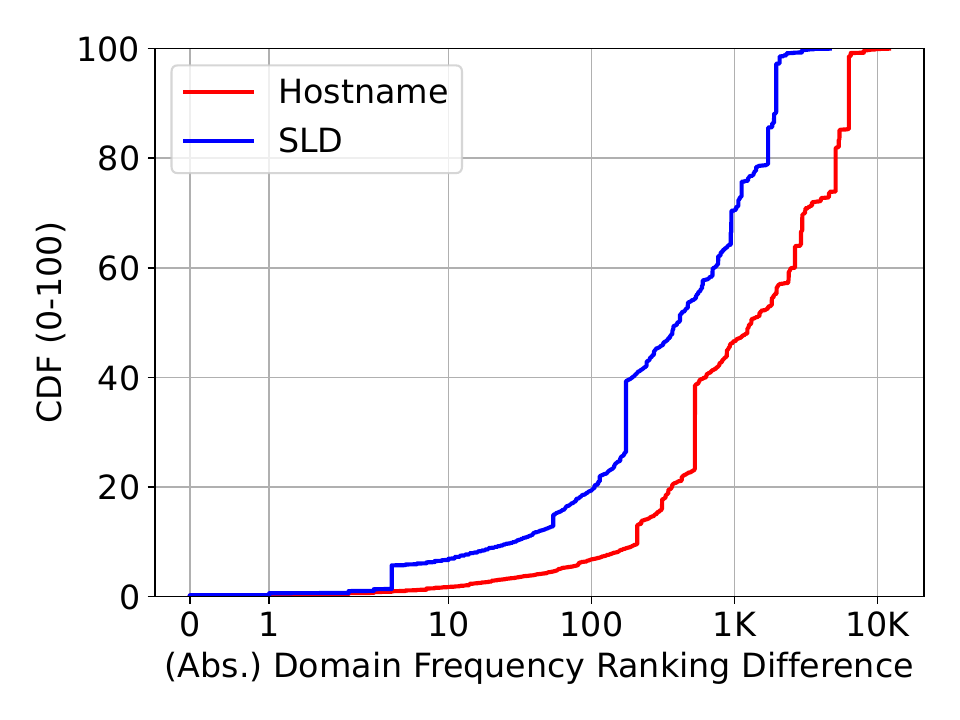, 
  width=0.24\linewidth, trim=5mm 6mm 5.5mm 6mm, clip=true}\label{fig:eval:domain-frequency-change}}
  \hspace{0.1em}
  \subfigure[{\centering Web performance of different DNS protocols with connection reuse.}]{\psfig{figure=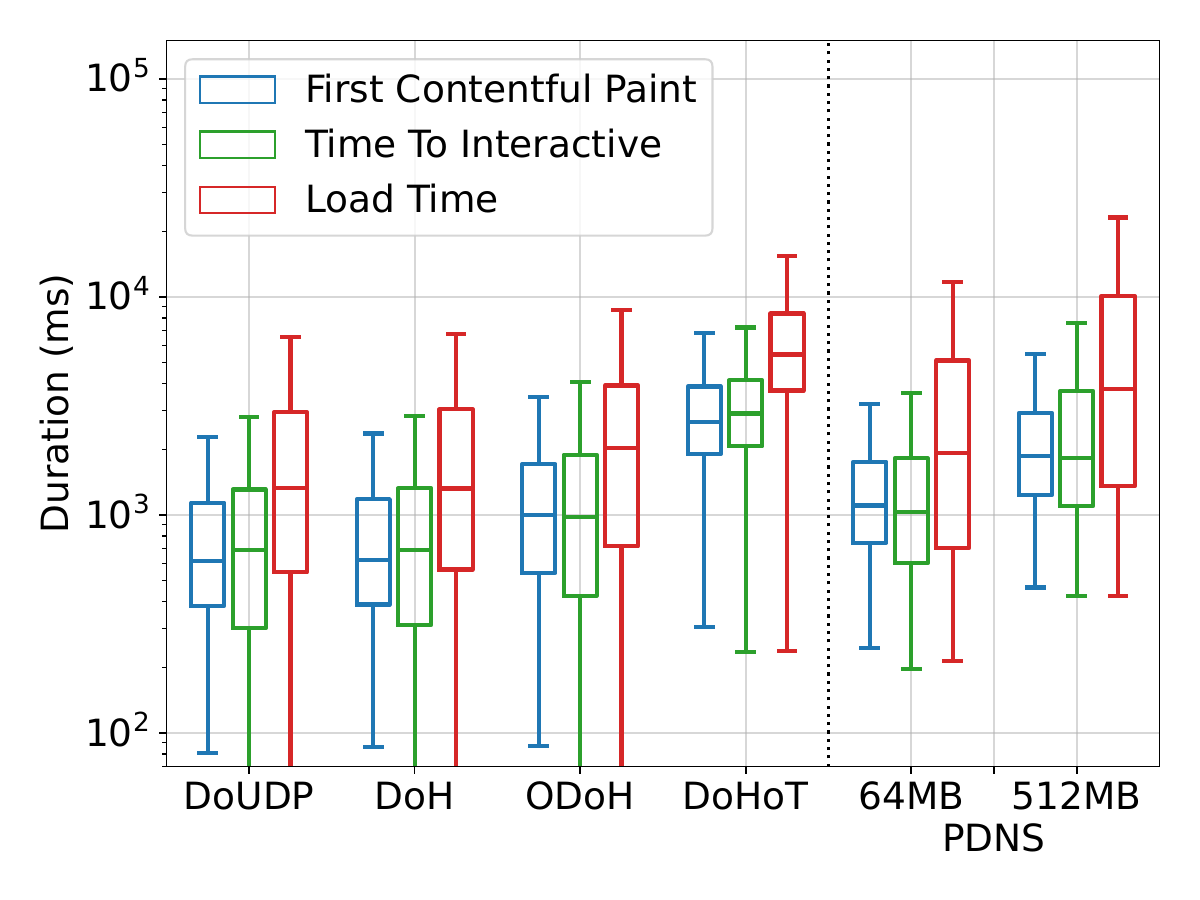, 
  width=0.235\linewidth, trim=7mm 8mm 6mm 6mm, clip=true}\label{fig:eval:web-load-time}}
  \vspace{-0.2in}
  \caption{Evaluation of \pirdns.} %
  \vspace{-0.18in}
  \label{fig:eval}
\end{figure*}

\subsection{Performance}
\label{sec:eval:performance}
Figure~\ref{fig:eval:query-time} shows boxplots of query duration per DNS protocol. For \pirdns, we differentiate between small (64MB, up to 1.6M entries) and large (512MB, up to 13M entries) caches, as well as a \textit{futuristic} implementation relying on F1 hardware~\cite{f1} with a large cache. %
While we cannot obtain such hardware yet, we simulate its performance assuming that the ${\sf Answer}$ primitive would only require 5ms.\footnote{With the anticipated \textit{at least} $10^3\times$ speedups to CPU by FHE accelerators, PIR computation time would be reduced to sub-milliseconds. However, data movement must also be considered. F1~\cite{f1} envisions a memory bandwidth of 1TB/s (with even higher bandwidths possible, \eg NVLink achieving 2.4TB/s~\cite{nvlink}). Transferring data for a large cache (e.g., 512MB plaintext expanded to 4GB due to FHE encoding) would take 4ms at 1TBps. We round up to 5ms for the complete operation.} %
We further distinguish between queries which triggered cache misses or not, and whether HTTPS connection reuse was adopted. %

Figure~\ref{fig:eval:query-time} shows that DoUDP, \ie DNS without any security or privacy guarantees, achieves the best performance with a median query time of 25~ms, which is slightly longer than the median RTT of 22~ms between our emulated users (via Mysterium as detailed in Appendix~\ref{sec:eval:meth}) and the \RR -- thus confirming negligible time spent at both client and \RR to handle a DNS query. DoH almost triples the query duration from DoUDP (median of 69~ms) due to the two additional RTTs ($\sim$\xspace45ms) required by TLSv1.3; as expected, the query duration in DoH drops to 28~ms in presence of connection reuse. ODoH -- which further enhances user privacy while sacrificing performance -- achieves a median query duration of 272~ms, which reduces to 63~ms with connection reuse. %

In the case of \RR-Less DNS, when utilizing DoUDP, the median latency is around 84 ms with considerable variability. When emulating%
ReR-Less DoH (emulation is required as ANSes currently do not support DoH), which is necessary to protect from pervasive DNS monitoring, latency significantly increases, with a median exceeding 250~ms and the third-quarter latency reaching above 450~ms. Our analysis assumes no HTTPS connection re-use, as \RR-Less DoH requires establishing new HTTPS connections with various ANSes, reducing the likelihood of benefiting from persistent HTTPS connections, except perhaps with root/TLD ANSes. 

Notably, these results reflect the performance of \RR-Less DoH from the perspective of a single user with relatively good bandwidth and location, and are limited to top-ranked domains. The performance for average users worldwide is likely to be significantly lower than what we have observed, due to an inflated network path to an ANS. Further, these results do not account for the extra load incurred on ANSes if \RRs are bypassed globally. Such loads can only be amortized via more hardware or by slowing down queries. 

\pirdns achieves median query durations of 208~ms and 450~ms for cache sizes of 64MB and 512MB, respectively. This means that \pirdns is slightly faster than DoH and \RR-Less DoH assuming a small cache, and no connection reuse. While HTTPS connection reuse is also beneficial to \pirdns, the benefit is less evident compared to DoH and ODoH because CPU processing is the main bottleneck of \pirdns. When considering a much bigger cache, \pirdns pay a penalty of 180ms when compared with ODoH. Still, the median query latency is 2x faster than DoHoT. %
This shows that the performance of \pirdns is already acceptable today, especially from a user-experience perspective \cref{sec:eval:page-load} and to the many people who care about their privacy.

\vspace{0.05in}
\noindent\textbf{Hardware acceleration} -- 
The most time-consuming component in PDNS is the homomorphic ciphertext evaluation in PIR. This bottleneck can be eliminated by dedicated hardware acceleration tools~\cite{f1,craterlake,cammarota2022intel,IntelHEXLFPGA,geelen2023basalisc}. Among these, F1 and Intel HERACLES~\cite{cammarota2022intel} report potential reductions of the processing time by four or five orders of magnitude, which will make PDNS's performance comparable with DoH and better than ODoH (Figure~\ref{fig:eval:query-time}) while providing better privacy protection. In contrast, existing solutions, \eg ODoH, ReR-Less DNS, and DoHoT, are constrained by network latency, hence leaving little space for performance improvements.

\subsection{Privacy and Security}
\label{sec:eval:privacy_sec}
\noindent
\textbf{Information exposure to ANSes} -- %
\pirdns utilizes the presence of \RRs to decrease the frequency of iterative DNS lookups, thereby reducing information exposure to ANSes. In our simulation, we observed a cache miss rate of less than 10\%. Within this rate, 90\% was caused by expired entries, and 10\% by uncached entries. While our cache miss rate is lower than some previous studies (30\% was reported~\cite{DBLP:journals/ccr/CallahanAR13}), the pattern of cache misses being primarily due to expired entries is consistent with these findings. Therefore, in 90\% of cache miss cases in \pirdns, users only need to contact the final ANSes, minimizing exposure to non-final ANSes. 

In short, \pirdns cuts down 70\% to 90\% of the ANS queries required by \RR-Less DNS. For the remaining traffic directed to ANSes, only roughly 10\% require interactions with non-final ANSes. Such limited exposure will likely disrupt the frequencies of sub-domains observed at non-final ANSes, similar to Figure~\ref{fig:eval:domain-frequency-change}. This achievement aligns with our goal to minimize user interactions with ANSes and reduce privacy risks, even though we consider these risks acceptable today. In contrast, \RR-Less DNS leads to the exposure of all user information to ANSes and, on average, causes a 4x to 16x increase in their workload compared to the current norms~\cite{cache-effect-19}. The impact is more pronounced at ANSes for popular domains, where the workload can increase by $\sim$100x. %

\vspace{0.05in}
\noindent
\textbf{Resilience to timing attacks} -- Next, we investigate the efficacy of employing random response delays as a strategic defense mechanism against timing attacks. Comprehensive findings outlined in Appendix~\ref{sec:appendix:timing-attack} substantiate the fact that our proposed solution effectively thwarts such attacks when the delay duration matches or exceeds the query interval. Additionally, these results unveil a trade-off dynamic between the average response forwarding delay and the magnitude of defense effectiveness. 

\vspace{0.05in}
\noindent
\textbf{Resilience to reflection attacks} -- We proceed to evaluate \pirdns resilience to reflection attacks. We simulate attackers launching reflection attacks by pretending to experience cache misses. As Figure~\ref{fig:eval:attack} shows, without the security feature introduced in \cref{sec:security}, a single attacker can generate over 100MB of traffic per second, and the traffic grows linearly as the number of attackers grows. In contrast, our security mechanism reduces significantly and puts a deterministic cap on the reflection traffic. Given the space constraints, we leave the full result analysis in Appendix~\ref{sec:appendix:security-effect}.

\vspace{0.05in}
\noindent
\textbf{Preserving regional access pattern} -- 
In addition to the collusion resistance guaranteed by its cryptographic theorems~\cite{regevfhe,gswfhe}, \pirdns features an (arguable) privacy benefit that might be ahead of its time: preserving regional access patterns from the \RR. %
To demonstrate this, please refer to Figure~\ref{fig:eval:domain-frequency-change} which shows the CDF of domain (hostname or SLD) popularity rank difference (absolute value) obtained with and without PIR (ground truth). For instance, ``cnn.com'' has a ranking difference of 23,270 since it is ranked 104,614th without \pirdns but 81,344th assuming \pirdns. This figure relies on a trace of 20 million DNS queries of 122K unique domains we collected. %
Overall, the figure confirms the preservation of regional pattern by \pirdns as the overall ranking has been disrupted. For example,  99\% of the 122K hostnames in our traces have an absolute ranking difference larger than 100 (and 40\% larger than 10K). Similar results apply to SLD.

\subsection{User Experience (Web Performance)}
\label{sec:eval:page-load}
We use web performance as a proxy to evaluate \pirdns's real-world performance and its impact on end-user experience. This is done via WebPageTest~\cite{webpagetest} to automate Chrome (on Linux) and both perform web page loads and collect telemetry, \eg classic web performance metrics: FirstContentful Paint (FCP), Time To Interactive (TTI), and PageLoadTime (PLT). We target the top 1,000 websites from Tranco~\cite{tranco} which are loaded 5 times per configuration and then report the median of each metric. We test each website using all DNS solutions studied so far except \RR-Less DNS since no DoH is supported by ANSes today, and assuming HTTPS connection reuse, since it is a realistic behavior of different OSes and browsers~\cite{doh-reuse-verify}. We use Cloudflare as a public \RR for existing DNS protocols, and our \RR for \pirdns while injecting the latency measured between testing client and Cloudflare (2-3ms). Note that our testing machine connects to the Internet with a symmetric upload/download bandwidth of $\sim$100 Mbps. 

Overall, the Web performance tests (Figure~\ref{fig:eval:web-load-time}) confirm the results of the query duration (Figure~\ref{fig:eval:query-time}), with DoUDP being the fastest protocol and DoHoT being the slowest. When using a small cache, \pirdns has performance comparable with ODoH, with a difference of less than 100~ms per metric which can hardly be perceived by the user. Still, ODoH outperforms \pirdns with a large cache, saving $\sim$900~ms for FCP/TTI which may indeed be perceived by the user, but at the cost of user privacy.

\section{Related Work}
\label{sec:related}
\noindent
\textbf{DNS} --  In addition to the mainstream DNS systems discussed in \cref{sec:background} and evaluated in \cref{sec:eval}, privacy-aimed amendments to these proposals exist. For example, %
Di Bella et al.~\cite{di2013secret} propose a secret-sharing scheme for anonymous DNS queries relying on a peer-to-peer proxy network. %
However, it is not robust when there are many malicious users in the system. EncDNS~\cite{herrmann2014encdns} has a similar system design to ODoH but is less flexible in %
key distribution mechanism, deployability, and compatibility. %
PINOT~\cite{wang2021programmable} proposes to obfuscate DNS traffic at the border of a trusted network to preserve user anonymity.
None of these solutions satisfy our privacy goals.

\vspace{0.05in}
\noindent
\textbf{PIR-based DNS systems} -- Few previous studies consider adopting PIR into DNS. Differently from \pirdns, they require significant modifications of the modern DNS architecture. 
A recent efficient stateful PIR benchmarks its performance on private DNS queries~\cite{piano-pir} assuming a single PIR server storing all the DNS records. This ignores the self-governing nature of the DNS across all ANSes. It is also not practical from the performance perspective. %
Lu and Tsudik~\cite{lu2009towards} propose a DNS system based on distributed hash tables and single-server PIR, where DNS records are stored in a table split into chunks %
assigned to different \RRs. %
The main drawback of this is that a query reveals the ``chunk'' of DNS records it is contained in, which can be used to analyze user preferences. Previous work also consider DNS with two-server PIR~\cite{zhao2007two,shi2021puncturable}, which does not offer %
collusion resistance. %

\section{Discussion and Limitations}
\label{sec:discussion}

\noindent\textbf{Limitation on anycast} -- \pirdns does not support \textit{anycast}~\cite{partridge1993host, rfc-anycast}, a network addressing and routing methodology which allows a single IP address to be shared by multiple \RRs around the world~\cite{google_dns, cloudflare_dns}. This is because a \RR's cache is populated by participating final ANSes. If anycast were to be implemented for \pRR, an ANS located in a different area than the user would route the DNS response to a different \RR than the original one queried by the user. %

\vspace{0.05in}
\noindent\textbf{Cluster of \pRR} -- In a realistic deployment, a \pRR consists of potentially multiple clusters of resolvers located at different locations. The user would select the closest location either manually or by using some self-configuring software. As this selection occurs, the user would register to a \RR which implies learning about the cache size $N$ and compute and share ${\sf pk}$ (see \cref{sec:overview:arch}). We assume that servers within the same cluster (\ie one single IP) share a synchronized cache, \ie having same size and content. This implies that ${\sf pk}$ can be shared among such servers, and load balancing of both queries and DNS record updates is straightforward.

\vspace{0.05in}
\noindent\textbf{DoH support at ANSes} -- As noted in \cref{sec:deep:discussion}, ANSes currently do not support DoH. This is mainly due to the lack of incentives: currently, ANSes only communicate with \RRs, where privacy exposure in this link is less concerned. Thus, the benefits of supporting DoH on ANSes do not outweigh the costs. However, with the introduction of \pirdns, ANSes will communicate directly with end-users. Implementing DoH at ANS would enhance the privacy protections for websites under ANS-controlled domains, which is particularly valuable for those offering sensitive content or for customers who prioritize privacy. This provides a much stronger incentive for ANSes to support DoH. While we cannot precisely predict future support of DoH by ANSes, we anticipate it will increase rapidly, like the history of HTTPS, once PDNS is deployed.

\vspace{0.05in}
\noindent
\textbf{\RR in a Trusted Execution Environment (TEE)} -- This approach allows to verify, through a hardware vendor, that the \RR's code corresponds to an open-source non-logging implementation. Relying on TEEs~\cite{tee-15} inherently introduces a reliance on hardware vendors, which contradicts our objective of eliminating non-collusion agreements. It also adds another point of failure, as TEEs are not free from bugs and security breaches~\cite{sgx-breaches}. 
Further, TEEs often have limited memory~\cite{tee-memory-1}, constraining the \RR functionality.
Finally, it imposes a great burden on TEE vendors while negatively impacting user experience, as every \RR attestation requires communication with an attestation server at TEE vendor.

\vspace{0.05in}
\noindent
\textbf{Multi-service Internet companies} -- 
In some cases, companies operate diverse businesses. For instance, Google and Cloudflare simultaneously provide \RR and (final) ANS services. This positions them as critical junctions in the Internet, aggregating a wealth of information from these different operations. DoHoT and \pirdns stand out in offering optimal joint privacy protection to users in such scenarios. %
Notably, \RR-Less DoH and \pirdns require ANSes to receive direct queries from users, potentially revealing their identities. As discussed earlier, we deem this acceptable since this information would be accessible to final ANSes through other means anyway. Still, we aim to minimize information exposure. \RR-Less DoH exposes 100\% of the DNS queries to all ANSes. In contrast, \pirdns limits this by directing only 3\% of the queries to non-final ANSes and 30\% to final ANSes (\cref{sec:eval:privacy_sec}). Additionally, \pirdns provides security guarantees to ANSes thanks to its query validation scheme (\cref{sec:security}), which is instead not viable in \RR-Less DNS. %
DoHoT, on the other hand, sacrifices performance but preserves information exposure.
We therefore conclude that \pirdns and DoHoT are the preferable choices here, minimizing information exposure jointly at both \RR and ANS for multi-service companies. %

\vspace{0.05in}
\noindent
\textbf{Third-party final ANSes} -- Many small/medium businesses rely on third-party final ANS services such as Route 53~\cite{awsroute53}. In this case, \pirdns direct ANSes queries -- in presence of cache misses -- would leak some information to a third party which does not already have access to this information via direct traffic, \eg HTTP(S) in case of a webpage. The solution to handle such privacy leaks is for Route 53 to implement support for the PIR protocol used by \pRR. In this setup, the \pRR's cache record for a domain name would include only the IP address of the final ANS and a distinct flag, signaling users to communicate directly with the final ANS using the PIR protocol. We leave this as future work.

\vspace{0.05in}
\noindent
\textbf{Resistance to Internet centralization} -- The trend of Internet centralization has been evident over the past decades, with a few hyper-giants controlling major portions of Internet resources. This centralization impacts the DNS as well~\cite{d-dns-1-20, d-dns-3-23}, with studies~\cite{dns-consolidation-20} showing that Google and Cloudflare handle about half of all \RR queries. Such centralization threatens user privacy. Existing DNS privacy solutions, like rotating \RRs or proxies, are also negatively affected by this trend due to the decreasing number of available options. In contrast, \pirdns resists centralization as it ensures \RRs are blind to the queries they receive. In fact, more population using the same \pRR improves user privacy by raising cache hit rates and complicating potential cache miss analysis for \RRs. Similar resistance applies to the centralization of final ANSes following the above discussion of ``Third-Party Final ANSes''. Overall, once scalability issues are addressed by near-future hardware accelerators, \pirdns will be more advantageous than other solutions in the trend of Internet centralization.

\vspace{0.05in}
\noindent
\textbf{User privacy beyond DNS} -- This paper focuses on DNS privacy, a critical aspect of online privacy, as DNS serves as a central point where all online activity converges. Unlike web requests, which are decentralized and require compromising multiple ANSes or web providers to gain comprehensive insights, DNS represents a unique vulnerability and therefore demands stronger privacy guarantees.

Nevertheless, we acknowledge that \pirdns is limited to DNS privacy and does not address privacy concerns beyond DNS. Users may still leak information with \pirdns when interacting with other online services, such as web browsing. Various solutions exist for different online activities, such as search and browsing~\cite{patel2016tor, private-browsing-10, private-browsing-18, snatch-24, lightweb-pir-web-23, tiptoe-private-search-23}. 
Notably, one alternative to PDNS -- DoHoT -- leverages Tor which extends privacy protection beyond DNS to general web activities. %
While this paper advocates for \pirdns, this does not imply opposition to DoHoT or Tor as a whole. On the contrary, \pirdns can complement Tor and other privacy-preserving technologies.

\vspace{0.1in}
\section{Conclusion}
\label{sec:conclusion}
DNS still suffers from privacy infringements~\cite{trump_case} despite the recent adoption of encryption (DoH), and new proposals to de-associate a user identity with her query (ODoH). The main culprit of such privacy violation is the \textit{recursive resolver} (\RR) whose role is to provide distributed caching to DNS enabling its scalability and fast speed. While some recent provocative proposals argue for its removal -- which would naturally address such privacy concerns -- such approaches bring up new security issues and thus are unfeasible. 
We instead proposed \pirdns, to our knowledge the first proof-of-concept to allow \RRs to operate privately leveraging single-server PIR.
The integration of DNS and PIR is not straightforward and our design navigates many performance and privacy trade-offs. Although \pirdns slightly relaxes the collusion assumptions concerning millions of authoritative name servers, it significantly enhances DNS privacy beyond existing methods.
\pirdns is financially viable to operate today and has significant potential for further performance improvements, despite that it already outperforms the only other solution, DoHoT, that offers the same level of privacy protection.

\newpage

\Urlmuskip=0mu plus 1mu\relax
\bibliographystyle{abbrv} 
\bibliography{main}

\begin{thebibliography}{100}

\bibitem{aws-pricing}
{Add Service - AWS Pricing Calculator}.
\newblock \url{https://calculator.aws/#/addService/ec2-enhancement}. Accessed
  in 2024.

\bibitem{awsroute53}
Aws route 53.
\newblock \url{https://aws.amazon.com/route53/}.

\bibitem{bind}
{Bind 9}.
\newblock \url{https://www.isc.org/bind/}.

\bibitem{dig-update}
{BIND DoH Update}.
\newblock \url{https://www.isc.org/blogs/bind-doh-update-2021/}.

\bibitem{blyss}
{Blyss (Previously Spiral-rs)}.
\newblock \url{https://github.com/blyssprivacy/sdk}.

\bibitem{dig}
{dig(1): DNS lookup utility - Linux man page - Die.net}.
\newblock \url{https://linux.die.net/man/1/dig}.

\bibitem{dnscrypt}
Dnscrypt.
\newblock \url{https://www.dnscrypt.org/}.

\bibitem{gcp-pricing}
{Google Cloud Pricing Calculator}.
\newblock \url{https://calculator.aws/#/addService/ec2-enhancement}. Accessed
  in 2024.

\bibitem{tee-memory-1}
{Intel® Software Guard Extensions (Intel® SGX) SDK for Linux* OS Developer
  Reference}.
\newblock
  \url{https://download.01.org/intel-sgx/sgx-linux/2.10/docs/Intel_SGX_Developer_Reference_Linux_2.10_Open_Source.pdf}.
  Accessed in 2024.

\bibitem{avx2}
{Intrinsics for Intel® Advanced Vector Extensions 2 (Intel® AVX2)}.
\newblock
  \url{https://www.intel.com/content/www/us/en/develop/documentation/cpp-compiler-developer-guide-and-reference/top/compiler-reference/intrinsics/intrinsics-for-avx2.html}.

\bibitem{linuxtc}
Linux traffic control (tc).
\newblock \url{https://man7.org/linux/man-pages/man8/tc.8.html}.

\bibitem{mysterium}
{Mysterium network: Censorship free Internet for all}.
\newblock \url{https://mysterium.network/}.

\bibitem{nvlink}
{NVIDIA NVSWITCH: The World’s Highest-Bandwidth On-Node Switch}.
\newblock
  \url{https://images.nvidia.com/content/pdf/nvswitch-technical-overview.pdf}.
  Accessed in 2024.

\bibitem{trump_case}
{\em \textup{USA v. Sussmann, 2022 WL 1124755 (D.D.C. 2022)}}.

\bibitem{tor-code-audit-17}
{Tor Code Audit Finds 17 Vulnerabilities}.
\newblock
  \url{https://www.securityweek.com/tor-code-audit-finds-17-vulnerabilities/}.
  Accessed in 2024.

\bibitem{tranco}
{Tranco: A Research-Oriented Top Sites Ranking Hardened Against Manipulation}.
\newblock \url{https://tranco-list.eu/}.

\bibitem{webpagetest}
{WebPageTest}.
\newblock \url{https://github.com/WPO-Foundation/webpagetest}.

\bibitem{cloudflare_blog}
{What is DNS? | How DNS works | Cloudflare}.
\newblock \url{https://www.cloudflare.com/learning/dns/what-is-dns/}.

\bibitem{bind-popular}
{When to replace BIND DNS}.
\newblock \url{https://bluecatnetworks.com/blog/when-to-replace-bind-dns/}.

\bibitem{rfc_dns}
{Domain names - implementation and specification}.
\newblock RFC 1035, Nov. 1987.

\bibitem{cloudflare_dns}
{Cloudflare DNS}, 2022.
\newblock \url{https://www.cloudflare.com/dns/}.

\bibitem{firefox_doh}
{Firefox DNS-over-HTTPS.}, 2022.
\newblock \url{https://support.mozilla.org/en-US/kb/firefox-dns-over-https}.

\bibitem{google_public_doh}
{Google Public DNS: DNS over HTTPS (DoH)}, 2022.
\newblock \url{https://developers.google.com/speed/public-dns/docs/doh}.

\bibitem{google_dns}
{Introduction to Google Public DNS}, 2022.
\newblock \url{https://developers.google.com/speed/public-dns/docs/intro}.

\bibitem{chrome_doh}
{The chromium projects: DNS over HTTPS (aka DoH)}, 2022.
\newblock \url{https://www.chromium.org/developers/dns-over-https/}.

\bibitem{sgx-breaches}
{Intel patches up SGX best it can after another load of security holes found},
  2023.
\newblock \url{https://www.theregister.com/2023/02/15/intel_sgx_vulns/}.

\bibitem{rfc-anycast}
J.~Abley and K.~E. Lindqvist.
\newblock Operation of anycast services.
\newblock {\em {RFC}}, 4786:1--24, 2006.

\bibitem{private-browsing-10}
G.~Aggarwal, E.~Bursztein, C.~Jackson, and D.~Boneh.
\newblock An analysis of private browsing modes in modern browsers.
\newblock In {\em 19th {USENIX} Security Symposium, Washington, DC, USA, August
  11-13, 2010, Proceedings}, pages 79--94. {USENIX} Association, 2010.

\bibitem{ali2021communication}
A.~Ali, T.~Lepoint, S.~Patel, M.~Raykova, P.~Schoppmann, K.~Seth, and K.~Yeo.
\newblock $\{$Communication--Computation$\}$ trade-offs in $\{$PIR$\}$.
\newblock In {\em 30th USENIX Security Symposium (USENIX Security 21)}, pages
  1811--1828, 2021.

\bibitem{dns-amplification-revisit}
M.~Anagnostopoulos, G.~Kambourakis, P.~Kopanos, G.~Louloudakis, and
  S.~Gritzalis.
\newblock {DNS} amplification attack revisited.
\newblock {\em Comput. Secur.}, 39:475--485, 2013.

\bibitem{seal-pir}
S.~Angel, H.~Chen, K.~Laine, and S.~Setty.
\newblock Pir with compressed queries and amortized query processing.
\newblock In {\em 2018 IEEE symposium on security and privacy (SP)}, pages
  962--979. IEEE, 2018.

\bibitem{rfc-dnssec}
R.~Arends, R.~Austein, M.~Larson, D.~Massey, and S.~Rose.
\newblock {DNS} security introduction and requirements.
\newblock {\em {RFC}}, 4033:1--21, 2005.

\bibitem{covert-adversery-10}
Y.~Aumann and Y.~Lindell.
\newblock Security against covert adversaries: Efficient protocols for
  realistic adversaries.
\newblock {\em J. Cryptol.}, 23(2):281--343, 2010.

\bibitem{dns-dos-ccs08}
H.~Ballani and P.~Francis.
\newblock Mitigating {DNS} dos attacks.
\newblock In {\em Proceedings of the 2008 {ACM} Conference on Computer and
  Communications Security, {CCS} 2008, Alexandria, Virginia, USA, October
  27-31, 2008}, pages 189--198. {ACM}, 2008.

\bibitem{rfc-query-minimization-21}
S.~Bortzmeyer, R.~Dolmans, and P.~Hoffman.
\newblock {DNS} query name minimisation to improve privacy.
\newblock {\em {RFC}}, 9156:1--11, 2021.

\bibitem{bottger2019empirical}
T.~B{\"o}ttger, F.~Cuadrado, G.~Antichi, E.~L. Fernandes, G.~Tyson, I.~Castro,
  and S.~Uhlig.
\newblock An empirical study of the cost of dns-over-https.
\newblock In {\em Proceedings of the Internet Measurement Conference}, pages
  15--21, 2019.

\bibitem{bgvfhe}
Z.~Brakerski, C.~Gentry, and V.~Vaikuntanathan.
\newblock (leveled) fully homomorphic encryption without bootstrapping.
\newblock In {\em Proceedings of the 3rd Innovations in Theoretical Computer
  Science Conference}, ITCS '12, page 309–325, New York, NY, USA, 2012.
  {ACM}.

\bibitem{brakerski2014leveled}
Z.~Brakerski, C.~Gentry, and V.~Vaikuntanathan.
\newblock (leveled) fully homomorphic encryption without bootstrapping.
\newblock {\em ACM Transactions on Computation Theory (TOCT)}, 6(3):1--36,
  2014.

\bibitem{burton2024respire}
A.~Burton, S.~J. Menon, and D.~J. Wu.
\newblock Respire: High-rate pir for databases with small records.
\newblock {\em Cryptology ePrint Archive}, 2024.

\bibitem{DBLP:journals/ccr/CallahanAR13}
T.~Callahan, M.~Allman, and M.~Rabinovich.
\newblock On modern {DNS} behavior and properties.
\newblock {\em Comput. Commun. Rev.}, 43(3):7--15, 2013.

\bibitem{doh-reuse-verify}
P.~Callejo, M.~Bagnulo, J.~G. Ruiz, A.~Lutu,
  A.~Garc{\'{\i}}a{-}Mart{\'{\i}}nez, and R.~Cuevas.
\newblock Measuring doh with web ads.
\newblock {\em Comput. Networks}, 2022.

\bibitem{cammarota2022intel}
R.~Cammarota.
\newblock Intel heracles: homomorphic encryption revolutionary accelerator with
  correctness for learning-oriented end-to-end solutions.
\newblock In {\em Proceedings of the 2022 on Cloud Computing Security
  Workshop}, pages 3--3, 2022.

\bibitem{castillo2017contributions}
S.~Castillo~P{\'e}rez.
\newblock {\em Contributions to privacy and anonymity on the Internet: domain
  name system and second-generation onion routing}.
\newblock 2017.

\bibitem{DBLP:conf/imc/ChhabraM0BW21}
R.~Chhabra, P.~Murley, D.~Kumar, M.~Bailey, and G.~Wang.
\newblock Measuring dns-over-https performance around the world.
\newblock In {\em {IMC} '21: {ACM} Internet Measurement Conference, Virtual
  Event, USA, November 2-4, 2021}. {ACM}, 2021.

\bibitem{pir:CKGM}
B.~Chor, E.~Kushilevitz, O.~Goldreich, and M.~Sudan.
\newblock Private information retrieval.
\newblock {\em J. {ACM}}, 45(6):965--981, 1998.

\bibitem{rfc-ecs}
C.~Contavalli, W.~van~der Gaast, D.~C. Lawrence, and W.~Kumari.
\newblock Client subnet in {DNS} queries.
\newblock {\em {RFC}}, 7871:1--30, 2016.

\bibitem{corrigan2022single}
H.~Corrigan-Gibbs, A.~Henzinger, and D.~Kogan.
\newblock Single-server private information retrieval with sublinear amortized
  time.
\newblock {\em Cryptology ePrint Archive}, 2022.

\bibitem{corrigan2020private}
H.~Corrigan-Gibbs and D.~Kogan.
\newblock Private information retrieval with sublinear online time.
\newblock In {\em Advances in Cryptology -- EUROCRYPT 2020}, pages 44--75,
  Cham, 2020. Springer International Publishing.

\bibitem{csikor2021privacy}
L.~Csikor, H.~Singh, M.~S. Kang, and D.~M. Divakaran.
\newblock Privacy of dns-over-https: Requiem for a dream?
\newblock In {\em 2021 IEEE European Symposium on Security and Privacy
  (EuroS\&P)}, pages 252--271. IEEE, 2021.

\bibitem{rfc_edns}
J.~Damas, M.~Graff, and P.~Vixie.
\newblock Extension mechanisms for {DNS} {(EDNS(0))}.
\newblock {\em {RFC}}, 6891:1--16, 2013.

\bibitem{lightweb-pir-web-23}
E.~Dauterman and H.~Corrigan{-}Gibbs.
\newblock Lightweb: Private web browsing without all the baggage.
\newblock In {\em Proceedings of the 22nd {ACM} Workshop on Hot Topics in
  Networks, HotNets 2023, Cambridge, MA, USA, November 28-29, 2023}, pages
  287--294. {ACM}, 2023.

\bibitem{davidson2023frodopir}
A.~Davidson, G.~Pestana, and S.~Celi.
\newblock Frodopir: Simple, scalable, single-server private information
  retrieval.
\newblock {\em Proceedings on Privacy Enhancing Technologies}, 2023.

\bibitem{query-minimization-1-19}
W.~B. de~Vries, Q.~Scheitle, M.~M{\"{u}}ller, W.~Toorop, R.~Dolmans, and R.~van
  Rijswijk{-}Deij.
\newblock A first look at {QNAME} minimization in the domain name system.
\newblock In D.~R. Choffnes and M.~P. Barcellos, editors, {\em Passive and
  Active Measurement - 20th International Conference, {PAM} 2019, Puerto Varas,
  Chile, March 27-29, 2019, Proceedings}, volume 11419 of {\em Lecture Notes in
  Computer Science}, pages 147--160. Springer, 2019.

\bibitem{di2013secret}
G.~Di~Bella, C.~Barcellona, and I.~Tinnirello.
\newblock A secret sharing scheme for anonymous dns queries.
\newblock In {\em AEIT Annual Conference 2013}, pages 1--5. IEEE, 2013.

\bibitem{rfc_dotcp}
J.~Dickinson, S.~Dickinson, R.~Bellis, A.~Mankin, and D.~Wessels.
\newblock {DNS Transport over TCP - Implementation Requirements}.
\newblock RFC 7766, Mar. 2016.

\bibitem{dingledine2004tor}
R.~Dingledine, N.~Mathewson, and P.~Syverson.
\newblock Tor: The second-generation onion router.
\newblock Technical report, Naval Research Lab Washington DC, 2004.

\bibitem{rhine-e2e-23}
H.~Duan, R.~Fischer, J.~Lou, S.~Liu, D.~A. Basin, and A.~Perrig.
\newblock {RHINE:} robust and high-performance internet naming with {E2E}
  authenticity.
\newblock In {\em 20th {USENIX} Symposium on Networked Systems Design and
  Implementation, {NSDI} 2023, Boston, MA, April 17-19, 2023}, pages 531--553.
  {USENIX} Association, 2023.

\bibitem{Fan_somewhatpractical}
J.~Fan and F.~Vercauteren.
\newblock Somewhat practical fully homomorphic encryption.
\newblock {\em IACR Cryptology ePrint Archive}, 2012.

\bibitem{fvhe}
J.~Fan and F.~Vercauteren.
\newblock Somewhat practical fully homomorphic encryption.
\newblock Cryptology ePrint Archive, Paper 2012/144, 2012.
\newblock \url{https://eprint.iacr.org/2012/144}.

\bibitem{cache-effect-19}
K.~Fujiwara, A.~Sato, and K.~Yoshida.
\newblock Cache effect of shared {DNS} resolver.
\newblock {\em {IEICE} Trans. Commun.}, 102-B(6):1170--1179, 2019.

\bibitem{geelen2023basalisc}
R.~Geelen, M.~Van~Beirendonck, H.~V. Lima~Pereira, B.~Huffman, T.~McAuley,
  B.~Selfridge, D.~Wagner, G.~Dimou, I.~Verbauwhede, F.~Vercauteren, et~al.
\newblock Basalisc: programmable hardware accelerator for bgv fully homomorphic
  encryption.
\newblock {\em IACR Transactions on Cryptographic Hardware and Embedded
  Systems}, 2023(4):32--57, 2023.

\bibitem{gentryfhe}
C.~Gentry.
\newblock {\em A Fully Homomorphic Encryption Scheme}.
\newblock PhD thesis, Stanford, CA, USA, 2009.
\newblock AAI3382729.

\bibitem{gswfhe}
C.~Gentry, A.~Sahai, and B.~Waters.
\newblock Homomorphic encryption from learning with errors:
  Conceptually-simpler, asymptotically-faster, attribute-based.
\newblock In {\em Advances in Cryptology -- CRYPTO 2013}, pages 75--92, Berlin,
  Heidelberg, 2013. Springer Berlin Heidelberg.

\bibitem{DBLP:conf/pet/GuhaF07}
S.~Guha and P.~Francis.
\newblock Identity trail: Covert surveillance using {DNS}.
\newblock In {\em Privacy Enhancing Technologies, 7th International Symposium,
  {PET} 2007 Ottawa, Canada, June 20-22, 2007, Revised Selected Papers}, volume
  4776 of {\em Lecture Notes in Computer Science}, pages 153--166. Springer,
  2007.

\bibitem{private-browsing-18}
H.~Habib, J.~Colnago, V.~Gopalakrishnan, S.~Pearman, J.~Thomas, A.~Acquisti,
  N.~Christin, and L.~F. Cranor.
\newblock Away from prying eyes: Analyzing usage and understanding of private
  browsing.
\newblock In M.~E. Zurko and H.~R. Lipford, editors, {\em Fourteenth Symposium
  on Usable Privacy and Security, {SOUPS} 2018, Baltimore, MD, USA, August
  12-14, 2018}, pages 159--175. {USENIX} Association, 2018.

\bibitem{hal-adot-operational-considerations-02}
K.~M. Henderson, T.~April, and J.~Livingood.
\newblock {Authoritative DNS-over-TLS Operational Considerations}.
\newblock Internet-Draft draft-hal-adot-operational-considerations-02, Internet
  Engineering Task Force, Aug. 2019.
\newblock Work in Progress.

\bibitem{tiptoe-private-search-23}
A.~Henzinger, E.~Dauterman, H.~Corrigan{-}Gibbs, and N.~Zeldovich.
\newblock Private web search with tiptoe.
\newblock In J.~Flinn, M.~I. Seltzer, P.~Druschel, A.~Kaufmann, and J.~Mace,
  editors, {\em Proceedings of the 29th Symposium on Operating Systems
  Principles, {SOSP} 2023, Koblenz, Germany, October 23-26, 2023}, pages
  396--416. {ACM}, 2023.

\bibitem{simplepir}
A.~Henzinger, M.~M. Hong, H.~Corrigan-Gibbs, S.~Meiklejohn, and
  V.~Vaikuntanathan.
\newblock One server for the price of two: Simple and fast single-server
  private information retrieval.
\newblock Cryptology ePrint Archive, Paper 2022/949, 2022.
\newblock \url{https://eprint.iacr.org/2022/949}.

\bibitem{herrmann2014encdns}
D.~Herrmann, K.-P. Fuchs, J.~Lindemann, and H.~Federrath.
\newblock Encdns: A lightweight privacy-preserving name resolution service.
\newblock In {\em European Symposium on Research in Computer Security}, pages
  37--55. Springer, 2014.

\bibitem{d-dns-2-20}
N.~P. Hoang, I.~Lin, S.~Ghavamnia, and M.~Polychronakis.
\newblock K-resolver: Towards decentralizing encrypted {DNS} resolution.
\newblock {\em CoRR}, abs/2001.08901, 2020.

\bibitem{rfc_doh}
P.~E. Hoffman and P.~McManus.
\newblock {DNS Queries over HTTPS (DoH)}.
\newblock RFC 8484, Oct. 2018.

\bibitem{ietf-dprive-opportunistic-adotq-02}
P.~E. Hoffman and P.~van Dijk.
\newblock {Recursive to Authoritative DNS with Unauthenticated Encryption}.
\newblock Internet-Draft draft-ietf-dprive-opportunistic-adotq-02, Internet
  Engineering Task Force, Apr. 2021.
\newblock Work in Progress.

\bibitem{d-dns-1-20}
A.~Hounsel, K.~Borgolte, P.~Schmitt, and N.~Feamster.
\newblock {D-DNS:} towards re-decentralizing the {DNS}.
\newblock {\em CoRR}, abs/2002.09055, 2020.

\bibitem{hounsel2021encryption}
A.~Hounsel, P.~Schmitt, K.~Borgolte, and N.~Feamster.
\newblock Encryption without centralization: distributing dns queries across
  recursive resolvers.
\newblock In {\em Proceedings of the Applied Networking Research Workshop},
  pages 62--68, 2021.

\bibitem{rfc_dot}
Z.~Hu, L.~Zhu, J.~Heidemann, A.~Mankin, D.~Wessels, and P.~E. Hoffman.
\newblock {Specification for DNS over Transport Layer Security (TLS)}.
\newblock RFC 7858, May 2016.

\bibitem{dpf-pir}
D.~Kales, O.~Omolola, and S.~Ramacher.
\newblock {Revisiting User Privacy for Certificate Transparency}.
\newblock In {\em 2019 IEEE European Symposium on Security and Privacy (EuroS
  P)}, pages 432--447, 2019.

\bibitem{katz2020introduction}
J.~Katz and Y.~Lindell.
\newblock {\em Introduction to modern cryptography}.
\newblock CRC press, 2020.

\bibitem{rfc9230}
E.~Kinnear, P.~McManus, T.~Pauly, T.~Verma, and C.~A. Wood.
\newblock {Oblivious DNS over HTTPS}.
\newblock RFC 9230, June 2022.

\bibitem{checklist-pir}
D.~Kogan and H.~Corrigan-Gibbs.
\newblock {Private Blocklist Lookups with Checklist}.
\newblock In {\em 30th USENIX Security Symposium (USENIX Security 21)}, pages
  875--892. USENIX Association, Aug. 2021.

\bibitem{kosek2022measuring}
M.~Kosek, T.~V. Doan, S.~Huber, and V.~Bajpai.
\newblock Measuring dns over tcp in the era of increasing dns response sizes: A
  view from the edge.
\newblock {\em ACM SIGCOMM Computer Communication Review}, 52(2), 2022.

\bibitem{dns-hijacking-19}
B.~Krebs.
\newblock {A Deep Dive on the Recent Widespread DNS Hijacking Attacks}, 2019.
\newblock
  \url{https://krebsonsecurity.com/2019/02/a-deep-dive-on-the-recent-widespread-dns-hijacking-attacks/}.

\bibitem{d-dns-3-23}
R.~Kumar and F.~E. Bustamante.
\newblock Reclaiming privacy and performance over centralized {DNS}.
\newblock {\em CoRR}, abs/2302.13274, 2023.

\bibitem{muODNS-23}
J.~Kurihara, T.~Tanaka, and T.~Kubo.
\newblock {\(\mu\)}odns: {A} distributed approach to {DNS} anonymization with
  collusion resistance.
\newblock {\em Comput. Networks}, 237:110078, 2023.

\bibitem{li2024hintless}
B.~Li, D.~Micciancio, M.~Raykova, and M.~Schultz-Wu.
\newblock Hintless single-server private information retrieval.
\newblock In {\em Annual International Cryptology Conference}, pages 183--217.
  Springer, 2024.

\bibitem{lu2009towards}
Y.~LU.
\newblock Towards plugging privacy leaks in domain name system, cornell
  university library.
\newblock {\em http://arxiv. org/abs/0910.2472}, 2009.

\bibitem{rlwe}
V.~Lyubashevsky, C.~Peikert, and O.~Regev.
\newblock On ideal lattices and learning with errors over rings.
\newblock {\em J. ACM}, 60(6), nov 2013.

\bibitem{incremental-pir}
Y.~Ma, K.~Zhong, T.~Rabin, and S.~Angel.
\newblock {Incremental {Offline/Online} {PIR}}.
\newblock In {\em 31st USENIX Security Symposium (USENIX Security 22)}, Boston,
  MA, Aug. 2022. USENIX Association.

\bibitem{k-anonymity-vulnerable-07}
A.~Machanavajjhala, D.~Kifer, J.~Gehrke, and M.~Venkitasubramaniam.
\newblock \emph{L}-diversity: Privacy beyond \emph{k}-anonymity.
\newblock {\em {ACM} Trans. Knowl. Discov. Data}, 1(1):3, 2007.

\bibitem{query-minimization-2-23}
J.~Magnusson, M.~M{\"{u}}ller, A.~Brunstr{\"{o}}m, and T.~Pulls.
\newblock A second look at {DNS} {QNAME} minimization.
\newblock In A.~Brunstr{\"{o}}m, M.~Flores, and M.~Fiore, editors, {\em Passive
  and Active Measurement - 24th International Conference, {PAM} 2023, Virtual
  Event, March 21-23, 2023, Proceedings}, volume 13882 of {\em Lecture Notes in
  Computer Science}, pages 496--521. Springer, 2023.

\bibitem{IntelHEXLFPGA}
Y.~Meng, S.~Butt, Y.~Wang, Y.~Zhou, S.~Simoni, et~al.
\newblock {I}ntel {Homomorphic Encryption Acceleration Library for FPGAs}
  (version 2.0).
\newblock \url{https://github.com/intel/hexl-fpga}, 2022.

\bibitem{spiral-pir}
S.~J. Menon and D.~J. Wu.
\newblock \textsc{Spiral}: Fast, high-rate single-server {PIR} via {FHE}
  composition.
\newblock In {\em {IEEE} {S\&P}}, 2022.

\bibitem{menon2024ypir}
S.~J. Menon and D.~J. Wu.
\newblock Ypir: High-throughput single-server pir with silent preprocessing.
\newblock {\em Cryptology ePrint Archive}, 2024.

\bibitem{michael2002high}
M.~M. Michael.
\newblock High performance dynamic lock-free hash tables and list-based sets.
\newblock In {\em Proceedings of the fourteenth annual ACM symposium on
  Parallel algorithms and architectures}, pages 73--82, 2002.

\bibitem{dohot-cloudflare}
A.~Muffett.
\newblock {DNS over Tor}.
\newblock
  \url{https://developers.cloudflare.com/1.1.1.1/other-ways-to-use-1.1.1.1/dns-over-tor/}.

\bibitem{dohot-muffett}
A.~Muffett.
\newblock {DoHoT: making practical use of DNS over HTTPS over Tor}.
\newblock
  \url{https://medium.com/@alecmuffett/dohot-making-practical-use-of-dns-over-https-over-tor-ef58d04ca06a}.

\bibitem{mughees2021onionpir}
M.~H. Mughees, H.~Chen, and L.~Ren.
\newblock {OnionPIR: Response Efficient Single-Server PIR}.
\newblock In {\em Proceedings of the 2021 ACM SIGSAC Conference on Computer and
  Communications Security}, pages 2292--2306, 2021.

\bibitem{mughees2022vectorized}
M.~H. Mughees and L.~Ren.
\newblock Vectorized batch private information retrieval.
\newblock {\em Cryptology ePrint Archive}, 2022.

\bibitem{mughees2023simple}
M.~H. Mughees, I.~Sun, and L.~Ren.
\newblock Simple and practical amortized sublinear private information
  retrieval.
\newblock {\em Cryptology ePrint Archive}, 2023.

\bibitem{partridge1993host}
C.~Partridge, T.~Mendez, and W.~Milliken.
\newblock Host anycasting service.
\newblock Technical report, 1993.

\bibitem{patel2016tor}
N.~Patel.
\newblock Tor networking vulnerabilities and breaches.
\newblock 2016.

\bibitem{encrypting-rr-auth}
L.~Pernal-Stoddart.
\newblock {The Camel’s Back: Recursive to Authoritative DNS with Encryption}.
\newblock \url{https://www.centr.org/news/blog/ietf110-camel-back.html}.

\bibitem{dns-consolidation-20}
R.~Radu and M.~Hausding.
\newblock Consolidation in the dns resolver market--how much, how fast, how
  dangerous?
\newblock {\em Journal of Cyber Policy}, 5(1):46--64, 2020.

\bibitem{dns-snoop-imc20}
A.~Randall, E.~Liu, G.~Akiwate, R.~Padmanabhan, G.~M. Voelker, S.~Savage, and
  A.~Schulman.
\newblock Trufflehunter: Cache snooping rare domains at large public {DNS}
  resolvers.
\newblock In {\em {IMC} '20: {ACM} Internet Measurement Conference, Virtual
  Event, USA, October 27-29, 2020}, pages 50--64. {ACM}, 2020.

\bibitem{regevfhe}
O.~Regev.
\newblock On lattices, learning with errors, random linear codes, and
  cryptography.
\newblock {\em J. ACM}, 56(6), sep 2009.

\bibitem{regev2009lattices}
O.~Regev.
\newblock On lattices, learning with errors, random linear codes, and
  cryptography.
\newblock {\em Journal of the ACM (JACM)}, 56(6):1--40, 2009.

\bibitem{rescorla2018rfc}
E.~Rescorla.
\newblock {Rfc 8446: The transport layer security (tls) protocol version 1.3},
  2018.

\bibitem{rivera2020leveraging}
S.~Rivera, V.~K. Gurbani, S.~Lagraa, A.~K. Iannillo, and R.~State.
\newblock Leveraging ebpf to preserve user privacy for dns, dot, and doh
  queries.
\newblock In {\em Proceedings of the 15th International Conference on
  Availability, Reliability and Security}, pages 1--10, 2020.

\bibitem{tee-15}
M.~Sabt, M.~Achemlal, and A.~Bouabdallah.
\newblock Trusted execution environment: What it is, and what it is not.
\newblock In {\em 2015 {IEEE} TrustCom/BigDataSE/ISPA, Helsinki, Finland,
  August 20-22, 2015, Volume 1}, pages 57--64. {IEEE}, 2015.

\bibitem{f1}
N.~Samardzic, A.~Feldmann, A.~Krastev, S.~Devadas, R.~Dreslinski, C.~Peikert,
  and D.~Sanchez.
\newblock F1: A fast and programmable accelerator for fully homomorphic
  encryption.
\newblock In {\em MICRO-54: 54th Annual IEEE/ACM International Symposium on
  Microarchitecture}, MICRO '21, page 238–252, New York, NY, USA, 2021.
  {ACM}.

\bibitem{craterlake}
N.~Samardzic, A.~Feldmann, A.~Krastev, N.~Manohar, N.~Genise, S.~Devadas,
  K.~Eldefrawy, C.~Peikert, and D.~Sanchez.
\newblock Craterlake: A hardware accelerator for efficient unbounded
  computation on encrypted data.
\newblock In {\em Proceedings of the 49th Annual International Symposium on
  Computer Architecture}, ISCA '22, page 173–187, New York, NY, USA, 2022.
  {ACM}.

\bibitem{schmitt2019oblivious}
P.~Schmitt, A.~Edmundson, A.~Mankin, and N.~Feamster.
\newblock {Oblivious DNS: Practical privacy for DNS queries}.
\newblock {\em Proceedings on Privacy Enhancing Technologies},
  2019(2):228--244, 2019.

\bibitem{no-recursive-resolver-14}
K.~Schomp, M.~Allman, and M.~Rabinovich.
\newblock {DNS} resolvers considered harmful.
\newblock In {\em Proceedings of the 13th {ACM} Workshop on Hot Topics in
  Networks, HotNets-XIII, Los Angeles, CA, USA, October 27-28, 2014}, pages
  16:1--16:7. {ACM}, 2014.

\bibitem{akamai-dns}
K.~Schomp, O.~Bhardwaj, E.~Kurdoglu, M.~Muhaimen, and R.~K. Sitaraman.
\newblock Akamai {DNS:} providing authoritative answers to the world's queries.
\newblock In {\em {SIGCOMM}'20, Virtual Event, USA, August 10-14, 2020}, pages
  465--478. {ACM}, 2020.

\bibitem{schomp2016towards}
K.~Schomp, M.~Rabinovich, and M.~Allman.
\newblock Towards a model of dns client behavior.
\newblock In {\em International Conference on Passive and Active Network
  Measurement}, pages 263--275. Springer, 2016.

\bibitem{shi2021puncturable}
E.~Shi, W.~Aqeel, B.~Chandrasekaran, and B.~Maggs.
\newblock Puncturable pseudorandom sets and private information retrieval with
  near-optimal online bandwidth and time.
\newblock In {\em Annual International Cryptology Conference}, pages 641--669.
  Springer, 2021.

\bibitem{odoh}
S.~Singanamalla, S.~Chunhapanya, J.~Hoyland, M.~Vavrusa, T.~Verma, P.~Wu,
  M.~Fayed, K.~Heimerl, N.~Sullivan, and C.~A. Wood.
\newblock {Oblivious {DNS} over {HTTPS} (ODoH): {A} Practical Privacy
  Enhancement to {DNS}}.
\newblock {\em Proc. Priv. Enhancing Technol.}, 2021(4):575--592, 2021.

\bibitem{staff2015ripe}
R.~N. Staff.
\newblock Ripe atlas: A global internet measurement network.
\newblock {\em Internet Protocol Journal}, 18(3):2--26, 2015.

\bibitem{doh-size}
D.~Vekshin, K.~Hynek, and T.~Cejka.
\newblock Doh insight: detecting {DNS} over {HTTPS} by machine learning.
\newblock In {\em {ARES} 2020: The 15th International Conference on
  Availability, Reliability and Security, Virtual Event, Ireland, August 25-28,
  2020}, pages 87:1--87:8. {ACM}, 2020.

\bibitem{wang2021programmable}
L.~Wang, H.~Kim, P.~Mittal, and J.~Rexford.
\newblock Programmable in-network obfuscation of dns traffic.
\newblock In {\em NDSS: DNS Privacy Workshop}. sn, 2021.

\bibitem{winter2016identifying}
P.~Winter, R.~Ensafi, K.~Loesing, and N.~Feamster.
\newblock Identifying and characterizing sybils in the tor network.
\newblock In {\em USENIX Security Symposium}, volume~12, 2016.

\bibitem{TOCS25-Snatch}
Y.~Xiao, Y.~Gu, Y.~Zhao, S.~Lin, and A.~Kuzmanovic.
\newblock {Enabling Anonymous Online Streaming Analytics at the Network Edge}.
\newblock {\em ACM Transactions on Computer Systems}, 2025.

\bibitem{dvpn}
Y.~Xiao, M.~Varvello, and A.~Kuzmanovic.
\newblock Monetizing spare bandwidth: The case of distributed vpns.
\newblock {\em Proceedings of the ACM on Measurement and Analysis of Computing
  Systems}, 6(2):33:1--33:27, 2022.

\bibitem{snatch-24}
Y.~Xiao, Y.~Zhao, S.~Lin, and A.~Kuzmanovic.
\newblock Snatch: Online streaming analytics at the network edge.
\newblock In {\em Proceedings of the Nineteenth European Conference on Computer
  Systems, EuroSys 2024, Athens, Greece, April 22-25, 2024}. {ACM}, 2024.

\bibitem{zhao2007two}
F.~Zhao, Y.~Hori, and K.~Sakurai.
\newblock {Two-servers PIR based DNS query scheme with privacy-preserving}.
\newblock In {\em The 2007 International Conference on Intelligent Pervasive
  Computing (IPC 2007)}, pages 299--302. IEEE, 2007.

\bibitem{piano-pir}
M.~Zhou, A.~Park, E.~Shi, and W.~Zheng.
\newblock Piano: Extremely simple, single-server pir with sublinear server
  computation.
\newblock Cryptology ePrint Archive, Paper 2023/452, 2023.
\newblock \url{https://eprint.iacr.org/2023/452}.

\bibitem{zhou2023piano}
M.~Zhou, A.~Park, E.~Shi, and W.~Zheng.
\newblock Piano: Extremely simple, single-server pir with sublinear server
  computation.
\newblock {\em Cryptology ePrint Archive}, 2023.

\end{thebibliography}

\appendix
\section{Ethics Considerations}
\label{sec:appendix:ethics}

Our methodology in the evaluation (Appendix~\ref{sec:eval:meth}) uses synthetic DNS traffic via a publicly available service (Mysterium and Tor), avoiding ethical concerns. However, in \cref{sec:eval:privacy_sec} we rely on DoUDP traces collected with a passive Mysterium node. To protect the privacy of Mysterium users, we discard user-related information such as their IP addresses and anonymize the queried domains by hashing each domain level to a random value at the time of data collection. Our institution's IRB has deemed this data collection as non-human research, as we only collect non-identifiable private information without any accompanying data that could reveal individuals' identities. And we have obtained Mysterium’s permission to collect data for research purposes.

\begin{table*}[t]
\footnotesize
\centering
\setlength\tabcolsep{3pt}
\caption{Summary and performance analysis of state-of-the-art single-server PIR solutions.}
\label{tab:pir_solution}
\begin{tabular*}{\linewidth}{@{\extracolsep{\fill}} c c c | c c c | c c c}
\toprule
    \multirow{2}{*}{\parbox{2.3cm}{\centering\textbf{Solution}}} & \multirow{2}{*}{\parbox{3cm}{\centering\textbf{When Cache Updates}}} &
    \multirow{2}{*}{\parbox{3cm}{\centering\textbf{Performance}}} &
    \multicolumn{3}{c|}{\parbox{3cm}{\centering\textbf{\# Number of Slots (S=64B)}}} 
    & \multicolumn{3}{c}{\parbox{3cm}{\centering\textbf{Slot Size (NumSlots=$2^{20}$)}}} \\
    & & & \textbf{$2^{16}$} & \textbf{$2^{18}$} &  \textbf{$2^{20}$} 
    & \textbf{128B} & \textbf{512B} &  \textbf{2,048B} \\
\toprule

    \multirow{3}{*}{\parbox{2.5cm}{\centering %
    SimplePIR~\cite{simplepir}
    }} 
    & \multirow{3}{*}{\parbox{3.4cm}{\centering Update required for server and every user}}
    & Update (MB/user) & 7.4 & 14.7 & 29.5 & 42.4 & 86.8 & 178.1 \\
    && Query Duration  (ms) & 4.04 & 8.51 & 19.07 & 30.29 & 80.87 & 256.38 \\
    && Query Comm. (KB) & 7 & 14 & 28 & 41 & 84 & 173 \\\cline{1-9}
     
    \multirow{3}{*}{\parbox{2.5cm}{\centering SealPIR~\cite{seal-pir}}}  
    & \multirow{3}{*}{\parbox{3.4cm}{\centering Server update only}}
    & Update ($\mu$s/slot) & 2.21 & 2.19 & 2.19 & 4.22 & 16.36 & 78.1 \\
    && Query Duration  (ms) & 117 & 301 & 902 & 1,636 & 5,831 & 25,338 \\
    && Query Comm. (KB) & 278 & 278 & 278 & 278 & 278 & 278 \\\hline
     
    \multirow{3}{*}{\parbox{2.5cm}{\centering Spiral~\cite{spiral-pir}}}  
    & \multirow{3}{*}{\parbox{3.4cm}{\centering Server update only}}
     & Update ($\mu$s/slot) & 37.62 & 62.28 & 31.36 & 31.34 & 63.63 & 298.31 \\
    && Query Duration  (ms) & 249 & 501 & 794 & 797 & 1,423 & 3,882 \\
    && Query Comm.   (KB) & 30 & 30 & 36 & 36 & 36 & 36 \\
\bottomrule
 \vspace{-0.2in}
\end{tabular*}
\end{table*}

\section{Deep Dive into PIR}
\label{sec:appendix:pir}

\subsection{PIR Construction}

\cref{sec:background:pir} concludes that \textit{single-server} \textit{stateless} PIR is best suited for \pirdns. Consequently, we here further elaborate on the construction of this scheme. Prior to delving into PIR, we introduce homomorphic encryption, which is the key component of PIR schemes within this category~\cite{spiral-pir,mughees2021onionpir,seal-pir}.

\vspace{0.05in}
\noindent\textbf{Homomorphic Encryption (HE)} -- It allows to perform computation on encrypted data~\cite{regevfhe, gswfhe, brakerski2014leveled}. We only focus on HE that relies on a cryptographic computational hard assumption known as learning-with-error (LWE)~\cite{regev2009lattices, rlwe}.
Define an HE scheme $\mathcal{HE}:\{{\sf KeyGen}, {\sf Enc}, {\sf Dec}, {\sf Eval}\}$ which contains the following algorithms:
\begin{enumerate}
    \item ${\sf KeyGen}(1^{\kappa}) \rightarrow ({\sf sk}, {\sf pk})$: Define a security parameter $\kappa$, $1^\kappa$ is a canonical notation to define the strength of a cryptographic scheme. This algorithm outputs a secret key ${\sf sk}$ and a public key ${\sf pk}$.
    \item ${\sf Enc}({\sf sk}, x) \rightarrow [x]$: On input the secret key ${\sf sk}$ and an integer $x$, the encryption algorithm outputs a ciphertext $[x]$, which hides the $x$. We denote an integer in ``$[]$'' as an encrypted value.
    \item ${\sf Dec}({\sf sk}, [x]) \rightarrow x$: On input the secret key ${\sf sk}$ and a ciphertext $[x]$, the decryption algorithm outputs a plaintext $x$.
    \item ${\sf Eval}({\sf pk}, [x], [y], g) \rightarrow [z]$: On input the public key ${\sf pk}$, ciphertexts $([x], [y])$, and an operator $g\in \{{\sf Add},{\sf Mult}\}$, the evaluation algorithm outputs a ciphertext $[z] = g([x],[y])$. Depends on $g$, it is either $z = x + y$ or $z = x * y$.
\end{enumerate}

For $\mathcal{HE}$ to be secure, it is essential that adversaries without access to the secret key are unable to obtain any information about the encrypted values. Homomorphism requires that for any operation $g \in \{{\sf Add},{\sf Mult}\}$ and a ciphertext $[z] = g([x],[y])$, anyone who holds a secret key can compute ${\sf Dec}({\sf sk}, [z]) \rightarrow z$ which satisfies $z = g(x, y)$. The significant benefit of this property is that individuals with the public key can perform addition and multiplication operations directly on ciphertexts without requiring knowledge of the underlying values. Furthermore, it also supports a relaxed evaluation algorithm ${\sf Eval}({\sf pk}, x, [y], g) \rightarrow [z]$ for relation $z = g(x, y)$. This implies the arithmetic operation between a plaintext and an encrypted value. %

\vspace{0.05in}
\noindent\textbf{Construct PIR from HE} --
Suppose a PIR server maintains a cache $\mathcal{C} = (c_1,\dots,c_N)$ and a user generates keys $({\sf sk}, {\sf pk})$ by invoking ${\sf KeyGen}$. The user shares ${\sf pk}$ with the server. To retrieve the $i$-th element $c_i$ from the cache, the user encrypts a one-hot vector\footnote{A one-hot vector is a binary representation of a categorical variable in which only one element is set to 1 (hot) and the rest are set to 0 (cold).} $\vec{q} = (q_1,\dots,q_N)$ in which only $q_i=1$ but $q_j=0$ for any $j \neq i$. An encrypted query $[\vec{q}]=([q_1],\dots,[q_N])$ is then transmitted to the server that performs the homomorphic evaluation by utilizing the algorithm ${\sf Eval}$. Specifically, it computes the inner product $[r] = \mathcal{C} * [\vec{q}] = \sum_{j\in[N]} c_j \cdot [q_j]$ by repeatedly invoking ${\sf Eval}$ with operators ${\sf Add}$ and ${\sf Mult}$, and then returns the response $[r]$ to the user. The user decrypts the response by ${\sf Dec}({\sf sk}, [r])\rightarrow r$. Due to the homomorphism, it satisfies that $r = \sum_{j\in[N]} c_j \cdot q_j = c_i \cdot q_i = c_i$.

\vspace{0.05in}
\noindent\textbf{Shrink Query} -- The concept described earlier enables the implementation of PIR at a cost of high communication overhead, as the entire encrypted query vector is transmitted. SealPIR~\cite{seal-pir} and Spiral~\cite{spiral-pir} adopt different techniques for query compression and expansion that shorten the query to a constant number of ciphertexts.

Specifically, define an algorithm ${\sf Expand}({\sf pk}, [i]) \rightarrow [\vec{q}]$ which takes input an encrypted index $[i]$ and outputs a length-$N$ query vector $\vec{q}$. The knowledge of ${\sf sk}$ is not needed to perform ${\sf Expand}$, thus the server is able to construct $\vec{q}$ by itself given $[i]$. Based on this, the user only needs to send one ciphertext instead of $N$. We refer the readers to Section 3 of \cite{seal-pir} and Section 2.1 of \cite{spiral-pir} for more details.

\vspace{0.05in}
\noindent\textbf{Optimize Query Processing} -- Note that the above approach is not feasible when $N$ is large, \eg $N=4,096$~\cite{seal-pir}. The problem is overcome by representing $\mathcal{C}$ as a multi-dimension hypercube~\cite{seal-pir,spiral-pir}. Take the two-dimension case as an example. Define parameters $m,\ell$ such that $m\ell = N$. The server constructs its cache $\mathcal{C} = (\vec{c}_1,\dots,\vec{c}_m)$ where each row contains $\vec{c}_i = (c_i^1,\dots,c_i^{\ell})$ for $i\in [m]$. To fetch $c_i^j$, the user constructs two ciphertexts $([i],[j])$. The server expands the queries to one-hot vectors $([\vec{q}_1],[\vec{q}_2])$, each having the $i$-th or $j$-th slot to be 1. The server  first performs an HE evaluation on $[\mathcal{C}]$ and $[\vec{q}_1]$ to extract the row $[\vec{c}_i]$, then computes another inner product on $[\vec{c}_i]$ and $[\vec{q}_2]$ to extract $[c_i^j]$.

Observe that the first inner product only involves the multiplication between plaintext and ciphertext, while the second operates purely on ciphertexts. The HE algorithm adopted in SealPIR, called FV~\cite{fvhe}, has limited ability to perform the latter one, which results in slow query processing and large response size. Spiral proposes a combination of Regev~\cite{regevfhe} and GSW~\cite{gswfhe} schemes which provides efficient ciphertext-ciphertext multiplication thus achieving better performance on query processing.

\subsection{PIR Scheme Selection}
\label{sec:appendix:pir-selection}
We further analyze our benchmark on recent single-server PIR schemes shown in Table~\ref{tab:pir_solution}. For fairness, we do not enable our optimizations to Spiral (see \cref{sec:impl:choice}). SimplePIR~\cite{simplepir},  the only stateful PIR in the table, is significantly more efficient on query processing, but it causes high traffic when the server updates the cache. For instance, each user needs to fetch a 29.5MB message from the server when an updated cache (with $2^{20}$ 64-byte slots) is queried, which is unrealistic for \pirdns. Further, its communication overhead grows as the cache size grows, and it is outperformed by Spiral when the slot size is $\geq 128$ bytes.

SealPIR~\cite{seal-pir} and Spiral~\cite{spiral-pir} are stateless PIR schemes. When updating the cache, the server only needs to encode the new slots and does not need to send any updates to the users. SealPIR is more efficient in encoding than Spiral, especially for small entries. Regarding the speed for query processing, SealPIR is better at dealing with small caches with small \rows but is less efficient in other cases. After further investigation, we find that Spiral's implementation pads all small \rows to 256 bytes, hence its actual performance for \row size smaller than 256 bytes is misleading. Finally, Spiral reduces the traffic needed to operate by $7.7\times$ compared to SealPIR. Note that a number of single-server PIR schemes are newly proposed as we were implementing \pirdns from Spiral~\cite{davidson2023frodopir,simplepir,mughees2023simple,li2024hintless,zhou2023piano,menon2024ypir,burton2024respire}. However, they are either stateful, or incur large communication overhead, or do not perform better than Spiral over small records. Hence, we do not switch from our underlying PIR to these schemes.   %

\section{Cache Miss Handling Alternatives}
\label{sec:appendix:cachs-miss}
We here discuss some alternatives to our mechanism to handle cache misses (see \cref{sec:cache-miss}), which we have discarded because they do not meet our privacy model (see \cref{sec:overview:models}).

A strawman solution is to retreat to regular DNS (no PIR) when a cache miss occurs, \ie letting the user send a second query in plaintext to the \RR. The benefit of this solution is that it keeps compatibility with the existing DNS system to the largest extent.  Nevertheless, it exposes some queries to the \RR which violates a user's privacy.

A more intriguing solution is to fallback to ODNS, \ie introduce a proxy to relay user's requests to a \RR (see \cref{sec:background:dns}). Adding such proxy in \pirdns, \eg provided by another corporation, would re-introduce the need for a non-collusion agreement, which downgrades the privacy guarantees. Alternatively, the proxy can be distributed, \eg integrating with Tor~\cite{dingledine2004tor} or allowing DNS users to act as each other ODNS proxies. The advantage of a distributed approach is to scatter private DNS traffic across many parties, making it harder to re-centralize the information if needed, \eg via subpoenas. Nevertheless, this introduces all security vulnerabilities of distributed systems; for example, privacy leaks are possible if a \RR controls some of the peers~\cite{winter2016identifying}.

\section{Additional \pirdns Benchmark}
\label{sec:appendix:benchmark}

\begin{figure}[t]
    \centering
    \includegraphics[width=0.8\linewidth, clip=true]{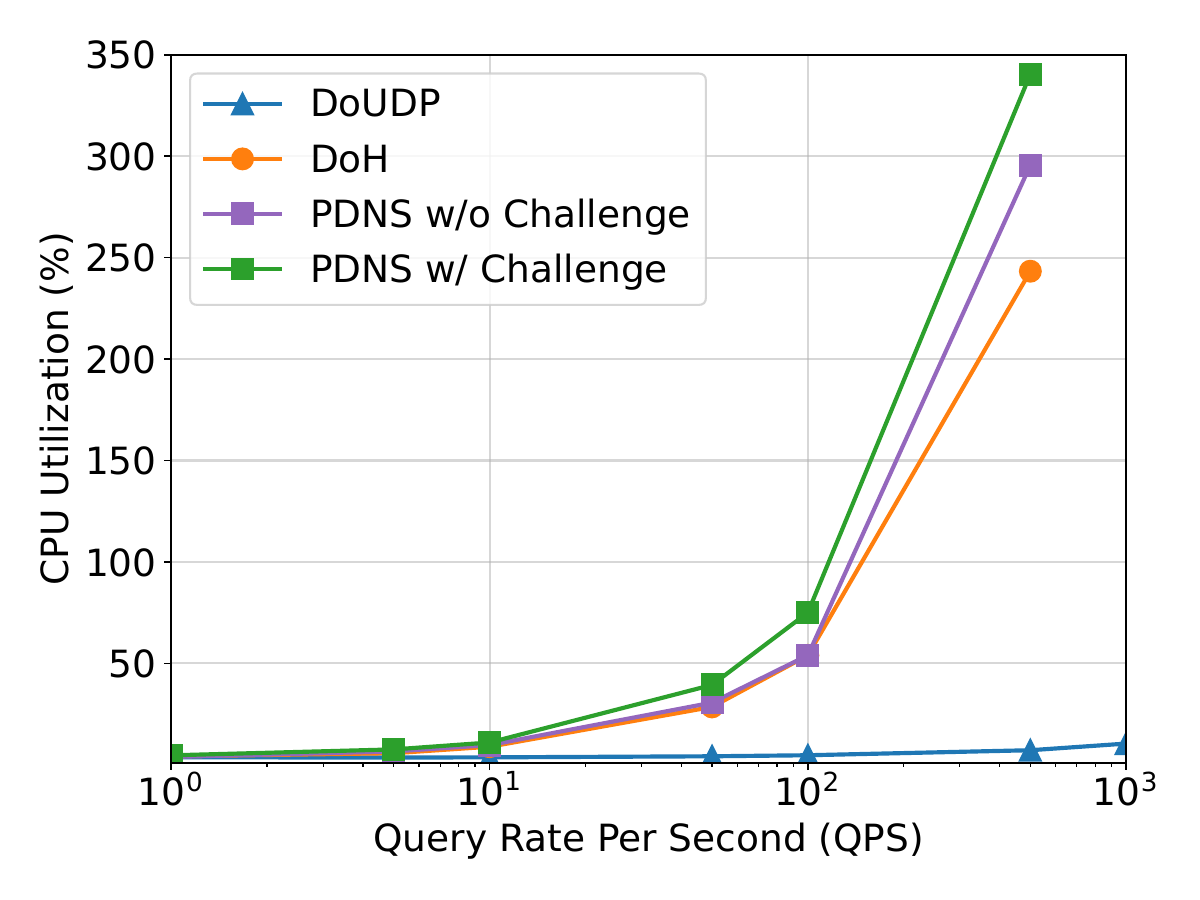}
    \vspace{-0.2in}
    \caption{ANS CPU usage as a function of the query rate.}
    \vspace{-0.15in}
    \label{fig:benchmark:cpu-authoritative}
\end{figure}

\begin{figure*}[t]
  \centering
  \subfigure[{\centering Initialization duration.}]{\psfig{figure=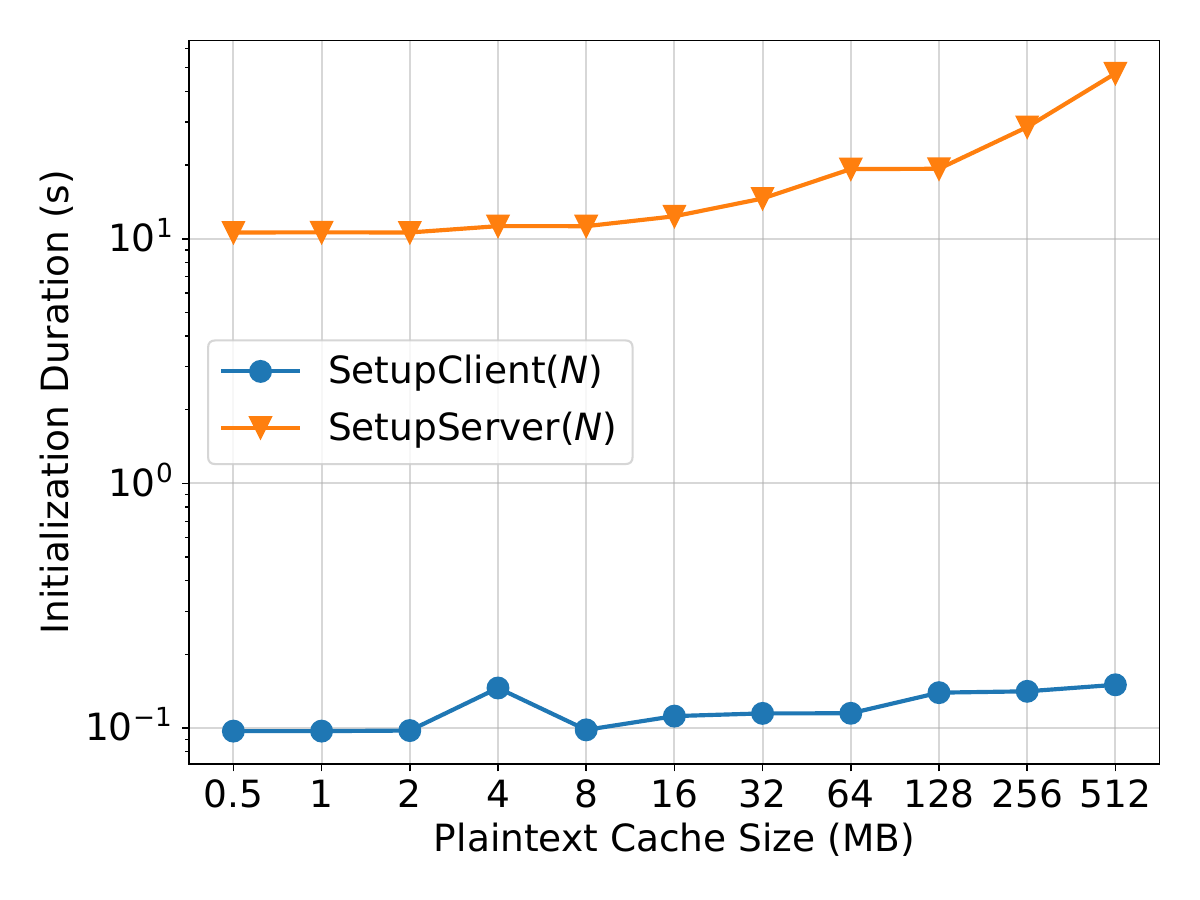, width=0.32\linewidth, trim=5mm 6mm 6mm 6mm, clip=true}\label{fig:benchmark2:init-time}}
  \hspace{0.2em}
  \subfigure[{\centering Initialization communication cost.}]{\psfig{figure=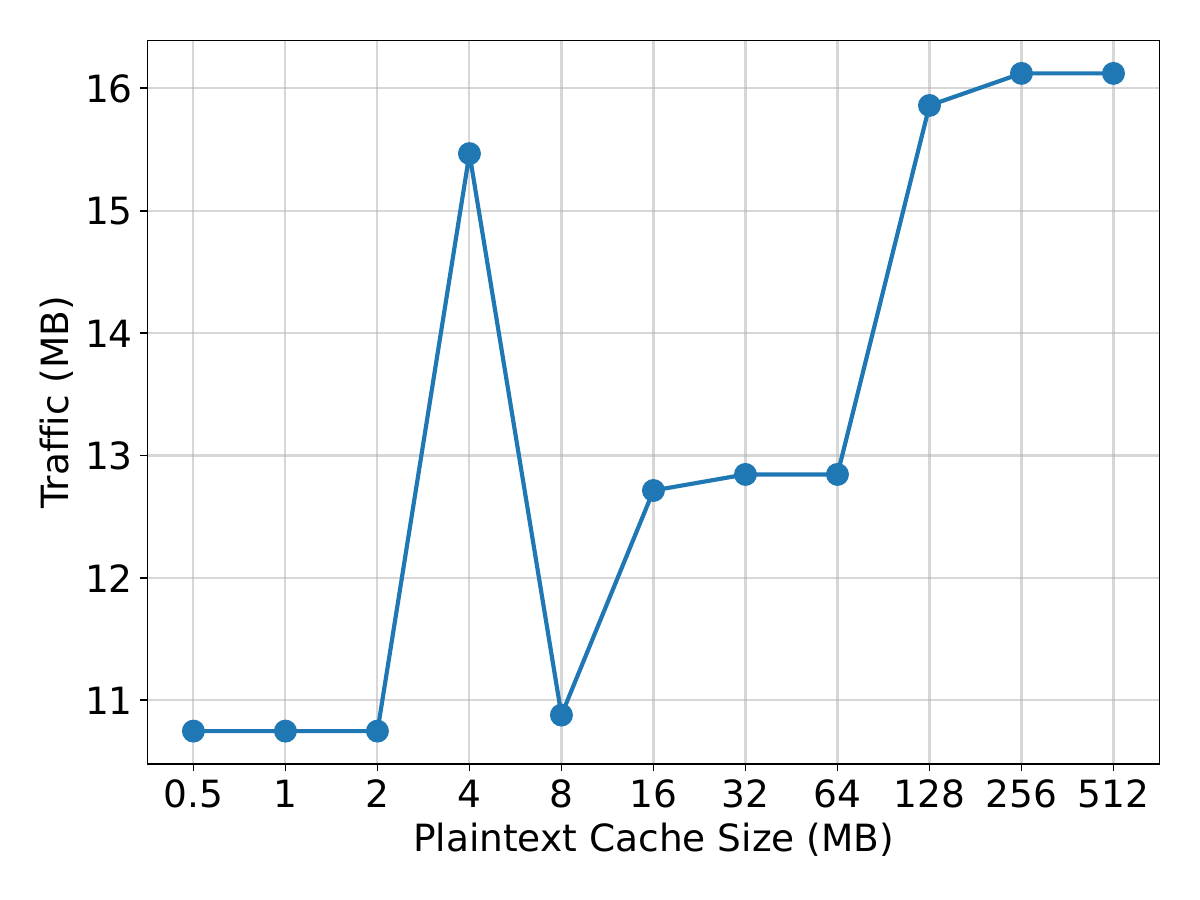, width=0.32\linewidth, trim=5mm 6mm 6mm 6mm, clip=true}\label{fig:benchmark2:init-time-entry-size}}
  \hspace{0.2em}
  \subfigure[{\centering Query and answer traffic.}]{\psfig{figure=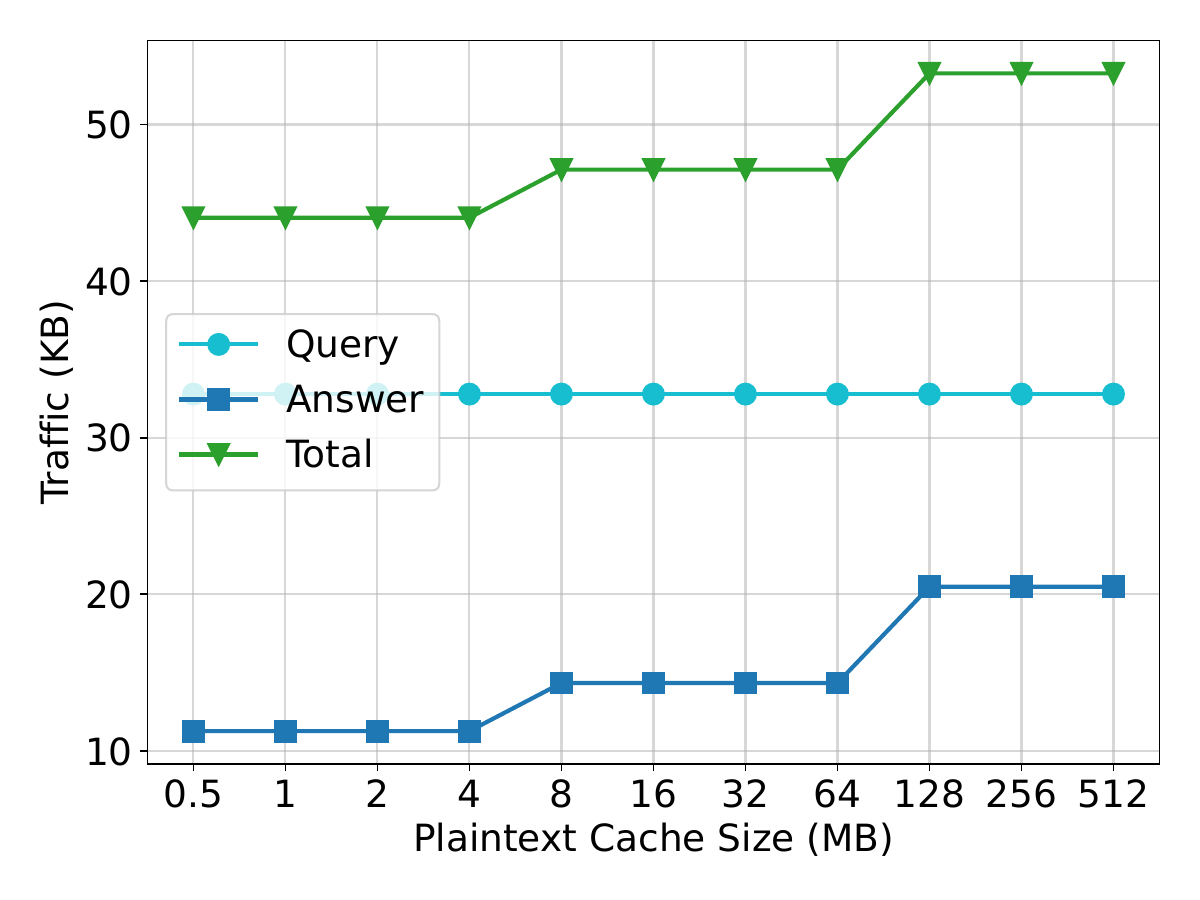, width=0.32\linewidth, trim=5mm 6mm 6mm 6mm, clip=true}\label{fig:benchmark2:query-comm-cache-size}}
  \vspace{-0.18in}
  \caption{Additional benchmarking results. (a) Initialization duration, (b) communication cost during initialization, and (c) query and answer traffic as functions of the plaintext cache size. }
  \vspace{-0.07in}
  \label{fig:benchmark2}
\end{figure*}

\begin{figure}[t]
    \centering
    \vspace{-0.15in}
    \includegraphics[width=0.8\linewidth, clip=true]{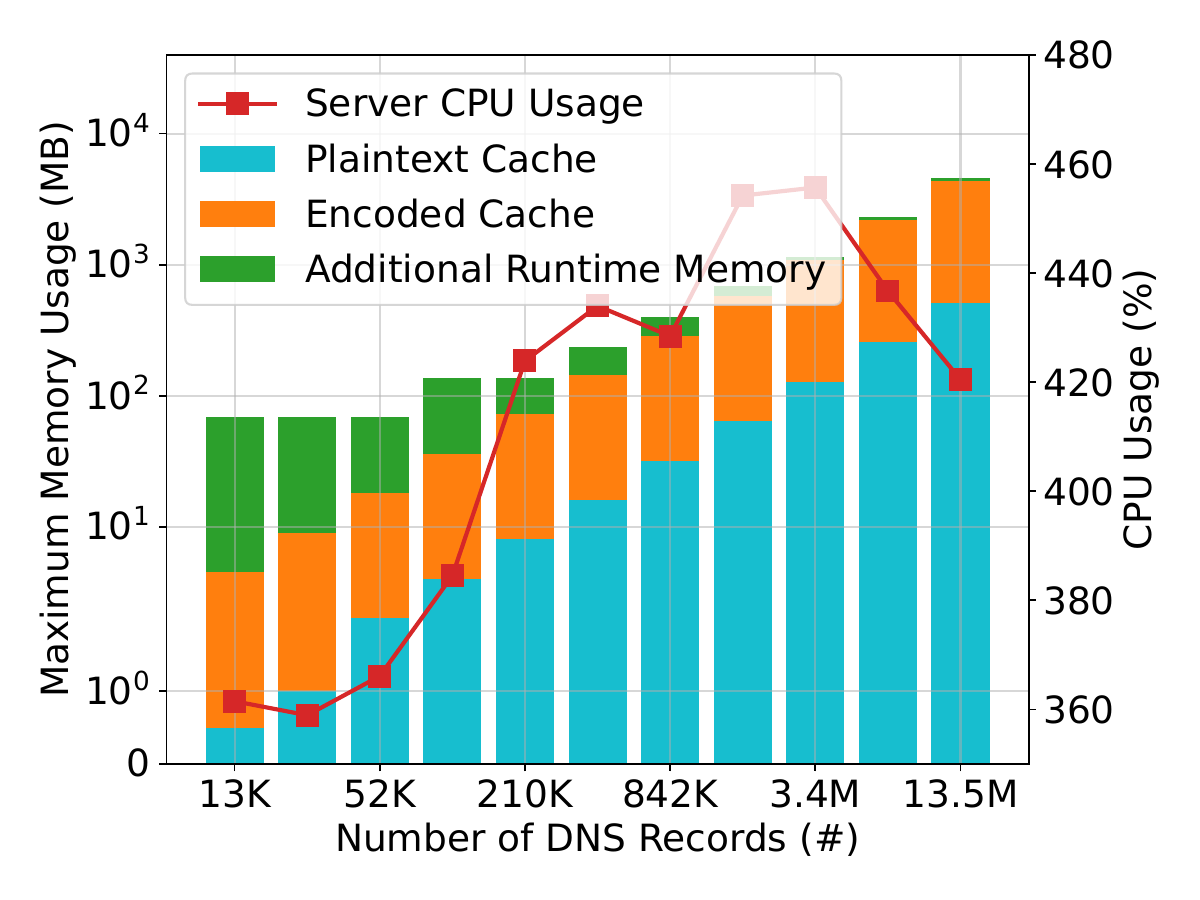}
    \vspace{-0.2in}
    \caption{Memory and CPU as a function of cache size.}
    \label{fig:benchmark:resource}
\end{figure}

\begin{figure*}[t]
  \centering
  \subfigure[{\centering CDF of RTT between 1,415 residential dVPN nodes and Google
and Cloudflare DNS \RRs.}]{\psfig{figure=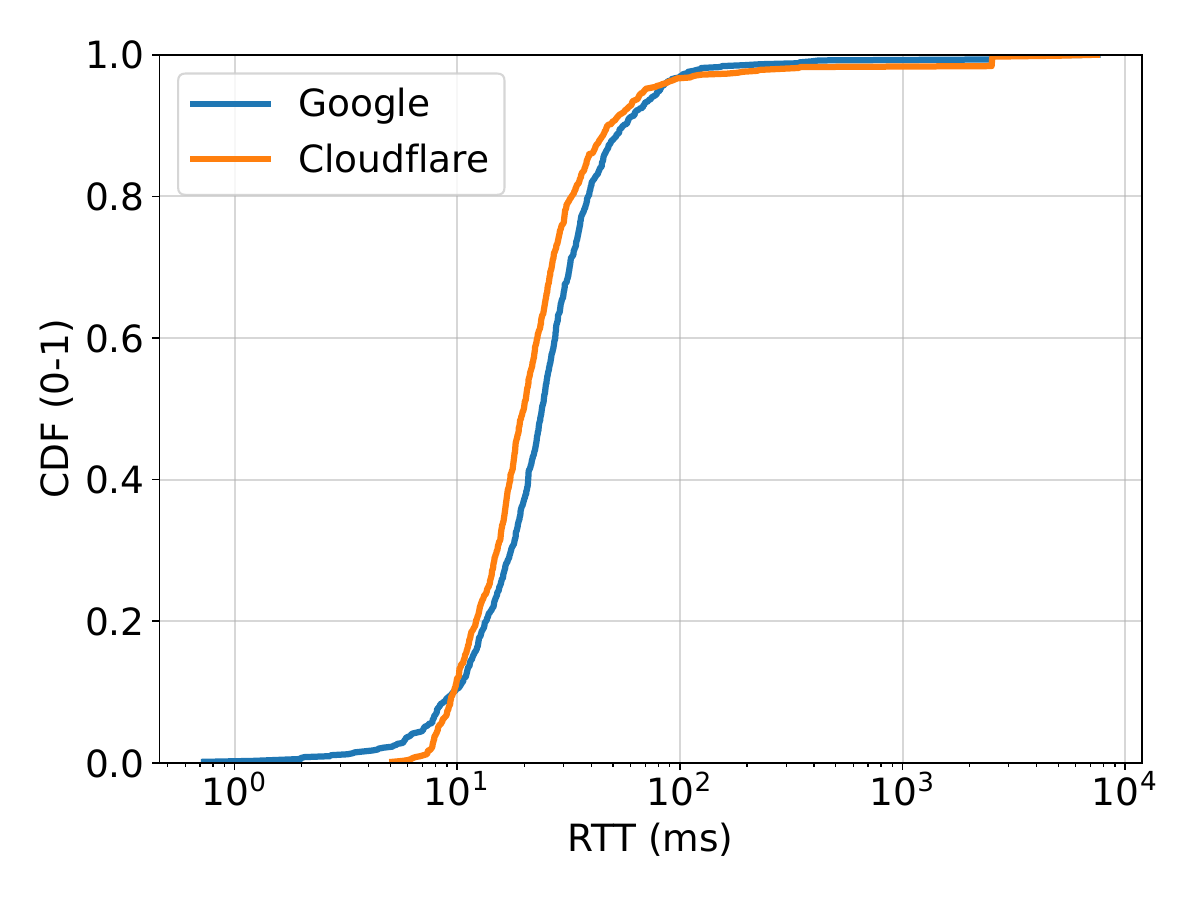, width=0.32\linewidth, trim=5mm 6mm 6mm 6mm, clip=true}\label{fig:measurement:global-rr}}
  \hspace{0.2em}
  \subfigure[{\centering CDF of the duration of iterative DNS queries distinguishing between root, TLD, and fianl ANSes.}]{\psfig{figure=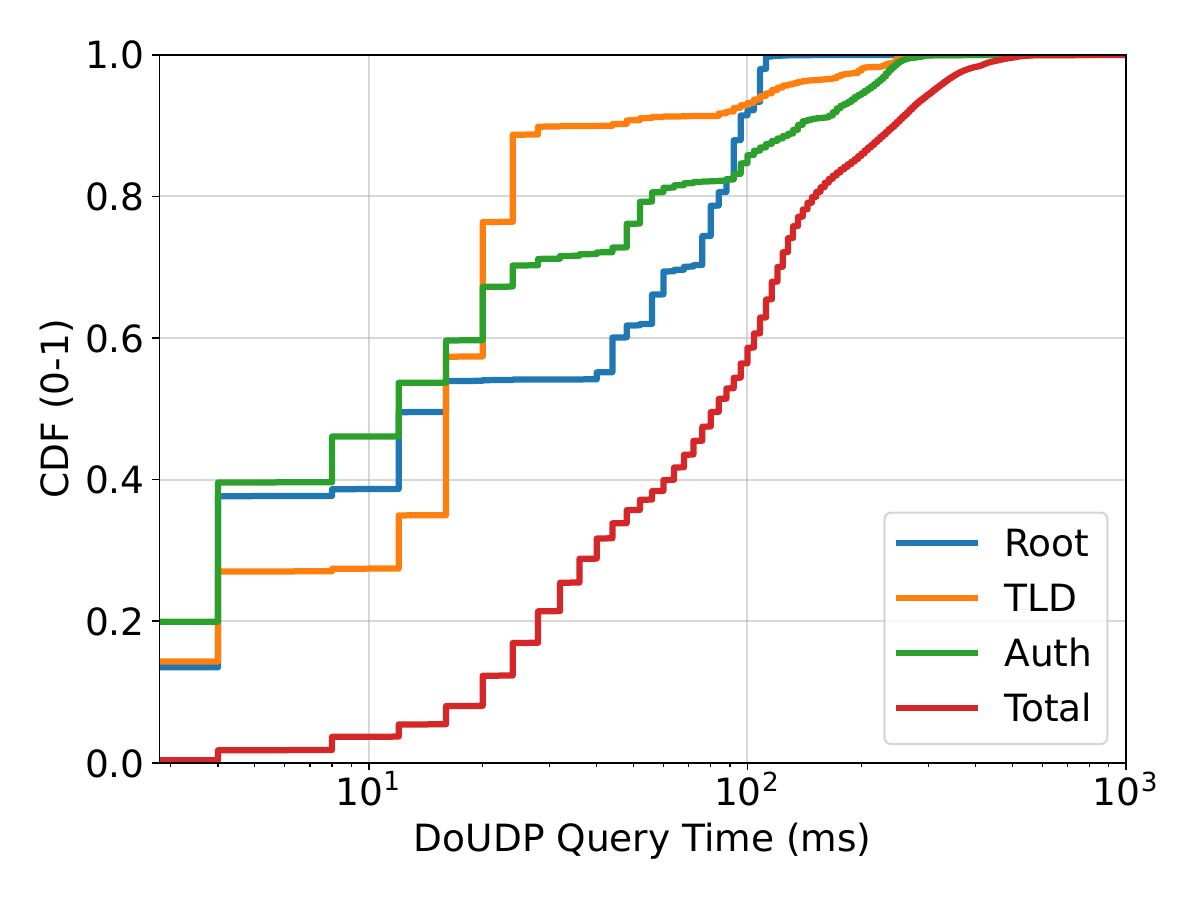, width=0.32\linewidth, trim=5mm 6mm 6mm 6mm, clip=true}\label{fig:measurement:local-iterative}}
  \hspace{0.2em}
  \subfigure[{\centering CDF of the duration of DoUDP queries per root ANS.}]{\psfig{figure=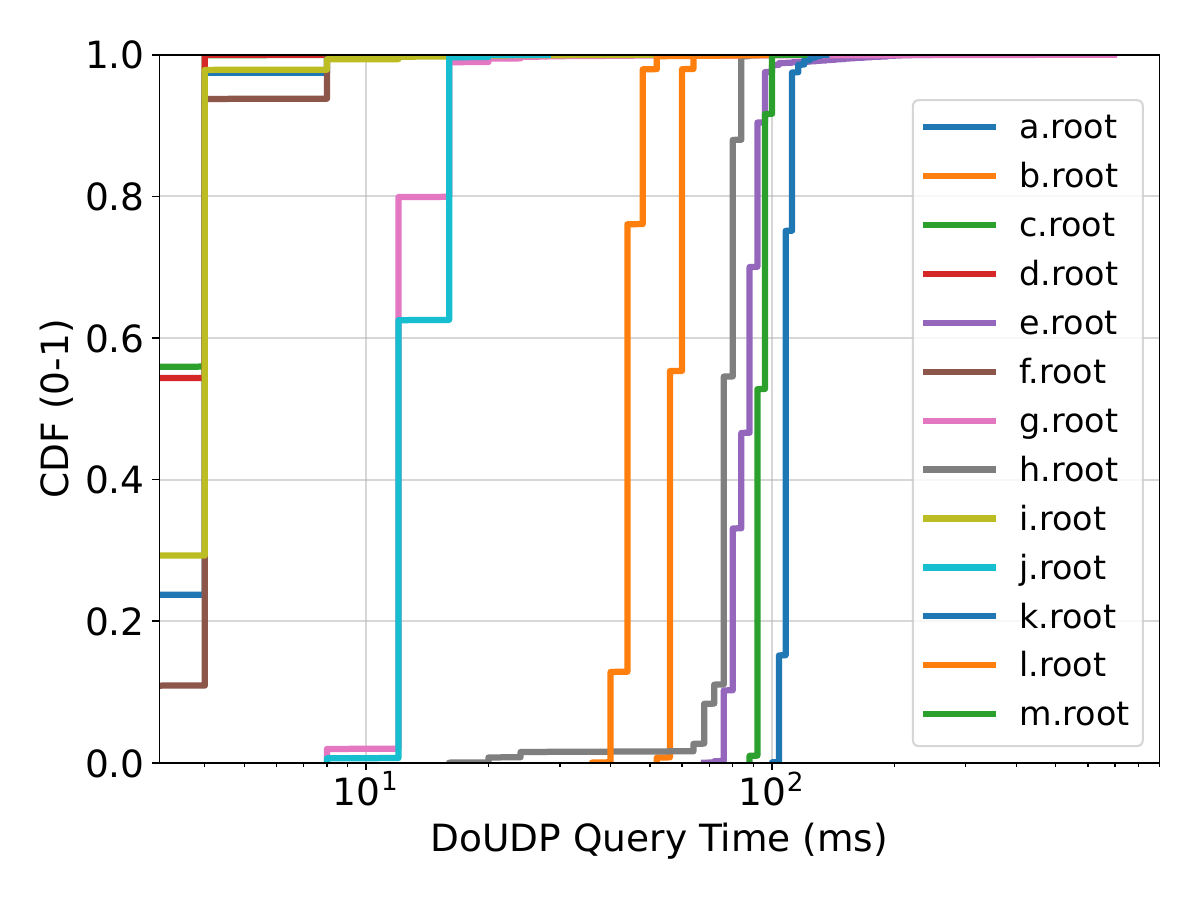, width=0.32\linewidth, trim=5mm 6mm 6mm 6mm, clip=true}\label{fig:measurement:query-time-root}}
  \vspace{-0.18in}
  \caption{Analysis of active DNS measurements.}
  \vspace{-0.07in}
  \label{fig:appendix:measurement}
\end{figure*}

\noindent\textbf{Scalability of Authoritative Name Server} -- 
Figure~\ref{fig:benchmark:cpu-authoritative} is a visualization of the ``Scalability'' discussion in \cref{sec:impl:benchmark}.

\vspace{0.05in}
\noindent\textbf{System Initialization} -- \pRR and client execute {\sf SetupServer} and {\sf SetupUser}, respectively, at each reboot. Further, the client re-runs {\sf SetupUser} when challenged by an ANS (see \cref{sec:security}) to regenerate its backup key pairs. Figure~\ref{fig:benchmark2:init-time} shows the duration of each initialization phase (at client and server) as a function of the cache size. The figure shows that, with a cache size of 0.5MB, the server initialization takes 10 seconds, while it takes over 50 seconds with a cache size of 512MB. While long, this duration is acceptable since it is only required during a reboot of the \RR. Conversely, the client initialization is not a significant burden for the user as it only takes up to 150ms, irrespective of the cache size.

{\sf SetupUser} also requires some traffic between \pirdns client and \RR to share a public key ${\sf pk}$. Figure~\ref{fig:benchmark2:init-time-entry-size} shows that this traffic increases as the cache size increases, \eg from 10.7MB with a small cache (0.5MB) up to 16MB when the cache size is larger than 128MB. An exception is observed when the cache size is 4MB where the traffic jumps to 15.5MB. This result is due to the selection of underlying cryptographic parameters by Spiral.

\vspace{0.05in}
\noindent\textbf{Query and Answer Traffic} --  Figure~\ref{fig:benchmark:query-comm-number-entry} shows the query and answer traffic as a function of the number of slots in a 512MB cache. We now further analyze such traffic as a function of the cache size. Figure~\ref{fig:benchmark2:query-comm-cache-size} confirms that query traffic is bound to 32KB since, as discussed in \cref{sec:impl:benchmark}, the query is the encryption of an index and hence independent from the cache size. The answer traffic only increases when the cache size grows from 4 to 8MB and from 64 to 128MB, but is constant otherwise. This, again, attributes to the selection of underlying cryptographic parameters by Spiral. 

\vspace{0.05in}
\noindent\textbf{Resource Usage} -- Next, we benchmark CPU and RAM usage for \pirdns. Figure~\ref{fig:benchmark:resource} shows the RAM required by the plaintext cache, the encoded cache, and the runtime \RR, as a function of the (plaintext) cache size. The figure shows that the encoded cache requires 8x the memory used by the plaintext cache, and that runtime memory usage is comprised between 60 and 100MB. As a result, the memory usage of \pirdns at a \RR ranges between 68MB and 4.3GB, assuming caches which can hold about 13K (512KB) and 13M (512MB) DNS records using IPv4 (\ie 38B as in Figure~\ref{fig:cache-design}). With respect to CPU usage\footnote{All numbers for CPU usage are concerning one core of single CPU.}, Figure~\ref{fig:benchmark:resource} shows that it ranges between 360\% and 460\%. This is a result of our multi-threading implementation of the {\em Answer} primitive, and the usage of 4 concurrent threads (see \cref{sec:impl:choice}).

\section{Additional \pirdns Evaluation}
\label{sec:eval:meth}

\subsection{Evaluation Methodology}
\label{sec:eval:meth}
We evaluate existing DNS solutions (DoUDP, DoH, ODOH, and DoHoT) using experiments in the wild. We select Google DNS~\cite{google_dns} and Cloudflare DNS~\cite{cloudflare_dns} as target \RRs for both DoUDP and DoH. To emulate a realistic access network of a DNS client, we resort to Mysterium~\cite{mysterium, dvpn}, a popular distributed VPN (dVPN) which provided us with 1,415 Internet residential dVPN nodes from 62 countries. To derive the duration of a DNS query between a Mysterium node and \RRs, we subtract the latency between our machine and the node for each RTT needed, \eg 1 RTT for DoUDP and 3 RTT for DoH given TLSv1.3~\cite{rescorla2018rfc} and no connection reuse. We preferred Mysterium over academic platforms like the popular RIPE Atlas~\cite{staff2015ripe} since it offers higher flexibility, \eg allowing to send DoH and ODoH queries. Appendix~\ref{sec:appendix:measurement} offers more insights into the data collected in these experiments. 

We use instead a single location (our lab) for DoHoT since we cannot force a Tor circuit between Mysterium nodes and a \RR. However, we restart Tor after each experiment which gives us 522 unique exit nodes over 24 hrs. For ODoH, we still rely on Mysterium but also iterate the three oblivious proxies provided by DNSCrypt Proxy~\cite{dnscrypt}, a popular and cross-platform local proxy which supports many DNS protocols. Finally, we perform iterative DNS lookups on our machine for 122K domains collected in Mysterium (see \cref{sec:eval:privacy_sec}), and estimate the duration of direct queries towards authoritative DNS servers, as used by \RR-Less DNS. 
This procedure was confined to our own machine due to the impracticality of executing all 122,000 iterative DNS lookups across each of the 1,415 dVPN nodes. It is crucial to mention that our machine, located in the United States, benefits from relatively high bandwidth and short latencies to ANSes. Additionally, our focus was exclusively on top-ranked sites. Consequently, our evaluation of the \RR-Less DNS performance should be viewed as a conservative estimation. Users with lower bandwidth and/or in developing countries might experience less favorable performance. %

When experimenting with DNS in the wild, there is no control on whether queried domain names are cached or not at a \RR. To study the effect of cache \textit{hit} or \textit{miss} at a \RR, we query for our own domain names -- registered at AWS Route 53~\cite{awsroute53} -- with TTLs of one second and one hour, respectively. The one-hour interval guarantees cache hits as long as our queries happen within such an interval. The one-second interval guarantees a cache miss as long as we perform queries slower than once a second. While many public \RRs would ignore such low TTL value, we have verified that Google and Cloudflare DNS both support it.

To evaluate \pirdns, we instead set up a test-bed composed of \pirdns client, \RR, and a participating final ANS. Each machine is equipped with the same hardware used in the benchmarking (see \cref{sec:impl:benchmark}). We then apply network delays between the machines using the Linux \texttt{Traffic Control} (tc) module~\cite{linuxtc} driven by the real latencies collected in the above experiments. We test a large (512MB) and small (64MB) cache using the best performing \textit{shape} as from our benchmarking experiments, \eg $2^{15}$ \rows with size $S=16KB$ for the 512MB cache.

\subsection{DNS Measurement Result}
\label{sec:appendix:measurement}

We here present high-level results from our DNS measurement study (see Appendix~\ref{sec:eval:meth}). We first analyze the network delay towards popular \RRs. To do so, we connect to 1,415 residential dVPN (Mysterium) nodes and send \texttt{ping} (ICMP packets) to both Google and Cloudflare public DNS \RRs, measuring the round trip time (RTT) of the path $<$client, dVPN node, \RR$>$. Then, we derive the latency of the path  $<$dVPN node, \RR$>$ by subtracting the latency from the path $<$client, dVPN node$>$ which we also obtain via \texttt{ping}. Figure~\ref{fig:measurement:global-rr} shows the CDF of the RTTs between 1,415 residential dVPN nodes and Google and Cloudflare DNS \RRs. Overall, faster RTTs are measured for Cloudflare, \eg a median RTT of 19.9ms compared to Google's median RTT of 24.0ms. %

Next, we analyze the iterative DNS lookups (DoUDP) we performed for 122K domains (40K SLDs) collected with a passive Mysterium node. Figure~\ref{fig:measurement:local-iterative} shows the CDF of DoUDP query duration per domain distinguishing between each step of the iterative lookup: root, TLD, and final ANSes. The figure shows similar results across ANSes, with a median duration of 16ms for both root and TLD ANSes, and 12ms for final ANSes. Further analysis shows that the highly variable query duration observed for both root and TLD ANSes is primarily due to \texttt{dig}~\cite{dig, dig-update} which iterates through different root and TLD ANSes. In fact, the closest root ANSes to our test-bed (c.root and d.root) consistently respond in less than 1ms, versus over 100ms for the furthest one (k.root) as shown in Figure~\ref{fig:measurement:query-time-root}. Similar results apply to TLD. 

\subsection{Security Effectiveness}
\label{sec:appendix:security-effect}
We evaluate \pirdns resilience to reflection attacks (see \cref{sec:overview:models}). We use our DNS traces, %
including 42K ANSes involved and 122K unique domains, to drive the distribution of the number of domains per ANS. Note that few ANSes control over 600 sub-domains, while over 90\% of the ANSes manage less than 4 sub-domains. We select the top N\% domains to be cached by \pRR and simulate attackers launching reflection attacks by pretending to experience cache misses for each domain. We assume an attacker has the list of domains supported by \pirdns and a maximum upload bandwidth of 1Gbps. 

Figure~\ref{fig:eval:attack} shows that, without the security feature introduced in \cref{sec:security}, a single attacker can generate over 100MB of traffic per second, and the traffic grows linearly as the number of attackers grows. The limit on the ``reflection traffic'', \ie from ANS to \RR, is dependent on the total bandwidth of all attackers and of all ANSes. When considering our security mechanism, Figure~\ref{fig:eval:attack} shows that the reflection traffic reduces to less than 12MB per TTL, \ie for how long a record stays in the \RR's cache, with little impact of the number of attackers and the amount of domains found in the cache. This is because our security mechanism ensures that each domain can only be populated once in the \pRR within its TTL, thus putting a deterministic cap on the reflection traffic. Having more attackers or fewer domains cached only allows to consume that cap faster within a TTL.

\section{Deployment Cost Analysis}
\label{sec:appendix:deployment-cost}
We here estimate the deployment costs of \pirdns and discuss whether it is viable to be provided as a subscription-based service as of \textit{today}. Our analysis is based upon a comprehensive study~\cite{schomp2016towards} that models the DNS client behavior with a dataset collected from a university campus network. %

We assume that \pirdns is deployed using a public cloud service. Upon consulting the pricing calculators provided by AWS~\cite{aws-pricing} and Google Cloud~\cite{gcp-pricing}, we determine that the costs associated with configuring an 8-core CPU  machine -- similar to the one used in the benchmarking (\cref{sec:impl:benchmark}) and evaluation (\cref{sec:eval}) sections -- are comparable across two cloud providers. Specifically, the estimated computing costs amount to approximately \$148.92 on AWS and \$148.37 on Google Cloud, per month. For simplicity, we will consider the rounded value of \$149 for the subsequent calculations. It is worth noting that both costs scale linearly with the number of CPU cores. As shown in  Figure~\ref{fig:benchmark:cpu-rr}, PDNS running on a machine equipped with an 8-core CPU can handle 8~QPS (assuming a small cache) or 4~QPS (assuming a large cache). In addition, the cloud also charges for data transfer from cloud machines to the Internet (the reverse is free). AWS and Google Cloud charge for at most \$0.09 per GB and \$0.085 per GB, respectively. We take the upper bound of \$0.09 per GB for the subsequent calculations.

According to the findings in~\cite{schomp2016towards}, users perform on average between 2,600 and 3,724 DNS queries every day. It follows that a PDNS using the configuration above can serve at least between 93 (large cache) and 186 users (small cache) in a day, \ie  $3600 \times 24 \times 8 / 3,724 = 186$. As a result, the computing cost per user amounts to approximately \$0.8 with a small cache or \$1.6 with a large cache. In addition, the data transfer will cost $3,724 \times 30 \times 40\text{KB} \times \$0.09/\text{GB} = \$0.4$ per user. This means that if a user is willing to pay more than \$2 per month to safeguard their privacy, \pirdns could be a viable business proposition.

The above analysis ignores potentially concurrent users as well as the bursty nature of DNS requests. In the study by \cite{schomp2016towards}, the bursty behavior of DNS queries is measured using ``clusters'', where each cluster consists of a minimum number of 3 queries and is separated from other clusters by an idle period of 2.5 seconds. In their research, they report the CDFs of the number of queries per cluster and cluster duration. Although direct ratios between the number of queries and duration within each cluster were not reported, we sample these values at various percentiles and approximate the bursty demand by calculating the ratios ourselves. Based on our analysis, we find that the bursty queries generated by up to 1,033 users reach a maximum of 126~QPS. Another study has reported similar peak QPS with many more users~\cite{cache-effect-19}. To support such capacity, machine(s) with 128 CPU cores are required with a small cache, or 256 CPU cores are needed with a large cache. Consequently, the monthly computing costs to serve one user would amount to approximately $128 / 8 \times 149 / 1,033 = \$2.3$ with a small cache, or \$4.6 with a large cache. Besides the data transfer cost of \$0.4 per user, we conclude that offering \pirdns as a service would be financially viable if the monthly subscription fee is set at \$5 or higher.

\section{Timing Attack Analysis}
\label{sec:appendix:timing-attack}

\begin{figure}[t]
    \centering
    \includegraphics[width=0.8\linewidth, clip=true]{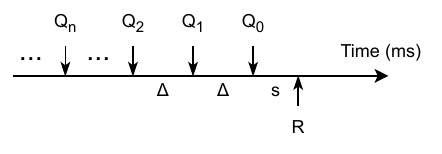}
    \vspace{-0.2in}
    \caption{Illustration of the modeling for delayed response forwarding.}
    \vspace{-0.15in}
    \label{fig:delay-response}
\end{figure}

\begin{figure*}[t]
  \centering
  \subfigure[{\centering Entropy as a function of $s$ with ${\mathbb{E}[X]=\Delta=31}$ms.}]{\psfig{figure=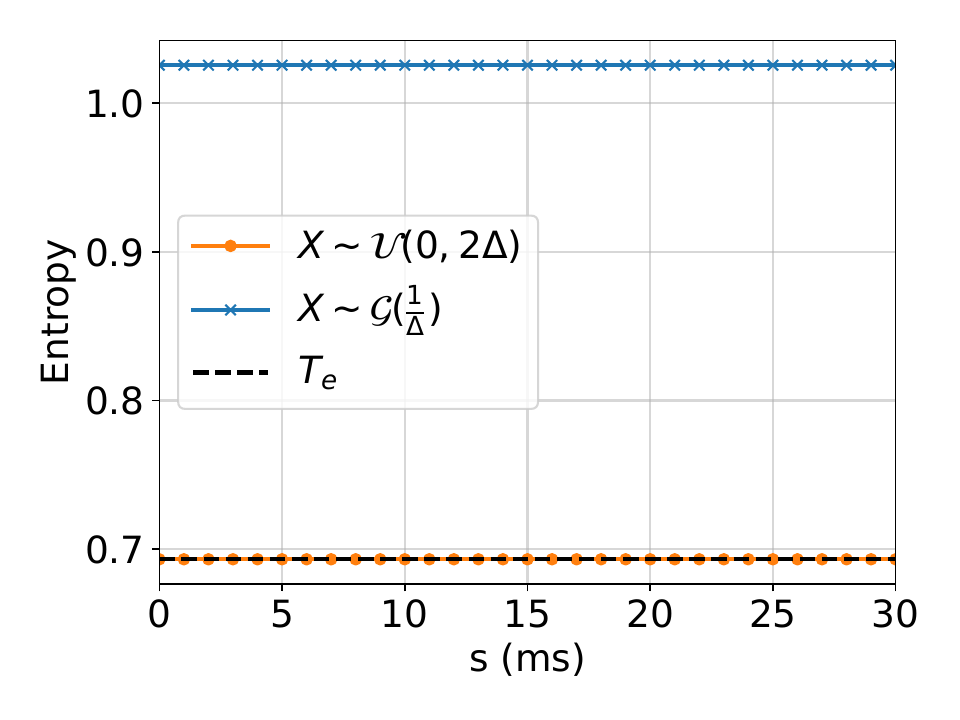, width=0.32\linewidth, trim=5mm 6mm 6mm 6mm, clip=true}\label{fig:entropy:s}}
  \hspace{0.2em}
  \subfigure[{\centering Entropy as a function of ${\mathbb{E}[X]}$ with ${\Delta=31}$ms.}]{\psfig{figure=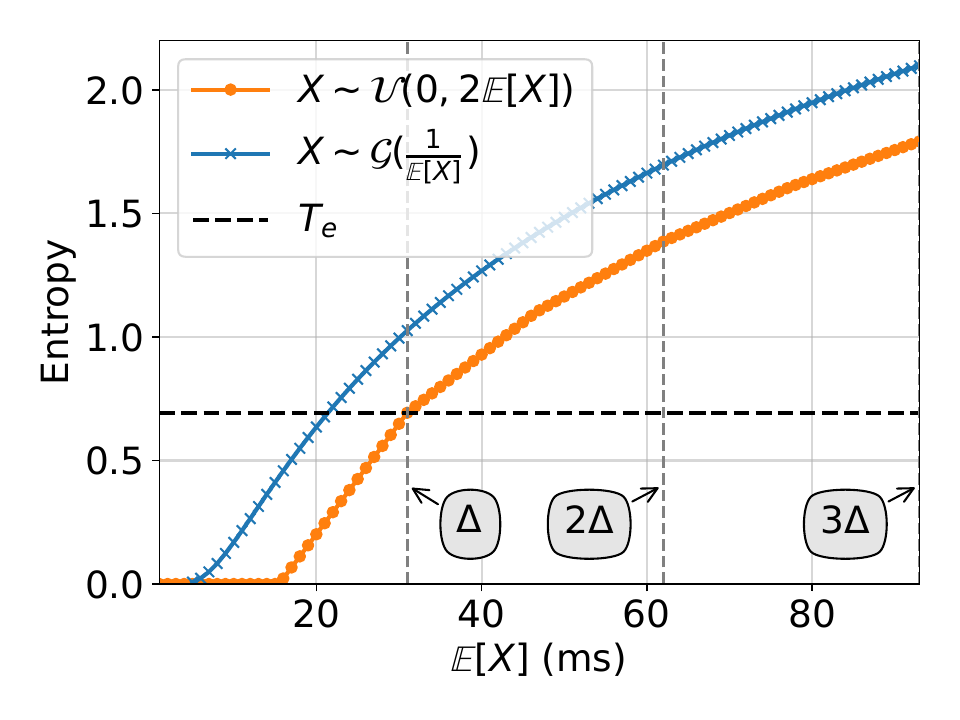, width=0.32\linewidth, trim=5mm 6mm 6mm 6mm, clip=true}\label{fig:entropy:ex}}
  \hspace{0.2em}
  \subfigure[{\centering Entropy as a function of ${\Delta}$ with ${\mathbb{E}[X]=\Delta}$.}]{\psfig{figure=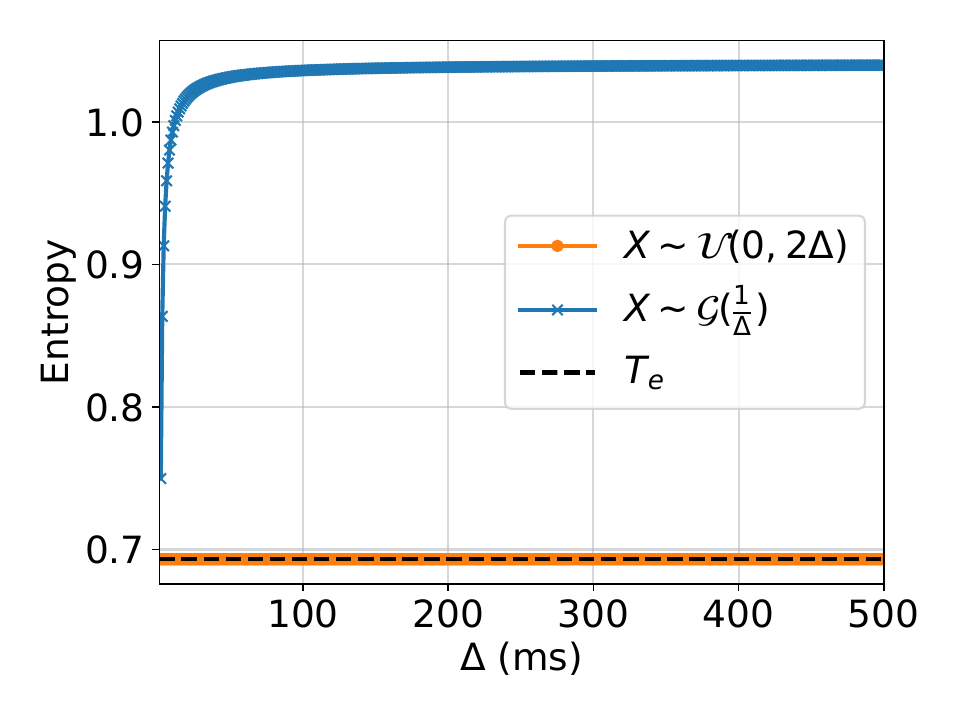, width=0.32\linewidth, trim=5mm 6mm 6mm 6mm, clip=true}\label{fig:entropy:delta}}
  \vspace{-0.18in}
  \caption{The effectiveness of delayed response forwarding to \pRR.}
  \vspace{-0.07in}
  \label{fig:appendix:entropy}
\end{figure*}

\noindent\textbf{Method} -- Assume an infinite series of queries with descending subscripts $\ldots, Q_{n}, Q_{n-1}, \ldots, Q_{1}, Q_{0}$ arrive at the \pRR, as shown in Figure~\ref{fig:delay-response}. Each pair of adjacent queries is made by different users. We assume that these queries finish with an equal interval $\Delta$, and each query has a probability of $m$ to trigger the cache miss. Let $Y_{i}$ be the random variable to denote whether a query $Q_{i}$ has triggered a cache miss. Then we have $\forall i \geq 0, {\sf Prob}(Y_{i}=1) = m, {\sf Prob}(Y_{i}=0) = 1 - m$. We further assume when a cache miss occurs, the response can be forwarded to \pRR instantaneously. However, to defend against timing attacks from \pRR, \ie correlating the forwarded response from ANSes and the queries where user identity is known to \pRR, the ANS should delay the response forwarding by a random duration. Let $X_{i}$ be the random variable of the delay of the response for query $Q_{i}$, sampled from distribution $\mathcal{D}$. For the sake of simplicity, we assume all ANSes have the same delay distribution. 

Consider a response $R$ arriving at time $s$ after $Q_{0}$. The probability that this response is for $Q_{i}$ is ${\sf Prob}(Y_{i} = 1 \cap X_{i} = s + i \cdot \Delta)$. Given that $X_{i}$ and $Y_{i}$ are two independent random variables, we have 
\begin{equation}
\begin{aligned}
    &\ {\sf Prob}(Y_{i} = 1 \cap X_{i} = s + i \cdot \Delta) \\
    = &\ {\sf Prob}(Y_{i} = 1) \cdot {\sf Prob}(X_{i} = s + i \cdot \Delta) \\
    = &\ m \cdot {\sf Prob}(X = s+ i \cdot \Delta).
\end{aligned}
\end{equation}

When ${\sf Prob}(X > \Delta) > 0$, meaning that the delay may be longer than $\Delta$, there are multiple possible queries for which $R$ might correlate to. It thus poses difficulty for the \pRR to perform the timing attack. To quantify the difficulty, we use Shannon entropy as the metric. Specifically, the entropy of a response arriving $s$~ms after $Q_{0}$ is the sum of the entropy of every past query correlating to the response, \ie
\begin{equation}
\label{eq:entropy}
\begin{aligned}
    {\sf Entropy}(s) = &\ \lim_{n \rightarrow \infty}\sum_{i=0}^{n} - \frac{{\sf Prob}(Y_{i} = 1 \cap X_{i} = s + i \cdot \Delta)}{\sum_{j=0}^{n}{\sf Prob}(Y_{j} = 1 \cap X_{j} = s + i \cdot \Delta)} \\
     &\ \cdot log\frac{{\sf Prob}(Y_{i} = 1 \cap X_{i} = s + i \cdot \Delta)}{\sum_{j=0}^{n}{\sf Prob}(Y_{j} = 1 \cap X_{j} = s + i \cdot \Delta)} \\ 
     = &\ \lim_{n \rightarrow \infty}\sum_{i=0}^{n} - \frac{{\sf Prob}(X_{i} = s + i \cdot \Delta)}{\sum_{j=0}^{n}{\sf Prob}(X_{j} = s + i \cdot \Delta)} \\
     &\ \cdot log\frac{{\sf Prob}(X_{i} = s + i \cdot \Delta)}{\sum_{j=0}^{n}{\sf Prob}(X_{j} = s + i \cdot \Delta)}
\end{aligned}
\end{equation}

Equation~\ref{eq:entropy} reveals a noteworthy point: the efficacy of the timing attack defense is not contingent upon the cache miss rate. This might seem counterintuitive. Nevertheless, the independence from the cache miss rate is rooted in the inherent inability of \pRR \ -- as guaranteed by the PIR -- to determine whether a query will indeed result in a cache miss.

\vspace{0.05in}
\noindent\textbf{Results} -- 
According to the findings in~\cite{schomp2016towards}, 1,000 users will perform queries at an average rate of 32~QPS, which translates to $\Delta = 31$ms. The cache miss chance for each query is 0.33, \ie $m=0.33$. Below, we calculate the entropy for a response arriving at $s$~ms after $Q_{0}$, where $s$ is an integer and $0 \leq s < \Delta$. 

To render the outcomes more comprehensible, we introduce a threshold denoted as $T_{e}$, which signifies the point at which the timing attack defense is considered effective. This effectiveness manifests when \pRR fails to differentiate between a response originating from two separate queries, both of which have an equal likelihood. In precise terms, we set $T_{e} = -\frac{1}{2}\log\frac{1}{2} -\frac{1}{2}\log\frac{1}{2} \approx 0.69$, delineating a criterion for the defense mechanism's adequacy. 

We consider two distributions in our evaluation: the uniform distribution $\mathcal{U}$ and geometric distribution $\mathcal{G}$. When $X \sim  \mathcal{U}(0, d)$, we have
\begin{equation}
    {\sf Prob}(X_{i} = k) = \left\{
    \begin{aligned}
        & \frac{1}{d}, \ \ k \leq d \\
        & 0, \ \ \text{otherwise}
    \end{aligned}
    \right.,
\end{equation}
where the average delay time is $\mathbb{E}[X] = \frac{d}{2}$.
When $X \sim  \mathcal{G}(p)$, we have 
\begin{equation}
    {\sf Prob}(X_{i} = k) = (1-p)^{k-1} \cdot p,
\end{equation}
where the average delay time is $\mathbb{E}[X] = \frac{1}{p}$. We leave other potential distributions as future work.

First, we explore the entropy across different $s$ ranging within the interval $[0, \Delta)$. In this experiment, we fix the $\Delta=31$ms. For both the uniform and geometric distributions where the delay is sampled from, we adjust the parameters so that $\mathbb{E}[X] = \Delta$.  Figure~\ref{fig:entropy:s} shows that the entropy of geometric distribution is much higher than our threshold $T_{e}$, whereas that of uniform distribution is equal to $T_{e}$. 
It further shows that the entropy for both distributions is barely changed given different $s$. This means that the \pRR will have roughly the same difficulty in correlating the response to the queries whenever the response arrives, illustrating the stable effectiveness of the delayed response. 

Next, we investigate the relationship between the average delayed duration $\mathbb{E}[X]$ and the entropy. We fix the $\Delta=31$ms, and calculate the average entropy across all $s \in [0, \Delta)$ for each $\mathbb{E}[X]$ value.
Figure~\ref{fig:entropy:ex} illustrates that the entropy for both distributions increases as the $\mathbb{E}[X]$ increases, where the geometric distribution is overall better than the uniform distribution. When $\mathbb{E}[X] \geq \Delta$, the entropy for both distributions is greater than or equal to $T_{e}$, demonstrating the effectiveness of the defense.

Finally, we investigate how $\Delta$ affects the defense effectiveness of delayed response forwarding. We adjust $\mathbb{E}[X] = \Delta$ and calculate the average entropy across all $s \in [0, \Delta)$ for each $\Delta$ value.
Figure~\ref{fig:entropy:delta} shows that the entropy is relatively stable regardless of the value of $\Delta$. 
This means that the efficacy of the defense mechanism is not notably impacted by the variation in $\Delta$. Instead, the crux lies in the proper selection of the average delayed duration. Remarkably, the frequency at which queries are made does not exert a substantial influence on the defense's effectiveness, provided that the average delay duration is thoughtfully determined. Consequently, the query frequency primarily functions as a determinant for determining the average delay duration itself.

While our model simplifies the intricacies inherent in real-world \pirdns scenarios, it undeniably underscores the potency of incorporating delayed response forwarding to \pRR as a robust mechanism to counter timing attacks. In actual deployments, ANSes would opt for diverse delay-sampling distributions. Moreover, the response delay would naturally incorporate random network-related delays. This layered complexity would invariably heighten the challenge for \pRR attempting timing attacks, thereby reinforcing the efficacy of the defense strategy.
We leave further explorations as future work.

\end{document}